\documentclass[10pt,aps,prd,twocolumn,showpacs,nofootinbib,eqsecnum,amsmath,amssymb]{revtex4-1}
\usepackage{epsf}
\usepackage[dvipsnames]{xcolor}
\usepackage[colorlinks]{hyperref}
\hypersetup{
	colorlinks=true, 
	linkcolor=blue,  
	citecolor=magenta,
	urlcolor=blue	
}

\usepackage{dcolumn}
\usepackage{bm}

\usepackage{dcolumn}
\usepackage{bm}
\everymath{\displaystyle}

\DeclareMathVersion{monobold}
\SetSymbolFont{letters}{monobold}{OML}{wcrum}{b}{n}
\SetSymbolFont{operators}{monobold}{OT1}{wcrr}{b}{n}
\SetSymbolFont{symbols}{monobold}{OMS}{wcry}{b}{n}

\DeclareMathSymbol{\singleGlyph}{\mathord}{symbols}{5}

\usepackage{cmap}

\usepackage{amssymb}
\usepackage{amsmath}
\usepackage{bm}
\usepackage{graphicx}
\usepackage{mathtools}
\newcommand{\matr}[1]{\mathbf{#1}}

	\def\exp{\mathop{\rm exp}\nolimits}

	\def\exp{\mathop{\rm exp}\nolimits}

	\def\exp{\mathop{\rm exp}\nolimits}
	
\begin{document}
	
	
\title{General relativity effects in precision spin experimental tests of fundamental symmetries} 
	
\author{\firstname{Sergey N.}~\surname{Vergeles}}
\email{vergeles@itp.ac.ru}
\affiliation{Landau Institute for Theoretical Physics, Russian Academy of Sciences, Moscow Region, Chernogolovka 142432, Russian Federation, \\
Moscow Institute of Physics and Technology, Department
of Theoretical Physics, Moscow region, 141707 Dolgoprudny, Russian Federation }

	
\author{\firstname{Nikolai N.}~\surname{Nikolaev}}
\email{nikolaev@itp.ac.ru}
\affiliation{Landau Institute for Theoretical Physics, Russian Academy of Sciences, Moscow Region, Chernogolovka 142432, Russian Federation, \\
Moscow Institute of Physics and Technology, Department
of Theoretical Physics, Moscow region, 141707 Dolgoprudny, Russian Federation }
	
\author{\firstname{Yuri N.}~\surname{Obukhov}}
\email{obukhov@ibrae.ac.ru}
\affiliation{Nuclear Safety Institute,
Russian Academy of Sciences, ul. Bolshaya Tulskaya 52, 115191 Moscow, Russian Federation }
	
\author{\firstname{Alexander J.}~\surname{Silenko}}
\email{alsilenko@mail.ru}
\affiliation{
Joint Institute for Nuclear Research, Bogoliubov Laboratory of Theoretical Physics, ul. Joliot-Curie 6, 141980 Dubna, Moscow Region, Russian Federation \\
Institute of Modern Physics, Chinese Academy of Sciences, Nanchang Rd. 509,  Lanzhou, 730000, China\\
Research Institute for Nuclear Problems, Belarusian State University, ul. Bobruiskaya 11, 220030 Minsk, Belarus}

\author{\firstname{Oleg V.}~\surname{Teryaev}}
\email{teryaev@jinr.ru}
\affiliation{Joint Institute for Nuclear Research,  Laboratory of Theoretical Physics and Veksler and Baldin
	Laboratory of High Energy Physics, ul. Joliot-Curie 6,
	141980 Dubna, Russian Federation \\
	National Research Nuclear University MEPhI, Kashirskoe Shosse 31, 115409 Moscow, Russian Federation\\
 Dubna University, ul. Uniuversitetskaya 19, 141980 Dubna, Moscow Region, Russian Federation }

\begin {abstract}
A search for the $P$- and $CP(T)$-violating electric dipole moments (EDM) of atoms,  particles, and nuclei with sensitivity up to $10^{-15}$ in units of the magnetic dipole moments, allowed by all discrete symmetries, is one of the topical problems of modern physics. According to Sakharov, $CP$ violation is one of the three key criteria of the baryogenesis in the generally accepted paradigm of the Big Bang cosmology. All three criteria are supported by the Standard Model, but it fails to describe quantitatively the observed baryon asymmetry of the Universe. This is regarded as a strong argument in favor of the existence of $CP$-symmetry breaking mechanisms beyond the minimal Standard Model, which can lead to measurable EDMs of atoms, particles, and nuclei. Searches for the EDM via the spin rotation in electric fields are currently underway in dozens of laboratories worldwide. Direct searches for the EDM of charged particles and nuclei are possible only in storage rings (COSY, NICA). After successful studies by the JEDI collaboration at the COSY synchrotron, at the forefront in the field is the search for the proton EDM in an electrostatic storage ring with the proton spin frozen at the magic energy with the projected sensitivity $d_p\sim 10^{-29}\,e\cdot$cm. A prototype PTR storage ring is proposed as a precursor to such a dedicated storage ring, with the prospect of the  frozen proton spin ring becoming a part of the physics at CERN beyond the Large Hadron Collider program. Following a brief introduction to  $CP$-violation physics and baryogenesis, the review presents a detailed discussion of significant contributions to the spin dynamics from the terrestrial gravity along with new effects of Earth's rotation in ultrasensitive  searches for the EDM of charged particles and neutrons. Quite remarkably, for the projected sensitivity to the proton EDM, these false EDM effects can exceed the signal of the proton EDM by one to two orders of magnitude, and become comparable to the EDM contribution in experiments with ultracold neutrons. We also discuss the role of a precessing spin as a detector of axion-like dark matter, and consider applications of quantum gravitational anomalies to the dense matter hydrodynamics and spin phenomena in non-central nuclear collisions.
\end{abstract}


\maketitle

\tableofcontents

\section{Introduction}\label{intro}

The gravitational interaction is the weakest one of those discovered in the microworld and macroworld. Its role in the high-energy processes is negligible at the available energies. The scale of energies at which gravity becomes significant is determined by the Planck mass, 
\begin{equation}
M_{\rm  P} = \sqrt{\frac{\hbar c}{G_N}} = 1.22 \cdot 10^{19}\ {\rm  GeV}/c^2\, ,
\end{equation}
where $G_N$ is Newton's constant, $c$ is the speed of light. All the more it's interesting that gravitational effects turn out to be quite appreciable in precision experiments, and the discussion of this new aspect of the particle physics will be the main subject of this review.

First of all, we have in mind the search for new mechanisms of violation of the combined $CP$ parity proposed by Landau in 1956 \cite{LandauEDM} (hereinafter, $P$ means spatial inversion, $C$ is the charge conjugation operation, i.e., the transition from particles to antiparticles, $T$ is the time reversal operation). As pointed out by Ioffe, Okun, and Rudik, by virtue of the $CPT$ theorem, the $CP$ noninvariance implies the simultaneous violation of the $T$
invariance in the particle physics \cite{IoffeOkunRudik}. The $CP$ violation was discovered experimentally in 1964 in decays of neutral $K$-mesons \cite{CroninCP}.

In the minimal Standard Model (SM) of electroweak interactions accepted today, the entire set of available experimental data on the $CP$ violation in particle decays can be described in terms of one parameter, the non-zero irremovable phase of the unitary $3\times 3$ Cabibbo-Kobayashi-Maskawa (CKM) matrix of quark mixing in weak currents \cite{Cabibbo,KobayashiMaskawa}. Corroborating the theoretical predictions \cite{AzimovCP1,AzimovCP2,BigiSanda}, the decays of particles with the beautiful $b$-quarks turned out to be especially rich in the detectable $CP$ violation, see reviews \cite{BondarPakhlov,EidelmanUFN} and the recent result of the LHCb collaboration \cite{LHCbCP}.

Despite the success of the Kobayashi-Maskawa (KM) mechanism, a search for deviations from the SM in the description of the $CP$ nonconservation remains one of the most topical problems. The point is that the SM is unable to explain one fundamental observable -- the density $n_B$ of the observed baryon matter in the Universe. When normalized to the cosmic microwave background radiation density, it is equal to \cite{AghanimPlanck} 
\begin{equation}
\eta_B = \frac{n_B}{n_\gamma} = (6.12 \pm 0.04)\times 10^{-10}\,. \label{eq:Bdensity}
\end{equation}
The interaction of cosmic protons and high-energy nuclei in the interstellar medium and in the Earth's  atmosphere explains the observed antiproton fluxes \cite{CosmicAntiprotonsAMS,CosmicAntiprotonsTheory}, and there are convincing arguments against the existence of galactic antimatter clusters in the observable  Universe \cite{CohenRujulaGlashow}. In the framework of the modern Big Bang theory, the baryon density (\ref{eq:Bdensity}) satisfactorily explains the data on the nucleosynthesis of light nuclei during the first minutes of the expansion of the Universe, although one still needs to refine the cross sections for a number of reactions \cite{FieldsBBN,yeh2021impact}.

The main open issue concerns the baryogenesis proper in the Big Bang paradigm with the zero initial baryon charge of the Universe. The questions of the matter burnup in the Universe and the freeze-out of the residual density of particles with conserved charges on the example of quarks as stable particles was for the first time raised by Zeldovich, Okun, and Pikelner in the article \cite{ZeldovichOkunPikelner}, written well before the discovery of the cosmic microwave background radiation. In 1966 A.D. Sakharov made the first in the literature attempt to explain baryogenesis in terms of the particle physics. He formulated three fundamental conditions for the baryogenesis \cite{SakharovCP}: (i) the violation of the baryon charge conservation (while maintaining the difference between the baryon and lepton charges) (ii) the violation of the $C$ and $CP$ invariance, (iii) the absence of the thermal equilibrium at the stage of processes with the nonconservation of baryon charge and the $CP$ parity. One should add to this the survival condition for the initial baryon asymmetry at the stage of the equilibrium expansion of the Universe. Sakharov also raised the question of the possible decay of protons. As we will discuss below, the Kobayashi-Maskawa mechanism in the SM  is unable by itself to explain the baryon asymmetry of the Universe. In the framework of the minimal SM, baryogenesis is possible due to phase transitions in the Higgs sector and the topological nonconservation of the baryon charge during the expansion of the Universe \cite{Kuzmin1985,RubakovUFN}.

There is still no generally accepted explanation for the observed baryogenesis. The main conclusion is that there should exist mechanisms of the $CP$ nonconservation beyond the Kobayashi-Maskawa phase in the SM, and experimental searches are in order for more $CP$-odd effects, which may prove to be appreciably larger  than those expected in the SM. An example of such a $CP$-odd observable is a permanent electric dipole moment (EDM) of particles with spin. As noted by Landau, it is possible only if the $CP$ invariance is violated \cite{LandauEDM}. The observed EDM signal is spin rotation in an electric field. Allowed by all discrete symmetries, the magnetic moment of nucleons $\mu$ is of order of the nuclear magneton, $\mu_N = e\hbar/(2m_N) \approx 10^{-14}\,e\cdot c \cdot$cm (we use the SI unit system). Inherent to the KM mechanism is a change of the flavor of quarks in the $CP$-odd transitions. Therefore, a flavor-diagonal nucleon EDM appears only to the second order in the weak interaction, and the dimensional estimates give \cite{OkunUFN1966,ShapiroEDM,Khriplovich_Springer} 
\begin{equation}
d_N^{\rm  SM}= \eta_N^{\rm  edm}\frac{\mu_N}{c}\sim 10^{-7} \times 10^{-10} \frac{\mu_N}{c} \sim 10^{-31}  e\cdot{\rm cm}\,. \label{eq:CKM_EDM}
\end{equation}
Here the factor $10^{-7}$ is the characteristic scale of the amplitudes of $CP$-even flavor-changing transitions, and similarly $10^{-10}$ is the scale for the amplitudes of $CP$-odd decays of neutral $K$-mesons. A more detailed analysis of the neutron EDM in the KM model by Shabalin  gave still smaller $d_N^{\rm SM} \sim 10^{-32} e\cdot$cm \cite{Shabalin1978,ShabalinUFN}. In many models, the EDM of nucleons is possible already in the first order in the $CP$-odd weak interaction and it can be of the order of \cite{OkunUFN1966,Khriplovich_Springer,ChuppRMPEDM} 
\begin{equation}
d_N \sim 10^{-10} \frac{\mu_N}{c} \sim 10^{-24}\,e\cdot{\rm cm}\,. \label{eq:BSMEDM}
\end{equation}

Experimental searches for the  EDM are extremely diverse and range from neutrons to neutral diamagnetic and paramagnetic atoms and molecules, to molecular ions and charged particles (protons, deuterons, helions...).  In the hadronic sector, the highest sensitivity was achieved in direct searches for the neutron EDM, $|d_n| < 1.8\times 10^{-26}$ $e\cdot$cm \cite{PSIEDM}. Already this limit is a record one in high-energy physics in terms of the number of excluded $CP$-nonconservation models \cite{ChuppRMPEDM,WirzbaEDM2016}. An increase of sensitivity by one or two orders of magnitude up to $\eta_n^{\rm edm}\sim 10^{-14}$ \cite{Serebrov2015,n2EDM} is being actively discussed. In principle, the possibility of $d_p \gg d_n$ is not ruled out, so the search for the proton EDM in dedicated electrostatic storage rings is on the agenda with an even higher projected sensitivity up to 
\begin{equation}
d_p \sim 10^{-29}\ e\cdot{\rm cm}\, , \label{pEDMtarget}
\end{equation}
i.e., the relative sensitivity $\eta_p^{\rm edm}\sim 10^{-15}$ \cite{srEDM,AbusaifCYR,YannisHybrid,YannisSnowmass2022}.

Such an ambitious sensitivity to the EDM of a single particle has already been achieved in experiments with diamagnetic mercury atoms: $|d_{\rm Hg}|<7.4\times 10^{-30}\,e\cdot$cm \cite{199HgEDM}. Assuming that the EDM of the atom is entirely due to the EDM of the nucleus, and making use of the formalism \cite{DmitrievProtonEDM} to evaluate the EDM of nucleons from the EDM of nuclei, the authors interpret their result as an indirect restriction on the EDM of the neutron $|d_n| < 1.6\times 10^{-26}\,e\cdot$cm. The same result for the $^{199}$Hg nucleus with a new calculation of Schiff's nuclear moments gives $|d_n| < 1\times 10^{-26}\,e\cdot$cm \cite{FlambaumDzubaNuclearEDM}. In the case of molecules, strong intramolecular electric fields \cite{SandarsAtom,SandarsMolecule} play an important role. A search for the EDM of a paramagnetic thorium monoxide ThO molecule gave the result $d_{\rm ThO}=(4.3\pm 3.1_{stat} \pm 2.6_{syst})\times 10^{-30}\,e\cdot$cm \cite{ACME2018EDM}. If the EDM of the molecule were completely determined by the EDM of the electron, then this result would have entailed the upper bound $|d_e| < 1.1\times 10^{-29}\,e\cdot$cm. When compared to the Bohr magneton following Eq. (\ref{eq:CKM_EDM}), this corresponds to the remarkably small $\eta_e^{\rm edm} < 5.7 \times 10^{-19}$. Of special interest is the experiment with  $^{180}$Hf$^{19}$F$^+$ ions, confined in a radio-frequency electric Paul trap, with the result $d_e = (0.9\pm 7.7_{ stat} \pm 1.7_{syst})\times 10^{-29}\,e\cdot$cm \cite{CairncrossHfF}.
 
This trap technique \cite{CairncrossHfF} is not applicable to the charged particles ($p,d,^3$He) though. Here the EDM searches are possible only in storage rings, where the EDM interacts with either the electric field in the comoving system on the orbit of magnetic storage rings, or the electric fields which are parts of the confinement of particles on the orbit. The search for the proton EDM with the declared sensitivity (\ref{pEDMtarget}) requires a control of systematic background effects at the same level. The only 
accelerator in the world at which precision experiments for the spin dynamics are possible today is the COSY (COler SYnchrotron) synchrotron at the Institute of Nuclear Physics in J\"ulich (now part of GSI, Darmstadt). After the completion of the MPD (Multi Purpose Detector) program of studies of the superdense baryon matter in heavy ion collisions, and the subsequent launch of the SPD (Spin Physics Detector) program, the leadership will be taken over by the NICA (Nuclotron based Ion Collider fAcility) collider at JINR (Joint Institute for Nuclear Research, Dubna) with the beams of polarized protons and deuterons \cite{KekelidzeUFN,SPD:NICA,Filatov2020PRL,Filatov2020EPJ}.

The record-breaking accuracy results obtained at COSY by the JEDI (J\"ulich Electric Dipole moment Investigations) collaboration motivated a proposal  of the  PTR (Prototype Test Ring) storage ring   by the CPEDM (Charged Particle Electric Dipole Moments) collaboration with the participation of the European Organization for Nuclear Research (CERN) \cite{AbusaifCYR}. The PTR will be the first storage ring in the world with an electric bending of protons with the kinetic energy of 30 MeV. It will primarily be used for the study of systematic effects in the spin dynamics for this new class of accelerators. In particular, PTR will enable the first  test of the  operation of such storage rings with the concurrent  clockwise (CW) and counterclockwise (CCW) beams rotating on the same orbit. Experiments at the PTR can be sensitive to the proton EDM down to $d_p \sim 10^{-24}\,e\cdot$cm \cite{AbusaifCYR}. In addition, the PTR is designed to operate at the energy 45 MeV with the hybrid electric and magnetic bending, with the  first implementation of the frozen spin mode. The PTR storage ring is important as a prologue to construction of a dedicated purely electrostatic proton storage ring with the spin frozen at the kinetic energy of 233 MeV to search for the proton EDM with sensitivity (\ref{pEDMtarget}) in the framework of the post-LHC (Large Hadron Collider) program of physics beyond the Standard Model at CERN. The document \cite{AbusaifCYR} is published by CERN as a monograph in the CERN Yellow Reports: Monographs series.

Terrestrial laboratories are located in the gravitational field of the rotating Earth. For protons with $\eta_p^{\rm edm} \sim 10^{-15}$, the EDM-induced spin angular velocity in the frozen spin electrostatic storage ring would be equal \cite{AbusaifCYR} 
\begin{equation}
\Omega_s \sim 10^{-9}\,{\rm rad/s}\, . \label{SpinBuildup}
\end{equation}
Let us cite the typical gravitational parameters for laboratories on the Earth with the radius $R_{\oplus}=6.378 \times 10^8\,$cm, rotating with the angular velocity of $\omega_\oplus=7.3\times 10^{-5}\,$rad :
\text{color}
\begin{itemize}
\item  the relative gravitational radius of the Earth is 
\begin{equation}
\eta_g=\frac{r_g}{R_\oplus}=\frac{2 G_N M_{\oplus}}{R_{\oplus}c^2}= 1.39\times 10^{-9} \, , \label{eq:GravRad}
\end{equation}
\item the equatorial velocity of the Earth's rotation in the units of the speed of light is 
\begin{equation}
\eta_{\oplus}=\frac{\omega_\oplus R_{\oplus}}{c}=1.55\times 10^{-6}\,,
\end{equation}
\item in a proton storage ring with the radius of $\rho \sim 80\,$m, the velocity on the orbit of the ring $\omega_\oplus\rho$ due to the rotation of the Earth is of the order of
\begin{equation}
\eta_\omega = \frac{\omega_\oplus\rho}{c} \sim 2\times 10^{-11}\,. \label{eq:EtaOmega}
\end{equation}
\end{itemize}

These small parameters are by no means negligible as compared to the relative value $\eta_p^{\rm edm}\sim 10^{-15}$ of our interest.

The influence of the gravity on the spin precession can be divided into direct and indirect effects. The direct effect is the geodetic precession of a classical gyroscope predicted by de Sitter in 1916 \cite{deSitter}. A century after de Sitter, the authors of this review were the first to point out an indirect gravitational effect of an immediate importance for the spin experiments with charged particles in storage rings \cite{SilenkoTeryaev2006}. Namely, one needs the focusing electromagnetic fields to compensate the gravitational attraction of the Earth in order to keep particles on a closed orbit. A contribution of these focusing fields to the spin precession proves to be comparable in magnitude to the de Sitter precession \cite{ObukhovSilenkoTeryaev}. In the planned all electric frozen spin proton rings, coupling of the proton magnetic moment to focusing fields produces the spin precession corresponding to a false EDM signal with $\eta_{\rm fake}^{\rm edm}\sim 2\times 10^{-14}$, significantly exceeding the projected sensitivity $\eta_{p}^{\rm edm}\sim 10^{-15}$, \cite{OrlovGravity,NikolaevFerrara,AbusaifCYR}. Remarkably, once the $CP$-odd EDM signal is separated from the $CP$-even contribution from gravity, the latter would become a  unique calibrating signal to identify  systematic effects in the search for the EDM.

A search for the $CP$-forbidden spin precession in an electric field with the sensitivity to $\eta_p^{\rm edm} \sim 10^{-15}$ requires a corresponding suppression of the spin precession in the background magnetic fields. From the point of view of an observer from distant stars, the static electric charges in the terrestrial laboratory rotate together with the Earth, creating currents and magnetic fields. Would a purely electrostatic laboratory be free from these magnetic fields for an observer resting in a terrestrial laboratory? According to \cite{VergelesJETP,VergelesJHEP}, such a geometric magnetic field proportional to the angular velocity of the Earth's rotation and electric field in the laboratory is possible. The peculiar feature of this magnetic field is a reversal of its sign upon the inversion of the electric field, so that the coupling of the magnetic moment with the geometric magnetic field imitates the interaction of the EDM with the electric field. In the proposed in 1968 by F.L. Shapiro approach to the search for the EDM of ultracold neutrons \cite{ShapiroEDM}, see also \cite{OkunUCN}, the false EDM signal can become significant at $d_n \sim 10^{-27}\,e\cdot$cm \cite{ VergelesJHEP}, i.e., already in the next generation of experiments on the neutron EDM \cite{n2EDM}. Already these two examples raise the role of gravity in the particle physics from the realm of purely academic discussions to the category of effects essential in the laboratory experiments.

A novel development in the subject is the use of the spin precession as a highly sensitive resonance detector of cosmic axion-like particles \cite{GrahamAxion2011,GrahamAxion,BudkerAxion,SikivieAxion}. Axions, like pseudo-Goldstone particles, and axion-like ultralight particles, are widely discussed as a plausible candidate for the dark matter (the search for weakly interacting massive dark matter particles is analysed in a recent review \cite{aleksandrov2021search}). Galactic field of the axion-like particles induces an oscillating EDM of atoms, molecules, and particles and simultaneously gives rise to  an oscillating pseudomagnetic field. The observed signal of axions will be an NMR-like rotation of the spin, provided the axion field oscillations are in resonance with the spin idle precession \cite{AbelAxion,PretzAxion,SikivieAxion}. We will also discuss the new interesting ideas on applications of the formalism of the gravitational quantum anomalies to the hydrodynamics of dense matter, formed in the non-central collisions of ultrarelativistic nuclei. Of particular interest here are the consequences for the polarization of produced particles, which can be studied at the NICA collider.

The further presentation is organized as follows. The review begins with two introductory sections devoted to an overview in Sec.~\ref{CPEDM} of the physics of the $CP$-{nonconservation} and consequences for the EDM of particles, and a discussion of baryogenesis in Sec.~\ref{baryon}. The principal conclusion from these necessarily brief sections is the incompleteness of the Standard Model and the relevance of the high-precision searches for new mechanisms of the $CP$ violation in spin experiments. 

We turn to the main topic of the review in Sec.~\ref{Spinclassic} on the dynamics of a classical spinning particles in external fields. Sec.~\ref{SpinQuantum} is devoted to the quantum spin dynamics based on the Foldy-Wouthuysen representation in external fields. In Sec.~\ref{spinSR} we discuss the derivation of gravitational corrections to the spin dynamics in cyclic accelerators. The role of gravitational corrections in the search for the EDM of charged particles in the practically interesting frozen spin mode is considered in Sec.~\ref{frozen}. Here, a brief review of the achievements of the JEDI collaboration on the spin dynamics at the COSY synchrotron is presented and the physics program of the planned PTR electrostatic storage ring is reviewed. The PTR in its hybrid magnetic and electric option will provide the  first ever implementation of the frozen proton spin regime. In Sec.~\ref{axion}, we focus on the use of the spin of particles in a storage ring as a detector of cosmic axion-like dark matter with an eye on experiments at NICA and PTR. The geometric magnetic field in electrostatic systems on the rotating Earth and its role in high-precision searches for the EDM of neutrons and charged particles are considered in Sec.~\ref{geo}. Finally, Sec.~\ref{anomaly} is devoted to applications of the formalism of the gravitational quantum anomalies to description of the hydrodynamic evolution of dense matter in ultrarelativistic heavy ion collisions. The concluding Sec.~\ref{conc} summarizes the main results. 

\section{$CP$ violation and electric dipole moments in Standard Model}\label{CPEDM}

Our presentation in this section will focus on the EDM of particles (electrons, nucleons, deuterons). The discussion of the subtleties of interpreting the data on the EDM of atoms and molecules in terms of the EDM of atomic electrons and nuclei, with an account of Schiff's shielding \cite{SchiffEDM}, and, in turn, the interpretation of the EDM of heavy nuclei in terms of the EDM of constituent nucleons, will be necessarily brief. To this end, we refer readers to a specialized review \cite{ChuppRMPEDM} and selected recent works \cite{FlambaumDzubaNuclearEDM,FlambaumPospelov,FlambaumSamsonovJHEP,FlambaumSamsonovInternucleon} with an extensive bibliography on the subject. 

\subsection{Kobayashi-Maskawa mechanism}\label{Kobayashi}

The standard electroweak model is constructed as a gauge theory with the $SU(2)_L\times U(1)_Y$ symmetry, with three doublets of left leptons and three doublets of left quarks $(u,d)_L,\ (c,s)_L, \ (t,b)_L$ (at this level, the quantum chromodynamic color symmetry of quarks is insignificant), with a doublet of complex scalar fields $\Phi$ and with right quarks and leptons in singlet representation\footnote{In sections \ref{CPEDM} and \ref{baryon} we use the system of units $\hbar = c =1$ as it is common in the high-energy physics.}. After the spontaneous symmetry breaking, the vacuum expectation value of the scalar field $\langle 0|\Phi|0\rangle = 246$ GeV appears, leaving the massive scalar Higgs particle $H$, the three vector mesons $W^+, W^-$ and $Z^ 0$ acquire masses and the photon with the electromagnetic gauge symmetry remains massless. The interaction $\propto \{ {\bar \Psi}_{L}\Phi \Psi_R + {\rm h.c.}\}$ of the initially massless quarks with a doublet of scalar bosons makes the quarks massive due to the vacuum expectation value $\langle 0|\Phi|0\rangle$ \cite{WeinbergSM,SalamSM}.

Grouping the quarks into the triplets $U_L = (u,c,t)_L$ and $D_L = (d,s,b)_L$, the weak interaction with the charged currents can be written as 
\begin{equation}
L_w = \frac{1}{\sqrt{2}}g_W W_{\mu}^+ \bar{U}_L V_{CKM} \gamma^{\mu} D_L \quad + \quad {\rm h.c.}\, ,
\end{equation}
where  $g_\text{W}$ is a coupling of $W$-boson to the isovector weak current, and $V_{CKM}$ is the $3\times 3$ unitary CKM quark mixing matrix. The CKM matrix admits one phase $\delta_{CKM}$, different from the zero and $\pi$, irremovable by unitary transformations. Such a complexity of the CKM matrix does not affect the renormalizability property of the electroweak interaction and leads to the $CP$-\textcolor{blue}{violation} in both semileptonic and non-leptonic weak decays of  strange and charmed particles and $B$-particles with the beauty $b$-quarks. To the first order in the weak interaction, all these are flavor changing weak transitions. An important consequence of the $CP$-{nonconservatyion} is the difference between the partial widths of decays of particles and antiparticles noticed by Okubo back in 1958 \cite{OkuboCP1958,OkunUFN1966,SakharovCP}.

In the rest frame, the Hamiltonian of the interaction of a particle with the spin $\bm{S}$ and the constant magnetic dipole moment (MDM) $\bm{\mu}= \mu\bm{S}/S$ and electric $\bm{d} = d \bm{S}/S$ dipole moments with an external electromagnetic field reads
\begin{equation}
H= -\,(\mu \bm{B}+d \bm{E})\cdot\bm{S}/S\, .
\end{equation}
The magnetic and electric fields $\bm{B}$ and $\bm{E}$ have opposite parities under both the time reversal ($T$) and the spatial reflection ($P$). The angular velocity of the spin precession is 
\begin{equation}
\Omega_s=  |\mu \bm{B}+d \bm{E}|/S \label{PrecessFreq}
\end{equation}
and the sought-for signal of a nonzero EDM is the change of $\Omega_s$ when the sign of the electric field is reversed.

In particle physics, the highest sensitivity to the EDM has been achieved in experiments with neutrons. The modern approach to the search for the EDM of ultracold neutrons (UCN) was laid down by F.L. Shapiro in 1968 \cite{ShapiroEDM}. The possibility of storing UCNs in storage cells was pointed out by Ya.B. Zeldovich in 1959 \cite{ZeldovichUCN}. A breakthrough in the EDM physics was the implementation in 1980 of the UCN approach at the Leningrad Institute of Nuclear Physics (now PNPI NRC Kurchatov Institute) at the WWR-M reactor in Gatchina, when the upper limit $d_n < 6\times 10^{-25}\,e\cdot$cm was obtained \cite{Altarev1980,Altarev1981} for the UCN storage with as yet modest UCN storage time of $\sim 5$ s. While this review was being written, in the experiment \cite{PSIEDM} with the UCN accumulation for 28 s and the subsequent spin precession in parallel and antiparallel electric and magnetic fields during 188 s, the sensitivity to the neutron EDM of $d_n < 1.8\times 10^{- 26}\,e\cdot$cm was reached. At present, the search for the neutron EDM is one of the main tasks of all laboratories worldwide where UCN are available; see \cite{ChuppRMPEDM,n2EDM} for a detailed history of neutron EDM searches and prospects for new experiments. The most promising are two-chamber UCN storage cells, practically free from systematic errors down to $d_n \sim 10^{-26}\,e\cdot$cm, developed by the A.P. Serebrov group at PNPI \cite{Serebrov2015}.

The magnetic moments of baryons are satisfactorily described as the sum of the magnetic moments of the constituent quarks \cite{BaryonMDM}. As pointed out by Shabalin in 1978, the KM mechanism of the $CP$ nonconservation predicts an extremely small quark EDMs, and the additive approximation would give $d_n \sim 10^{-34}\,e\cdot$cm \cite{Shabalin1978}. This was confirmed in a later paper \cite{Czarnecki} with a result for the EDM of the valence quarks 
\begin{equation}
\begin{split}
  d_{u} &\approx \frac{F_u}{108\pi^5} G_F^2 \alpha_S \bar{\delta} m_s^2 m_u
        \approx -0.15 \times 10^{-34} \ e\cdot\rm{cm}\, ,\\
    d_{d} &\approx \frac{F_d}{108\pi^5} G_F^2 \alpha_S \bar{\delta} m_c^2 m_d
  \approx -0.7 \times 10^{-34} \ e\cdot\rm{cm}\, .      \label{eq:quarkEDM}
\end{split}
\end{equation}
where $G_\text{F}$ is the Fermi weak interaction coupling, $\alpha_\text{s}= g_\text{s}^2/(4\pi)$, $g_\text{s}$ is the quantum chromodynamic (QCD) color charge, and the parameter of $CP$ parity violation is the Jarlskog invariant \cite{Jarlskog} 
\begin{equation}
{\bar{\delta}} = \rm{Im}\,(V_{us}V^*_{cs}V_{cb}V^*_{ub}) \approx 5\times 10^{-5}\, ,
\end{equation}
which does not change under unitary rotations in the quark basis. As dictated by the generalized Glashow-Illiopoulos-Maiani (GIM) mechanism \cite{IoffeShabalin1967,IoffeShabalin1968,Mohapatra,GIM}, the dimensionless constants $F_{u,d}$ are functions of the logarithms of the $b-, c-$, and $s$-quark  mass ratios and the ratio of the $W$-boson mass to the mass of the $b$-quark. This is a reflection of the fact that the $CP$ violation can be eliminated if there is a degeneracy of quark masses. Namely, the complete Jarlskog determinant is equal to 
\begin{equation}
\begin{split}
 J_{CP} = {\bar{\delta}}&\times (m_b^2-m_c^2)(m_b^2-m_d^2)(m_s^2-m_d^2)\\
&\times (m_t^2-m_c^2)(m_t^2-m_u^2)(m_c^2-m_u^2) \,, \label{eq:JarlskogCPV}
\end{split}
\end{equation}
but the specific observables, as in the above example of the EDM of quarks, include a truncated determinant.
  
Quantitatively, much more important are the essentially nonperturbative second-order multiquark mechanisms with weak interaction complemented by quantum chromodynamic (QCD) exchange currents, including the so-called penguin diagrams \cite{Penguin} with the gluon exchange between quarks in a nucleon, which give \cite{Shabalin1980,ShabalinUFN} 
\begin{equation}
d_n^{\rm  SM}\sim 10^{-32} e\cdot\rm{cm}\, . \label{eq:neutronEDM2}
\end{equation}
Similar results for the EDM of nucleons were obtained by Khriplovich and Zhitnitsky in their first evaluation of the hadronic nonperturbative large-distance contributions \cite{Khriplovich1982}, and in recent calculations of contributions of the one-loop meson-baryon diagrams with estimates of the $CP$-odd $\pi \Sigma N$ vertices from the chiral perturbation theory (see \cite{Seng2015} and the cited literature).
    
The valence quark EDM estimates (\ref{eq:quarkEDM}) should be treated as an appropriate illustration of the possibility of a strong difference between the EDM of the neutron and the proton, and thus as an illustration of the importance of the planned searches for the proton and deuteron EDM \cite{AbusaifCYR}. To this end, it is useful to remind  of the many open issues in our understanding of the spin structure of nucleons \cite{JiNature, Teramond,Efremov:1989sn}.

Let us recall that the $CP$-odd transitions in the CKM matrix are flavor-nondiagonal ones. Therefore, the EDM of leptons can only arise from the quark loop diagrams with the weak interaction to at least the second order. Just like in the case of quarks (\ref{eq:quarkEDM}), the EDM will be proportional to the lepton mass and the Jarlskog invariant. Omitting details, we give the commonly cited estimate for the electron EDM \cite{PospelovLeptonEDM} 
\begin{equation}
d_e \sim 10^{-44}\ e\cdot\rm{cm}\, . \label{eq:electronEDM}
\end{equation}
The contribution of the hadron loop large-distance corrections  has been discussed in a recent paper \cite{YamaguchiElectronEDM}. The effect of GIM cancellations and proportionality to the Jarlskog determinant are preserved in the contribution of the hadron loops, but the arguments are given in favor of small loop momenta, which can increase the electron EDM by 4 orders of magnitude as compared to (\ref{eq:electronEDM}). 

\subsection{$CP$ violation in quantum chromodynamics}\label{CP}

The $CP$ nonconservation in the SM is not limited to the KM mechanism in the electroweak sector. In the QCD sector of the strong interactions proper, a renormalizable $CP$-odd $\bar{\theta}$-term in the Lagrangian density is allowed,  
\begin{equation}
L_{\bar{\theta}} = -\,{\frac{1}{32\pi^2}}\,\bar{\theta} g_S^2\,G^{a\mu\nu} \tilde{G}^a_{\mu\nu}\,,\label{QCDCPV}
\end{equation}
where $\tilde{G}^a_{\mu\nu}=\frac{1}{2} \epsilon_{\mu\nu\rho\sigma}G^{a\rho\sigma}$ is the dual stress tensor of the octet of colored gluon fields $A^a_\mu$ with $a =1,\cdots,8$. In terms of the field strengths, the $\bar{\theta}$-term has the form of an explicitly $P$- and $T$-odd scalar product of the electric and magnetic fields $\propto (\bm{E}\cdot \bm{B})$ (the analogy is appropriate here with the  electrodynamics of gyrotropic media \cite{ShapiroGirotropy,Fedorov1,Fedorov2,Agranovich}). It is noteworthy that the expression $G^{a\mu\nu} \tilde{G}^a_{\mu\nu}$ can be rewritten as the total derivative 
\begin{equation}
\begin{split}
  &G^{a\mu\nu} \tilde{G}^a_{\mu\nu} = \partial_\mu K^{\mu}\,, \\
  &K^\mu = \epsilon^{\mu\nu\rho\sigma} \left(  A^a_\nu G^a_{\rho\sigma} -
  \frac{1}{3} g_s f^{abc} A^a_\nu A^b_\rho A^c_\sigma \right)\, , \label{Chern-Simons}
\end{split}
\end{equation}
where $K^\mu$ is the topological textcolor{magenta}{Chern-Simons-Pontryagin} current. Consequently, in the framework of the perturbation theory under the usual assumption that the fields disappear fairly fast at infinity, the $\bar{\theta}$-term can be omitted, and the problem of the $CP$-{nonconservation} in QCD would not exist at all.
  
Everything was changed with the discovery by Belavin, Polyakov, Schwartz, and Tyupkin (BPST) of the instanton nonperturbative solutions of the QCD equations of motion \cite{Instanton}, initially called pseudoparticles, which correspond to the topologically nonequivalent vacua \cite{HooftBarrier2,JackiwRebbi,Drinfeld}. Referring for a pedagogical introduction into the subject to the review \cite{NovikovUFN}) and the textbook \cite{RubakovClassicalFields}, we recall only the basic facts.
  
By the Gauss theorem, the contribution of the $\bar{\theta}$-term to the action in the Euclidean space can be rewritten as a flux of the current $K^\mu$ through the three-dimensional hypersphere $S_3$ 
\begin{equation}
\int d^4 x\,G^{a\mu\nu} \tilde{G}^a_{\mu\nu} = \int d^4 x\,\partial_\mu K^{\mu}=\int_{S_3} d\sigma_\mu K^\mu\,,  \label{Kflux}
\end{equation}
where $d\sigma_\mu$ denotes an element of the hypersurface. In the temporal gauge, $A_0^a=0$, the instanton is a nontrivial self-dual solution of the Yang-Mills equations of the purely gauge form 
\begin{equation}
 A_i^a T^a =\frac{i}{g_s} U^{-1}\partial_i U 
\end{equation}
with a $t$-time independent gauge transformation matrix $U$, where $T^a$ are the generators of $SU(3)$. The one-instanton solution of BPST corresponds to the finite minimum $8\pi^2/g_s^2$ of the action \cite{Instanton}, a wider class of multi-instanton solutions was found by 't Hooft \cite{HooftBarrier1}, an algorithm for constructing solutions of a general form is given in \cite{Drinfeld}. The common term {\it instanton} emphasizes the point that in the Euclidean space these field configurations are localized in all four dimensions. The meaning of instantons is best clarified by an example of fields from the $SU(2)$ subgroup of the $SU(3)$ color group, when the flux (\ref{Kflux}) can be recognized as a mapping of the sphere $S_3$ in the 4-dimensional Euclidean space onto the sphere $S_3$ in isotopic space. The integer-valued winding number (mapping degree)
\begin{equation}
 \nu=\frac{g_s^2}{32\pi^2} \int_{S_3} d\sigma_\mu K^\mu  \label{Pontryagin}
\end{equation}
is the  Chern-Simons-Pontryagin (CSP) index, with the BPST solution corresponding to the winding number $\nu=1$. Since in the gauge $A^a_0=0$ we have $K^i=0$, then in the Minkowski space the mapping degree can be written as 
\begin{equation}
\begin{split}
 &\frac{g_s^2}{32\pi^2} \int d^4x\,\partial_\mu K^{\mu} 
 = \frac{g_s^2}{32\pi^2} \int d^4 x\,\partial_0 K^{0}\\
&=\frac{g_s^2}{32\pi^2} \left.\left[\int d^3\bm{x}\,K^{0}(t,{\bm x})\right]\right|_{t=-\infty}^{t=+\infty} =\nu\,. \end{split}
\end{equation}
This is interpreted as a tunneling between the periodic vacuum configurations of the pure-gauge fields with a change in the mapping degree $n(t=+\infty)-n(t=-\infty)=\nu$. The physical $\bar{\theta}$-vacuum is a superposition 
\begin{equation}
|\bar{\theta}\rangle = \sum_{n=-\infty}^{+\infty}\,e^{in\bar{\theta}}\,|n\rangle\,,
\end{equation}
which provides a definition of the angle $\bar{\theta}$ \cite{JackiwRebbi,NovikovUFN,RubakovClassicalFields}.
 
The first principles of QCD do not put any restrictions on $\bar{\theta}$. We note now that the $CP$-odd $L_{\bar{\theta}}$ is related to the generalization of the Adler-Bell-Jackiw anomaly \cite{AdlerAnomaly,BellJackiwAnomaly} to the unitary-singlet $U(1)_A$ axial current in QCD 
\begin{equation}
\partial_\mu J_A^\mu = -\frac{N}{32\pi^2} \bar{\theta} g_S^2 G^{a\mu\nu} \tilde{G}^a_{\mu\nu}  + 2i \bar{\Psi}_R
\matr{M} \Psi_L\, ,
\label{eq:Anomaly}
\end{equation}
where $\matr{M}$ is the quark mass matrix. In the general case, in accordance with the axial anomaly Eq. (\ref{eq:Anomaly}), one can use the chiral rotation of fermion fields $\psi\rightarrow\exp(-i\gamma_5 \rho)\psi$ to remove $L_{\bar{\theta }}$ in favor of the complex mass matrix of the current quarks 
\begin{equation}
M_{ab}= \delta_{ab}\,m_a\,e^{-i\bar{\theta}}\,.
\end{equation}

Referring to the original sources \cite{BaluniEDM,CrewtherEDM} for the further technical details, we only quote the explicit form of the $CP$-odd Lagrangian $L_{CPV}$ in the quark sector 
\begin{equation}
L_{CPV} = 3 m^*\bar{\theta}  (\bar{\Psi}i\gamma_5 \Psi)\, . \label{QCDCPVquark}
\end{equation}
If at least one of the quarks is massless, then the reduced mass (one can neglect the contribution of heavy quarks) 
\begin{equation}
m^* = \frac{m_u m_d m_s}{m_u m_d  +m_u m_s + m_d m_s } \approx
\frac{m_u m_d}{m_u  +m_d}
\end{equation}
vanishes, i.e., to eliminate the $CP$-{nonconservation} due to the QCD $\bar{\theta}$ term, it is sufficient to make a chiral rotation of the massless quark field only (see also the useful discussion in \cite{BsaisouEDM}). This yields the dimensional estimate of the EDM of nucleons \cite{CrewtherEDM,BaluniEDM,PospelovRitz} 
\begin{equation}
d_N\sim\bar{\theta}\frac{m^*}{\Lambda_{QCD}}\mu_N\approx \bar{\theta}\times 10^{-16}\,e\cdot\rm{cm}\,,
\end{equation}
where $\Lambda_{QCD}\approx$~330 MeV is the QCD scale \cite{Teramond}.

The estimate of the EDM of diamagnetic atoms and molecules in terms of the EDM of the nucleus requires a careful account of Schiff's mechanism of the shielding of the external electric field on the nucleus by the electron shell of the atom \cite{SchiffEDM}. A conversion of the upper limit on the EDM of the nucleus to the EDM of the constituent nucleons of the nucleus also contains its own uncertainties \cite{DmitrievNuclearEDM, DmitrievSchiffEDM}. With these reservations, the result for the EDM of the mercury atom $d_{\rm Hg} < 7.4\times 10^{-30}\,e\cdot$cm \cite{Hg199EDM} can be converted into the restrictions on the EDM of the neutron, $d_{n} < 1.6\times 10^{-26}\,e\cdot$cm, and of the proton $d_{p} < 2\times 10^{-25}\,e\cdot$cm. If there are no competing sources of the EDM, then following \cite{deVriesTheta, BsaisouTheta,Hg199EDM,FlambaumDzubaNuclearEDM}, the upper bound on the neutron EDM \cite{PSIEDM} can be interpreted as an anomalously low upper bound $\bar{\theta} \sim 10^{-10}$.

We started with the statement that QCD allows for the strong $CP$ violation with $\bar{\theta} \sim 1$ and ended with a mysteriously low upper limit $\bar{\theta} \sim 10^{-10}$. A possible solution to the riddle was proposed as early as 1977 by Peccei and Quinn \cite{PecceiQuinn1,PecceiQuinn2} and it has already been mentioned above: this is the existence of an exact $U(1)_{PQ}$ chiral symmetry in QCD when one of the quarks is massless. Namely, $\bar{\theta}$ in the Lagrangian (\ref{QCDCPV}) is replaced by a dynamical pseudoscalar field $a(x)$, 
\begin{equation}
\bar{\theta} \to \frac{1}{f_{(a)}} a(x)\, . \label{ThetaToAxion}
\end{equation}
After the spontaneous $U(1)_{PQ}$ symmetry breaking by instantons, $a(x)$ acquires a vacuum expectation value and a very light pseudo-Goldstone boson, called an axion, is generated. With an account of (\ref{ThetaToAxion}), axions interact with gluons, 
\begin{equation}
L_{a} = -\,{\frac{1}{32\pi^2}}\,{\frac{a(x)}{f_{(a)}}}\,g_S^2\,G^{a\mu\nu} \tilde{G}^a_{\mu\nu}\,,
\label{Laxion}
\end{equation} 
Soon after that, in the 1978 paper, Weinberg gave an estimate of the coupling constant of an axion with fermions in a gradient interaction of the dipole type 
\begin{equation} 
L_{a\bar{\psi}\psi} = -\,{\frac{1}{2f_{(a)}}}\,g_\psi
\,\overline{\psi}\gamma^\mu\gamma_5\psi\,\partial_{\mu} a(x)\,, \label{AxionFermion}
\end{equation}
with the dimensionless constant $g_\psi \sim 1$ which depends on the specific model, and related the mass of the axion to the constant $f_{(a)}$ \cite{WeinbergAxion} 
\begin{equation}
m_{(a)} \approx m_\pi \frac{f_\pi}{f_{(a)}}\frac{\sqrt{m_u m_d}}{m_u+m_d}\,,\label{AxionMass}
\end{equation}
where $m_\pi$ and $f_\pi$ are the pion mass and decay constant.

A discussion of different scenarios of the axion phase transition in an inflationary Universe, the question of the constant $f_{(a)}$ and the axion mass $m_{(a)}$, and the possible contribution of axions to the dark matter can be found in the recent comprehensive reviews with an extensive bibliography on subject \cite{DiLuzioAxion,SikivieAxion}. Here we only mention that the most discussed Kim-Shifman-Vainstein-Zakharov (KSVZ) \cite{SVZAxion,KimAxion} and Dine-Fischler-Srednicki-Zhitnitsky (DFSZ) \cite{EricAxion,DineAxion} models allow for $f_{(a)}$ as large as the Planck mass, making  ultralight axions invisible. DFSZ axions directly couple to leptons, while in the KSVZ option the axion-lepton coupling are possible only via radiative corrections. In Sec.~\ref{axion}, we will dwell in more detail into the specific use of the precessing spin as an antenna for the search of the relic axions and axion-like particles.

It is worthwhile to notice that $CP$-odd $L_{\bar{\theta}}$ is an isoscalar one. However, that does not entail the equality of proton and neutron EDMs, since the electromagnetic current operator contains both isoscalar and isovector components.  In the framework of the chiral perturbation theory, a natural realization of the $CP$-odd sector of the low-energy QCD appears in the form of the isospin conserving $P$- and $T$-odd $\pi NN$ vertex \cite{CrewtherEDM, Khriplovich1982}. Then the pion-nucleon loop would contribute to the EDM of the proton and neutron with an opposite sign. A more detailed analysis of consequences of the chiral perturbation theory for the EDM of both nucleons and deuterons and helions was carried out in \cite{WirzbaEDM,BsaisouEDM} with the conclusion that the modern theory is unable to reliably predict the ratio of the proton and neutron EDM. The arising $CP$-odd potentials \cite{Khriplovich_Springer,Ulf1995,BsaisouEDM,deVriesCPV} lead to the deviation of the EDM of light nuclei from the additivity of the EDM of the nucleons constituting the nucleus. Therefore, the searches for the EDM of both neutrons and protons, and light nuclei \cite{AbusaifCYR} are imperative for unraveling mechanisms of $CP$ violation.

The expectations laid on the lattice QCD calculations of the neutron and proton EDM were not met so far. As noted in \cite{AbramczykLattice}, a finite lattice spacing introduces chiral mixing akin to the above discussed chiral rotations in the quark mass matrix. For this reason the earlier lattice calculations of the EDM were not free of mixing with the magnetic moments of nucleons. In the modern calculations, this mixing is under better control. Still another problem is that in the lattice QCD  the nucleon EDM, as well as other static characteristics of the nucleon, are extracted from fitting a Euclidean time dependence of the corresponding lattice correlators by the sum of decaying exponents. In principle, the decrease of the nucleon contribution should be the slowest decaying one. A proximity of masses of nucleons and of the $\pi N$ continuum makes the background contribution from the continuum a non-negligible one \cite{LatticeEDM}. On the one hand, this is interpreted as an indirect confirmation of the adequacy of the chiral perturbation theory. On the other hand, it suggests that a reliable separation of the nucleon contribution and a lattice measurement of the nucleon EDM requires an increase of statistics by at least one more order of magnitude, since with the existing lattice data, the estimates for the nucleon EDM change by several times depending on the modeling of the contribution of the excited states by $N^*$ resonance or the $\pi N$ continuum \cite{LatticeEDM}.

The QCD-motivated non-renormalizable models of the $CP$-\textcolor{blue}{violation} are broadly  discussed in the literature. For instance, one can endow quarks with permanent chromoelectric dipole moments with an obvious interaction $\bar{\Psi}\sigma^{\mu\nu}\lambda_a\tilde{G}^a_{\mu\nu}\Psi$. Weinberg proposed \cite{WeinbergGluonEDM} the $CP$-odd three-gluon interaction $f_{abc}G^a_{\mu\nu}G^b_{\nu\rho}G^c_{\rho\mu}$, which for massive gluons would correspond to the chromoelectric dipole moment of the gluons. Such an interaction could arise, in the spirit of the Heisenberg-Euler Lagrangian in QED, as the low-energy limit of the loop diagrams with the heavy particles. For a detailed discussion of the possible contribution of such interactions to the EDM of nucleons and light nuclei, we refer to \cite{BsaisouEDM}.

\subsection{Beyond the Standard Model: extension of the Higgs sector and supersymmetry}\label{beyondSM}

Historically, the first   renormalizable gauge model of $CP$ violation beyond the SM was proposed by Weinberg in 1976 \cite{WeinbergCPV}. He introduced the $CP$-odd phase into the Higgs sector of the SM with the then two generations of quarks postulating an extension of the scalar field sector to the two Higgs boson isodoublets. In the modern version of the model, the lightest of the Higgs particles is identified with the Higgs boson discovered at the LHC. For the EDM of neutrons, Weinberg obtained the estimate $d_n \sim 3\times 10^{-24}\,e\cdot$cm (see also \cite{KhriplovichWeinbergCPV}). The model is interesting in that it can generate the baryon asymmetry of the Universe \cite{CohenBaryonAsymmetry}. The original version of the Weinberg model was repeatedly rejected, in particular, by restrictions on the neutron EDM (see \cite{McKellarWeinbergCPV} and the cited literature). Nevertheless, even in the two-doublet version of the Weinberg model, the parameter space with the strong compensation of the loop contributions to the neutron EDM is not ruled out, and tests of such a model at the LHC are proposed \cite{TwoHiggsCPV}.

Tests of the three-doublet Weinberg model of $CP$-violation in the $t$-quark decays were considered in \cite{IvanovTopCPV}. In the three-doublet Weinberg model, of particular interest are the $CP$-odd asymmetries in the rare radiative decays of $B$-mesons, which are quite large in the Weinberg model and negligible in the Kobayashi-Maskawa SM \cite{ThreeHiggsCPV}. In general, models with an extended Higgs sector can give a noticeable EDM of electrons due to the loop diagrams with $t$-quarks \cite{EspinosaMultiHiggs}. A very interesting discussion of the group-theoretic properties of the $CP$ nonconservation in the multi-doublet Higgs sector is presented in a series of publications by Ivanov (see \cite{IvanovMultiHiggs1,IvanovMultiHiggs2} and references therein).

As Weinberg \cite{WeinbergSummary1992} has emphasized in his summary report at the XXXI International Conference on High Energy Physics in 1992, ``Also endemic in supersymmetry theories are CP violations that go beyond the CKM matrix, and for this reason it may be that the next exciting thing to come along will be the discovery of a neutron or atomic or electron electric dipole moment''. In renormalizable supersymmetric (SUSY) theories, a finite EDM is possible due to the one-loop diagrams, and unlike the KM mechanism, the answer is of the same structure for quarks and leptons (see \cite{PospelovRitz} and the cited literature) 
\begin{equation}
d_i \sim \frac{g^2}{16\pi^2} \cdot\left( \frac{m_i}{\Lambda^2}\right)^2  \cdot \frac{e_i} {m_i} \sin\phi\,,
\end{equation}
where $g^2 \sim 1$ is the coupling constant (recall that the ordinary quantum chromodynamics is an integral part of the theory), $\Lambda$ is the mass scale of supersymmetric particles in the loop diagram, and in the penultimate factor one recognizes the magnetic dipole moment of the quark (lepton). As a matter of fact, this form of the result was anticipated by Berestetsky, Krokhin and Khlebnikov back in 1956 \cite{BeresteskyMDM} and subsequently confirmed by calculations of the electroweak correction to the magnetic anomaly of the electron and muon \cite{JackiwMDM,FujikawaMDM,BarsMDM}. We reiterate that the Kobayashi-Maskawa model of CP-\textcolor{blue}{violation} predicts very strong suppression of the EDM of leptons vs. the EDM of quarks, cf. the estimates (\ref{eq:neutronEDM2}) for nucleons and (\ref{eq:electronEDM}) for electrons.

For the average mass $m_{u,d} \sim 5$ MeV of the light current quarks, one finds an estimate for neutrons
\begin{equation}\label{eq:EDMSUSY}
d_n \sim 10^{-24}\left(\frac{1\ {\rm TeV}}{\Lambda}\right)^2 \sin\phi\,~e\cdot{\rm cm}\,.
\end{equation}
There are no compelling reasons for the smallness of the $CP$-violating phase $\phi$ (there may be several such phases in different supersymmetric models). Therefore, the experimental upper bound on the neutron EDM $d_n < 1.8 \times 10^{-26}\,e\cdot$cm \cite{PSIEDM} can be interpreted as a lower bound for the mass of supersymmetric particles of the order of $\Lambda >$ 7 TeV. The proposed proton EDM searches with a sensitivity of $d_p \sim 10^{-29}\,e\cdot$cm \cite{AbusaifCYR} could set the lower bound $\Lambda \sim 300$ TeV, so that the potential of the high-precision low-energy experiments could largely exceed the potential of the direct searches for new particles at colliders. Here it is worthwhile to note that hopes of theorists that the Large Hadron Collider (LHC) from the first days of operation will become a factory of supersymmetric particles with masses of hundreds of GeV did not come true. A belief in the forthcoming era of supersymmetry has been shaken by the persistent increase of the lower bound on masses of squarks and gluinos: the recent results of the ATLAS collaboration exclude gluinos with masses below 2.3 TeV and squarks with masses below 1.85 GeV \cite{ATLAS:SUSY}. Suppression of the predicted neutron EDM at such masses of  SUSY particles by the small $CP$-violating phase $\phi \sim 10^{-2}$ is as unattractive as by the aforementioned still larger and larger  masses $\Lambda$. 
		
For a more detailed discussion of the allowed parameter space in various SUSY models and in the Weinberg two-doublet model, as well as an extensive bibliography on the subject, we refer to the exhaustive review \cite{ChuppRMPEDM}. 

\subsection{Millistrong $CP$ violation beyond Standard Model}\label{CPbeyond}
 
In 1965, Okun \cite{OkunMillistrong}, Prentki and Veltman \cite{PrentkiMillistrong} and Lee and Wolfenstein \cite{LeeMillistrong} noticed that the $CP$-{nonconservation} observed in the system of neutral $K$-mesons can be explained if, alongside with the $CP$-invariant weak interaction, there existed a flavor-diagonal, $P$-even, but $T$-noninvariant  \textcolor{magenta}{and $C$-odd} millistrong interaction with the dimensionless constant $\sim 10^{-3}$. In a fundamental distinction from the SM, the dimensional estimates suggest the $T$-odd effects of this order of magnitude in a wide spectrum of nuclear and hadronic processes. In the recent years, the millistrong model of the $CP$-{nonconservation} has been little discussed by theorists since, due to its symmetry properties, it has not been implemented in renormalizable generalizations of the SM.

Numerous experimental searches for the millistrong $T$-{nonconservation} have been carried out in the $\beta$-decay of the neutron \cite{Mumm:2011nd}, in the nuclear $\gamma$-transitions of mixed multipolarity \cite{GimlettAtomicFSI}, in the comparison of cross sections for direct and inverse nuclear reactions \cite {BlankeDetBalance,WeidenmullerDetBalance}, in the comparison of polarization parameters in the initial and final states in nucleon-nucleon scattering \cite{DavisPPspin}, and in the search for the polarization null effect -- the $T$-forbidden spin asymmetry in the total cross section for polarized neutron scattering by a tensor-polarized nucleus \cite{HuffmanHolmium}.

A significant relative phase $(-4.7+0.3)\times 10^{-3}$ rad of the ratio of $M1$ to $E2$ amplitudes was found in the $\gamma$-transition with the energy of 129 keV in the $^{191}$Ir nucleus \cite{GimlettAtomicFSI}. However, the final state interaction of $\gamma$-quanta with atomic electrons  gives the phase $ -(4.3\pm 0.4)\times 10^{-3}$ 
\textcolor{magenta}{rad}\cite{GoldwireAtomicFSI}, so that the experimental data yield only the upper limit for the $T$-violating phase $< 0.9\times 10^{-3}$ rad. The role of the interaction with atomic electrons in the scattering of $\gamma$-quanta in the magnetized ferromagnets was reliably established by Lobashev et al. back in 1971 \cite{LobashevFaraday}.

In the nonperturbative phenomenology of the meson-baryon interactions, the $P$-even millistrong $NN$ interaction at low and intermediate energies is modeled by the $T$- and $C$-odd $\rho NN$ vertex in the exchange of charged $\rho$ mesons \cite{SimoniusRhoNN}, 
\begin{equation}
L_{\rho NN}^{TV} = i \sqrt{2} g_{TV} g_{\rho NN} \frac{k_V}{2m_n} \bar{N}\sigma^{\mu\nu}
(\tau^- \partial_\mu \rho^+_{\nu} -\tau^+ \partial_\nu \rho^-_\mu )N\,,\label{eq:TVrhoNN}
\end{equation}
where $\tau^\pm$ denote the isospin matrices and $g_{TV}$ is the reduced $T$-odd (TV) amplitude. In addition to the expected smallness of the coupling $g_{TV}\sim 10^{-3}$, one finds an extra numerical suppression in the contribution from such an interaction to the spin observables of the $NN$ elastic scattering at intermediate energies (see \cite{UzikovPD,UzikovNullTest} and cited papers). Even the high accuracy $(P - A)=0.0047\pm 0.0025_{\rm stat} \pm 0.0015_{\rm sys}$ \cite{DavisPPspin}, achieved in the standard verification of the equality of the analyzing power and the polarization of scattered protons in the $pp$ elastic scattering is as yet insufficient for a critical check of the millistrong model.

The millistrong $T$-nonconserving interaction would generate the EDM of the nucleon in conjunction with the flavor-diagonal $P$-odd weak interaction in the ballpark of the aforementioned dimensional estimate (\ref{eq:BSMEDM}). But in the spirit of the dimensional counting rules in the chiral perturbation theory, the $T$-odd and $P$-even quark-quark interaction belongs to the class of higher dimension interactions. According to Kurylov et al. \cite{SimoniusRhoNN}, the dimensional dressing analysis of the Simonius-type interaction in the spirit of the chiral perturbation theory leaves room for a strong suppression of the nucleon EDM as the low-energy parameter (\cite{Kurylov}, for another example of such a suppression see \cite{El-Menoufi:2016cfo}). 

With reservations about the uncertainty of evaluations of the $T$-odd nuclear optical potential, the experimental result $A_{TV}=(8.6\pm7.7)\times 10^{-6}$ for the $T$-odd vector-tensor asymmetry in the total cross section of interaction of a polarized neutron with the tensor polarized $^{199}$Ho nucleus, corresponds to ${g_{TV}} = (2.33\pm2.1)\times 10^{-2}$ \cite{HuffmanHolmium}. A similar numerical suppression of the $T$-odd null effect in the doubly polarized proton-deuteron scattering was found in \cite{UzikovNullTest, UzikovPD,UzikovPDTV}. Still the doubly polarized $pd$ scattering has a higher sensitivity to the $T$-{nonconservation} \cite{UzikovNullTest} and it is feasible to lower the upper bound on the vector-tensor asymmetry to $A_{TV} \sim 10^{-6}$ in an accelerator experiment with a polarized beam and an internal polarized target \cite{ValdauTV,EversheimTV,LenisaTIVOLI}, thus realizing the first crucial test of the millistrong model.
  
Experiments on $pd$ scattering with static polarizations are subject to systematic errors due to the hard-to-eliminate vector polarization in a tensor-polarized deuteron target. As pointed out in \cite{NikolaevPrecessingD}, in the inverse kinematics with deuteron polarization oscillating in the plane of the accelerator ring, the $T$-odd polarization null effect in the $pd$ interaction cross section has a unique Fourier component with twice a precession frequency of the vector polarization, and thus can be reliably extracted without any systematic effects. It's up to the experiment, which can be performed at the COSY accelerator in J\"ulich \cite{ValdauTV, EversheimTV, LenisaTIVOLI} or at the NICA accelerator complex of JINR \cite{SPD:NICA}. 

As far as the EDM is concerned, the flavor conserving millistrong interaction in conjunction with the $P$-odd component of the weak interaction, generates the $P$- and $T$-violating nucleon-nucleon interaction. The prototype of such an interaction, usually modeled by a scalar $\pi NN$ vertex, was introduced in \cite{Khriplovich1982,Flambaum1984NuclearTV} and became a standard one in the popular chiral perturbation theory \cite{Dekens,WirzbaEDM2016,BsaisouEDM,FlambaumPospelov}. Due to the $P$- and $T$-odd intranuclear $NN$ interaction, the EDM of nuclei would not reduce to the sum of the EDM of constituent nucleons, in close similarity to the effect of exchange currents in the case of magnetic moments. For the light nuclei, this is discussed in detail in \cite{BsaisouEDM}. The EDM of heavy nuclei can be enhanced by the proximity of the levels of the nucleus with opposite parity \cite{FlambaumKhriplovichSushkovNuclearTV,Flambaum1984NuclearTV}. An incomplete shielding of the nuclear EDM in atoms and molecules requires a difference between the nuclear charge and EDM  densities \cite{SchiffEDM,Flambaum1984NuclearTV,KhriplovichParity}. Both the Schiff shielding considerations and the parity degeneracy of the nuclear levels favor the nuclei with the octupole deformation \cite{FlambaumDzubaNuclearEDM,Auerbach:1996zd}. A selection of the optimal atoms and molecules from the point of view of the EDM signal remains an artful task \cite{FlambaumDzubaNuclearEDM,FlambaumSamsonovJHEP,FlambaumPospelov}. There exists an exhaustive review literature on the chiral perturbation theory for the $P$- and $T$-violating nuclear forces \cite{JordyTVPVreview,WirzbaEDM2016}, but the issue of extraction of the $P$-even and $T$-violating millistrong coupling constant from such an analysis remains as yet open. 

\section{Baryon asymmetry of the Universe}\label{baryon}

To properly assess the result (\ref{eq:Bdensity}) for the baryon asymmetry $\eta_B$, let us start with the question of survival of the baryon matter in a $CP$-invariant theory with the zero net baryon charge at the start of the Big Bang \cite{ZeldovichOkunPikelner}. The conservation of entropy allows for a reliable extrapolation of the ratio of the density of baryons and antibaryons to the density of relic photons and ensures its constancy during the expansion stage with the thermodynamic equilibrium \cite{DolgovZeldovichRMP1981,DolgovITEPSchool1997}. The mutual annihilation of baryons and antibaryons stops at the density \cite{ZeldovichBdensity1965,ChiuBdensity1966,DolgovITEPSchool1997} 
\begin{equation}
n_{\bar{B}} =n_B \approx \frac{n_\gamma}{\sigma_{ann} m_B M_P} \approx 10^{-19}\,\label{eq:NaiveBassimmetry}\,,
\end{equation}
where $\sigma_{ann}$ is the annihilation cross section at the freeze-out. To this we should add the problem of separation of matter and antimatter in the Universe, unsolvable in this scenario. The irrefutable conclusion is that the baryon asymmetry had to be generated already in the early Universe according to Sakharov's scenario.

A possibility of a purely electroweak baryogenesis within the framework of the well established interaction mechanisms and the known particle mass spectrum is undoubtedly quite attractive. As mentioned above, apart from the 125  GeV Higgs boson, no new particles have been discovered so far. As first noted by Kirzhnits and Linde in 1972, electroweak phase transitions \cite{KirzhnitsLinde} are expected in the SM. The Big Bang paradigm assumes that the Universe develops from the initial state with the zero net baryon charge and with unbroken $SU(2)_L\times U(1)_Y$ symmetry with the vanishing vacuum expectation value of the Higgs field. The original electroweak Lagrangian has the $U(1)_B$ symmetry with conserved baryon and lepton currents $J^\mu_B$ and $J^\mu_L$ \cite{WeinbergSM,SalamSM}. The so-called sphaleron baryogenesis proposed by Kuzmin, Rubakov and Shaposhnikov, is based on the topological baryon charge nonconservation at the stage of phase transitions in the Higgs sector \cite{Kuzmin1985, RubakovUFN}. These ideas go far beyond the scope of the minimal electroweak model per se  and are worthwhile of a brief exposition.

The subsequent presentation repeats, with minor modifications, the discussion of instantons in Sec.~\ref{CP}. In the electroweak SM \cite{WeinbergSM,SalamSM} with non-Abelian symmetry $SU(2)_L \times U(1)$ one has a periodic series of classical vacua with integer CSP mapping indices $N_{CSP}= 0, \pm 1, \pm 2 ,\cdots$ coinciding with the baryon charge, realized by the gauge nonequivalent instanton solutions of the classical Yang-Mills equations for isovector fields. In the temporal gauge $W_0^a=0$, and the corresponding mapping degree of the coordinate space  mapped onto the isospin space is equal to 
\begin{equation}\label{eq:CScharge}
N_{CSP} = \frac{1}{96\pi^2} \,g_W^2 \int d^3{\bm x}\,\epsilon^{ijk} \epsilon_{abc}W^a_i W^b_j W^c_k\,,
\end{equation} 
where $g_\text{W}$ is the Weinberg electroweak coupling and $\epsilon_{abc}$ is the totally antisymmetric Levi-Civita symbol.
The isovector gauge bosons $W_{\mu}^a$ interact only with the left-handed quarks and leptons, and the conservation of classical currents is violated by the Adler-Bell-Jackiw quantum anomaly 
\begin{equation}
\partial_\mu J^\mu_B = \partial_\mu  J^\mu_L =
-\,{\frac{n_F}{32\pi^2}}\,g_W^2 W_{\mu\nu}^a W_a^{\mu\nu}\,, \label{eq:ABJ}
\end{equation}
where $n_F$ is the number of fermions and $W_{\mu\nu}^a$ is the corresponding field strength tensor. Obviously, the difference between the baryon and lepton charges $B-L$ is conserved.
  
We are interested in the time dependence of the baryon charge $\Delta B(t)=B(t)-B(0)$ per fermion 
\begin{equation}\label{eq:dBdt}
\Delta B(t) =-\int_0^t dt' \int d^3{\bm x}\,{\frac{1}{32\pi^2}}\,g_W^2 W_{\mu\nu}^a W_a^{\mu\nu}\,.
\end{equation}
It is expressed in terms of the divergence of the current 
\begin{equation}
K^{\mu} = \frac{g_W^2}{16\pi^2} W_\nu^a\epsilon^{\mu\nu\rho\tau} \left(\partial_\rho W_\tau^a+
\frac{1}{3}g_W \epsilon_{abc}W_{\rho}^b W_{\tau}^c\right)\, . \label{eq:CScurrent}
\end{equation}
As shown in Sec.~\ref{CP}, the change of the baryon charge during a tunneling between different vacua is related to the CSP index  
\begin{equation}
\Delta B = \Delta N_{CSP}\,.
\end{equation}
The tunneling probability at the zero temperature and energy can be computed exactly \cite{HooftBarrier1,HooftBarrier2} 
\begin{equation}
w_{inst} \propto \exp\left(-\,{\frac{4\pi}{\alpha_W}}\right) \sim 10^{-164}\, ,\label{eq:Penetration}
\end{equation}
and, due to its extreme smallness, has no practical consequences.

In the early Universe, before the electroweak transition with the spontaneous breaking of the $SU(2)_L\times U(1)_Y$ symmetry by the Higgs mechanism, particles remain massless and there is no barrier. After the symmetry breaking during the expansion, a jump through the barrier with the Boltzmann factor $\exp(-E_{sph}/T)$ is possible by means of the thermal fluctuations \cite{Kuzmin1985}. The electroweak constants and masses of the electroweak vector bosons are known from experiment. Therefore, the height of the barrier at the unstable saddle point $E_{sph}$, described by the so-called sphaleron static solution with the half-integer \textcolor{magenta}{ $N_{CSP}$ } of the classical equations of motion \cite{MantonSphaleron}, as well as the critical temperature $T_c \sim 100$ GeV, both depend on a single parameter -- the self-coupling constant of the Higgs boson, i.e., on its mass. Such a minimalism makes the electroweak baryogenesis extremely attractive, and the ideas of the paper \cite{Kuzmin1985}, which gathered about 3000 citations, are still in the center of attention. The first decade of the development of the theory and the main scenarios of the electroweak phase transition are described in the classic review \cite{RubakovUFN}, the subsequent development of the approach is covered in the reviews \cite{BernreutherBaryogenesis,Dine:2003ax,BodekerBaryogenesis}.

The crucial issue is the degree to which Sakharov's non-equilibrium criterion is fulfilled, i.e.  the rate of processes with a change of the baryon charge in comparison to the rate of expansion. It is clear that at the beginning of the phase transition, the order parameter -- the vacuum expectation value of the Higgs field -- starts from zero and only with a further decrease of the temperature it takes its value for the zero temperature. From the point of view of Sakharov's criterion, an ideal scenario would be the highly nonequilibrium first order phase transition with the formation of seeds from fluctuations and then bubbles with a nonzero condensate. That looked realistic for light Higgs particles with a mass below 70 GeV, but it was ruled out already by the upper limit $m_H > 114$ GeV established at the LEPII electron-positron collider \cite{HiggsLEPII}, and even more so by the further discovery of the Higgs boson with the mass of 125 GeV at the LHC \cite{HiggsMass1,HiggsMass2}. The analysis of the Higgs mass region $m_H > m_W$ began back in 1996 with the pioneering work of Shaposhnikov \cite{Shaposhnikov1} -- this is a practically interesting smooth crossover mode. Basic points of the notable progress in the analytic understanding of this mode \cite{Shaposhnikov1,Shaposhnikov2,Shaposhnikov3} have  been confirmed by the the recent $32^3$ lattice  simulation of the crossover transition in the minimal Standard Model with the experimentally known mass of the Higgs boson \cite{DOnofriCrossover}. Here the dynamics of the CPS topological charge was studied in a three-dimensional effective model truncated by neglecting the insignificant contribution of the Abelian vector boson. A transition to the crossover was found to begin at $T_c = 159\pm 1$ GeV. The freeze-out of the baryon asymmetry begins at the temperature $T_* = 132\pm 2.3$ GeV.

As noted above, in the minimal SM with the Kobayashi-Maskawa mechanism, the $CP$ violation is proportional to the Jarlskog determinant (\ref{eq:JarlskogCPV}). This implicitly assumes that the momentum scale $F$ in loop diagrams for the $CP$-odd transitions is larger than the quark masses. Then one could  have taken \cite{ShaposhnikovCPV, RubakovUFN} 
\begin{equation}
\delta_{CP} \sim \frac{J_{CP}}{F^{12}}\,
\end{equation}
for a dimensionless measure of the $CP$ nonconservation in the phase transition region. In the high-temperature phase transition with $T_c \sim 100$ GeV, we have $F\approx T_c$, that would give $\delta_{CP} \sim 10^{-19}$, which is entirely insufficient to explain the observed baryon asymmetry (see also \cite{ExcludingCKM}). With allowance for  the complex dynamics of the nucleation of bubbles filled by the $CP$-odd phase and their subsequent percolation into the large bubbles during the phase transition, this qualitative estimate can well be amplified \cite{FarrarShaposhnikovPRDCKM}. One mechanism of amplification of the effects of the $CP$ violation, based on the formation after the inflationary phase of bound states (bags) of a large, $O(1000)$, number of heavy $t$-quarks with the $W$- and $Z$-bosons with a suppressed vacuum expectation value of the Higgs field, was suggested by Flambaum and Shuryak \cite{FlambaumShuryak}. Such a scenario corresponds both to a decrease of the height of the sphaleron barrier and to an effective reduction of the scale $F$ to the mass of a $b$-quark with a truncated Jarlskog determinant (recall the similar role of the Jarlskog determinant in the calculations of the EDM of quarks in Sec.~\ref{Kobayashi}).

As pointed out by Shaposhnikov \cite{Shaposhnikov3} and discussed in detail in the review \cite{BodekerBaryogenesis}, the baryon asymmetry generated by a smooth crossover is insufficient to explain the observed result (\ref{eq:Bdensity}). A popular solution to the problem is to extend the Higgs sector towards a strong coupling \cite{BaldesHighScale,NiemiTwoStep,BiekotterEarlyUniverse}, which allows to shift the phase transition to higher temperatures. Various leptogenesis scenarios at the energies in the Grand Unification region are widely discussed (see the recent review \cite{BodekerBaryogenesis} and the cited literature). Here it is worthwhile to note that whereas the hypotheses about new particles in the generalizations of the low-energy baryogenesis allow for a direct experimental verification, various scenarios of leptogenesis remain to a large extent of the academic value.

On the whole, the problem of baryogenesis remains as yet open. The discussion in the literature is mainly focused on the predominantly perturbative analysis of renormalizable models that allow for a consistent extrapolation throughout the entire period of the expansion of the Universe. The uniqueness of the millistrong $P$-preserving and $T$-violating interactions is that it is $C$-noninvariant. The role of such interactions in the baryogenesis has not yet received a due attention.

We conclude this introduction to the subject with the main thesis about the undoubtedly important role of highly sensitive EDM searches for understanding the $CP$ nonconservation beyond the SM. This subject is actively developing, and encompasses the areas of the atomic physics, the particle physics from the low energies to collider experiments, and the modern cosmology. We move on to the discussion of the main topic of the role of the effects of the General Relativity theory in precision spin experiments.

\section{Relativistic spinning particle in external fields}\label{Spinclassic}

After Uhlenbeck and Goudsmit \cite{Uhlenbeck1,Uhlenbeck2} had introduced in 1925 the concept of spin to explain atomic spectra, Frenkel \cite{Frenkel1,Frenkel} and Thomas \cite{Thomas,Thomas:1926dy} had simultaneously developed the first models for a particle with spin and magnetic moment, and a year later Dirac \cite{Dirac} had formulated the relativistic quantum theory of a particle with spin \begin{footnotesize}${\frac 12}$\end{footnotesize}. The classical Frenkel-Thomas theory, which was further developed by Mathisson, Papapetrou and Dixon, gives an adequate description of a particle with spin, and it underlies the analysis of the dynamics of polarized particles in accelerators and storage rings, see Bargmann, Michel and Telegdi \cite{BMT}, Froissart and Stora \cite{Froissart}, Derbenev and Kondratenko \cite{derb1,derb2,derb3}; the history of the issue is presented in Ternov's review \cite{ternov}. Since a coherent review of the subject is still missing in the literature, the current Section~\ref{Spinclassic} and, to some extent, the next Section~\ref{SpinQuantum} will give a quite technical exposition of the relevant formalism.

\subsection{Classical theory of spin}\label{spinMPD}

The motion of classical particles with spin in a gravitational field is consistently described by the generally covariant theory of Mathisson-Papapetrou-Dixon \cite{Mathisson,Papapetrou,Dixon}. In the framework of this theory, a test particle is characterized by a 4-velocity $U^\alpha$ and a spin tensor $S^{\alpha\beta} = - S^{\beta\alpha}$. In general, the total 4-momentum is not collinear with the velocity. In \cite{KhriplovichPomeranskyJETP,PomeranskyUFN} a non-covariant approach was developed in which the main dynamical variable is the three-dimensional spin defined in the particle's rest frame. One can show \cite{SilenkoTeryaev2005,SilenkoTeryaev2006,OST,OSTRONG,OSTgen} that the Mathisson-Papapetrou theory is fully compatible with the non-covariant approach. Using the Frenkel supplementary condition $U_\alpha S^{\alpha\beta} = 0$, which means that the spin is a purely space-like variable in the comoving reference frame, one can introduce the 4-vector of spin 
\begin{equation}\label{Sa}
S_\alpha = {\frac 1{2c}}\,\varepsilon_{\alpha\beta\gamma\delta} U^\delta S^{\beta\gamma},
\end{equation}
where $\varepsilon_{\alpha\beta\gamma\delta}$ is the fully antisymmetric Levi-Civita tensor.

Thus, in the most general formulation of the Frenkel-Thomas model, the motion of a test spinning particle is characterized by the 4-velocity $U^\alpha$ and the 4-vector of spin $S^\alpha$, which satisfy the normalization $U_{\alpha} U^\alpha = c^2$ and the orthogonality condition $S_{\alpha} U^\alpha = 0$. Both variables are considered with respect to an orthonormal basis in which the indices are raised and lowered with the help of the Minkowski metric $g_{\alpha\beta} = {\rm diag}(c^2, -1, -1, -1)$. Neglecting second-order spin effects \cite{chicone,dinesh}, the dynamic equations for these variables can be written as 
\begin{eqnarray}
{\frac {dU^\alpha}{d\tau}} &=& {\cal F}^\alpha,\label{dotU}\\
{\frac {dS^\alpha}{d\tau}} &=& \Phi^\alpha{}_\beta S^\beta.\label{dotS}
\end{eqnarray}
External fields of different physical nature (electromagnetic, gravitational, scalar, etc.) determine the forces ${\cal F}^\alpha$ acting on the particle, as well as the spin transfer matrix $\Phi^\alpha{}_\beta $. The normalization and orthogonality of the velocity and spin vectors impose conditions on the right-hand sides (\ref{dotU}), (\ref{dotS}): 
\begin{equation}
U_\alpha {\cal F}^\alpha = 0,\qquad U_\alpha\Phi^\alpha{}_\beta S^\beta =
-\,S_\alpha {\cal F}^\alpha.\label{cc}
\end{equation}
Evidently, the spin transfer matrix should be skew-symmetric, $\Phi_ {\alpha\beta} = - \,\Phi_{\beta\alpha}$, which automatically guarantees $S_\alpha S^\alpha =\,$const.

When the particle is at rest, its spatial\footnote{Hereinafter, the letters from the beginning of the Latin alphabet are used for spatial indices: $a, b, c,\dots = 1,2,3$.} 3-velocity disappears $\widehat{v} ^a = 0$, and thus the 4-velocity 
\begin{equation}\label{U}
U^\alpha = \left\{\gamma, \gamma \widehat{\bm{v}}\right\},\quad
\gamma = {\frac 1{\sqrt{1 - \widehat{v}^2/c^2}}},
\end{equation}
where $\widehat{v}^2 = \delta_{ab}\widehat{v}^a\widehat{v}^b$, reduces to 
\begin{equation}\label{Ur}
u^\alpha = \delta^\alpha_0 = \left\{1, \bm{0}\right\}.
\end{equation}
The 4-velocity vector $U^\alpha$ in the laboratory frame (\ref{U}) is related to its value in the rest frame (\ref{Ur}) via the local Lorentz transformation $U^\alpha = \Lambda^\alpha {}_\beta u^\beta$, where in the block representation 
\begin{equation}\label{Lambda}
\Lambda^\alpha{}_\beta = \left(\begin{array}{c|c}\gamma & \gamma \widehat{v}_b/c^2 \\
\hline \gamma\widehat{v}^a & \delta^a_b + (\gamma - 1)
\widehat{v}^a\widehat{v}_b/\widehat{v}^2\end{array}\right).
\end{equation}
Substituting (\ref{U}) into the orthogonality relation $S_{\alpha} U^\alpha = 0$, we find the zeroth component of the spin 4-vector in terms of 3-spatial components: 
\begin{equation}
S^0 = {\frac 1{c^2}}\,\widehat{v}_a S^a.\label{S0}
\end{equation}
The components of the vector $S^\alpha$ in the laboratory reference frame do not describe the physical spin of the particle: we recall that the spin, as the ``internal angular momentum'' of the particle, is defined with respect to the rest frame (= comoving frame). This physical spin will be denoted by $s^\alpha$ (in the general case, the lower case letters will be used also for any other objects in the rest frame). Since the transition to the rest frame ($U^\alpha \longrightarrow u^\alpha$) is performed by means of the Lorentz transformation (\ref{Lambda}), we have $S^\alpha = \Lambda^\alpha{}_\beta s^ \beta$. Inverting this, we find the relation between the physical spin and the 4-vector in the lab frame: 
\begin{eqnarray}\label{sa}
s^\alpha &=& (\Lambda^{-1})^\alpha{}_\beta S^\beta =
\left\{0, s^a\right\},\\
s^a &=& S^a - {\frac {\gamma}{\gamma + 1}}\,{\frac {\widehat{v}^a\widehat{v}_b}{c^2}}\,S^b.
\end{eqnarray}

Substituting $S^\alpha = \Lambda^\alpha{}_\beta s^\beta$ into (\ref{dotS}), we find the dynamic equation for the {\it physical spin}: 
\begin{equation}
{\frac {ds^\alpha}{d\tau}} = \Omega^\alpha{}_\beta
s^\beta.\label{dsdt}
\end{equation}
Here we introduced
\begin{equation}\label{Omab1}
\Omega^\alpha{}_\beta = \phi^\alpha{}_\beta + \omega^\alpha{}_\beta\,,
\end{equation}
where $\phi^\alpha{}_\beta = (\Lambda^{-1})^\alpha{}_\gamma\Phi^\gamma{}_\delta\Lambda^\delta{}_\beta $ is the value of the spin transfer matrix $\Phi^\alpha{}_\beta$ in the rest frame, and 
\begin{equation}
\omega^\alpha{}_\beta := -\,(\Lambda^{-1})^\alpha{}_\gamma
{\frac d{d\tau}} \Lambda^\gamma{}_\beta.\label{Thomas}
\end{equation}
After substituting (\ref{Lambda}) into (\ref{Thomas}), with the help of the matrix algebra we derive 
\begin{eqnarray}
\omega^\alpha{}_\beta &=& \left(\begin{array}{c|c}0 & -f_b/c^2 \\
\hline -f^a & \omega^a{}_b\end{array}\right),\\
\omega^a{}_b &=& {\frac {\gamma^2}{\gamma + 1}}\left({\frac {\widehat{v}^a}{c^2}}
{\frac {d\widehat{v}_b}{d\tau}}
- {\frac {\widehat{v}_b}{c^2}}{\frac {d\widehat{v}^a}{d\tau}}\right).\label{Thomas1}
\end{eqnarray}
Here the components $f^\alpha = (\Lambda^{-1})^\alpha{}_\beta{\cal F}^\beta$ of the 4-vector of the force in the rest frame have the form 
\begin{equation}
f^0 = 0,\qquad f^a = {\cal F}^a - {\frac {\gamma}{\gamma + 1}}
\,{\frac {\widehat{v}^a\widehat{v}_b}{c^2}}\,{\cal F}^b,
\end{equation}
and we used (\ref{dotU}) to find the off-diagonal components in (\ref{Thomas1}).

The formula (\ref{Thomas}) provides perhaps the simplest derivation of the Thomas precession, which is explicitly computed in (\ref{Thomas1}). For a more detailed discussion of the Thomas precession, see Silenko \cite{SilenkoThomas,SilenkoThomas1}.

Calculation of the components of the spin transfer matrix in the rest frame 
\begin{equation}
\phi^\alpha{}_\beta = \left(\begin{array}{c|c}0 & \phi^0{}_b \\
\hline \phi^a{}_0 & \phi^a{}_b\end{array}\right)\label{phirest}
\end{equation}
is simple: we need to evaluate the product of the three matrices, $\phi^\alpha{}_\beta = (\Lambda^{-1})^\alpha{}_\gamma\Phi^\gamma{}_\delta\Lambda^ \delta{}_\beta$. As a result, we find $\phi^0{}_b = \delta_{ab} \phi^a{}_0/c^2$ and 
\begin{eqnarray}
\phi^a{}_0 &=& \gamma\left(\Phi^a{}_0 - {\frac {\gamma}{\gamma +
		1}}\,{\frac {\widehat{v}^a\widehat{v}_b}{c^2}}\,\Phi^b{}_0 +
\Phi^a{}_b\widehat{v}^b\right),\label{phi0a}\\
\phi^a{}_b &=& \Phi^a{}_b + {\frac 1 {c^2}}\left(\varphi^a\widehat{v}_b
- \varphi_b\widehat{v}^a\right),\label{phiab}\\
\varphi^a &=& \gamma\left(\Phi^a{}_0 + {\frac {\gamma}{\gamma + 1}}
\,\Phi^a{}_b\widehat{v}^b\right).\label{phia}
\end{eqnarray}

The physical spin is characterized by the three non-trivial spatial components (\ref{sa}), and one can show that the 0th component (\ref{dsdt}) is identically zero (this is equivalent to the second compatibility condition (\ref{cc})). As a result, the dynamic equation for the spin (\ref{dsdt}) reduces to the 3-vector form 
\begin{equation}
{\frac {d{\bm s}}{d\tau}} = \gamma\bm{\Omega}\times{\bm s}.\label{ds1}
\end{equation}
Here the components of 3-vectors are introduced via ${\bm s} = \{s^a\}$ and ${\bm \Omega} = \left\{-\,\epsilon^{abc}\Omega_{bc}/ 2\gamma\right\}$. Recalling (\ref{Omab1}), we find the angular velocity of the spin precession 
\begin{equation}
\bm{\Omega} = \bm{\phi} + \bm{\omega},\label{Omab2}
\end{equation}
where ${\bm \phi} = \left\{-\,\epsilon^{abc}\phi_{bc}/2\gamma\right\}$ and ${\bm \omega} = \left\{- \,\epsilon^{abc}\omega_{bc}/2\gamma\right\}$. The presence of the Lorentz factor in (\ref{ds1}) is a technical feature which is explained by the parametrization of the spin dynamics with the help of the laboratory time $t$ used in accelerator experiments, in contrast to the generally covariant form of the equations (\ref{dotU}), (\ref{dotS}) and (\ref{dsdt}), where the proper time $\tau$ is used.

The general equations (\ref{Thomas})-(\ref{ds1}) are valid for a spinning particle interacting with any external fields. The actual dynamics of the physical spin depends on the forces acting on the particle and on the law of the spin transfer.

\subsection{Interlude: gravity and inertia in particle physics}\label{inter}

For a better understanding of the dynamics of a spinning particle on arbitrary manifolds in curvilinear coordinates, we need to recall the necessary geometrical tools of the general relativity (GR) theory where the basics structures are the spacetime metric $g_{ij}$, the coframe (tetrad) $e^\alpha_i$ and the connection $\Gamma_{i\alpha }{}^\beta$.

From the point of view of geometry, the role of the metric is to determine the lengths and angles on the curved manifold $M$, the connection determines the parallel transport of geometric objects from one point of the manifold $M$ to another, and the fields of the frame and coframe define the bases in the tangent and cotangent spaces at any point $x\in M$. From the point of view of physics, the metric $g_{ij}$ is the potential of the gravitational field, the connection provides a realization of the principles of general covariance and equivalence and introduces the covariant derivatives $D_i$ of physical variables, while the (co)frame introduces the reference system of a physical observer (since the spacetime is four-dimensional, the (co)frame is usually called a tetrad). The choice of a local observer's frame is determined by the motion of the observer and the conditions for conducting physical measurements, and orthonormal frames are particularly convenient (although other options are also possible, such as, for example, isotropic or semi-isotropic tetrads), with respect to which the local Lorentz symmetry is realized, that underlies the relativistic quantum theory and the particle physics.

Let $x^i = (t, x^a)$ be the local coordinates on a four-dimensional curved manifold $M$. The spacetime interval 
\begin{equation}
ds^2 = g_{ij}dx^i dx^j = g_{\alpha\beta}\vartheta^\alpha\vartheta^\beta\label{ds}
\end{equation}
can be written equivalently either in terms of the holonomic $dx^i$ coframe or in terms of a anholonomic (tetrad) one: $\vartheta^\alpha = e^\alpha_idx^i$. Thus, from the formal mathematical point of view, the tetrad can be viewed as the ``square root'' of the metric $g_{ij} = e^\alpha_i e^\beta_j g_{\alpha\beta}$, where the flat Minkowski metric is $g_{\alpha \beta} = {\rm diag}(c^2, -1, -1, -1)$, however, from a physical point of view, it is important to remember that the choice of a (co)frame determines a reference system that, in general, moves in a non-trivial way, and, in particular, is non-inertial.

The tetrad is defined up to a local Lorentz transformation, and this arbitrariness is eliminated by the choice of a physical gauge. The most convenient is the Schwinger gauge, who was the first to use it \cite{Schwinger1,Schwinger2} (and independently Dirac \cite{diracS} did the same). In this gauge, the coframe matrix $e^\alpha_i$ and its inverse matrix $e^i_\alpha$ are both characterized by the trivial elements in the upper right block: 
\begin{equation}\label{Sgauge}
e^\alpha_i = \left(\begin{array}{c|c} e^{\widehat 0}_0 & 0 \\
\hline e^{\widehat a}_0 & e^{\widehat a}_b\end{array}\right),\qquad
e_\alpha^i = \left(\begin{array}{c|c} e_{\widehat 0}^0 & 0 \\
\hline e^a_{\widehat 0} & e_{\widehat b}^a\end{array}\right).
\end{equation}
In order to distinguish between coordinate and tetrad indices, we will mark the latter with a hat.

It should be noted that other gauges are also used in the literature, among which we mention the Landau-Lifshitz choice \cite{LLvol2}, in which the lower left block vanishes 
\begin{equation}\label{Lgauge}
e^\alpha_i = \left(\begin{array}{c|c} e^{\widehat 0}_0 & e^{\widehat 0}_b \\
\hline 0 & e^{\widehat a}_b\end{array}\right),\qquad
e_\alpha^i = \left(\begin{array}{c|c} e_{\widehat 0}^0 & e^0_{\widehat b} \\
\hline 0 & e_{\widehat b}^a\end{array}\right).
\end{equation}
Finally, yet another option arises if, with the help of the Minkowski metric $g_{\alpha\beta} = {\rm diag}(c^2, -1, -1, -1)$, we move the anholonomic index down: $e_{\alpha i} := g_{\alpha\beta}e^\beta_i$. A tetrad is called symmetric if the resulting matrix does not change when transposed, 
\begin{equation}
e_{\alpha i} = e_{i\alpha}.\label{Kgauge}
\end{equation}
The spin dynamics in the symmetric gauge was studied by Pomeransky and Khriplovich \cite{KhriplovichPomeranskyJETP,PomeranskyUFN} and Dvornikov \cite{dvornikov2006neutrino}.

It is convenient to parametrize the components of the coframe in the Schwinger gauge (where $e_a^{\,\widehat {0}} = 0$, and $e_{\widehat {a}}^{\, 0} = 0,\ a = 1,2, 3$), as follows: 
\begin{equation}\label{coframe}
e_i^{\,\widehat{0}} = V\,\delta^{\,0}_i,\qquad e_i^{\widehat{a}} =
W^{\widehat a}{}_b\left(\delta^b_i - cK^b\,\delta^{\,0}_i\right). 
\end{equation}
Here the functions $V = V(x^i)$ and $K^a = K^a(x^i)$, as well as the components of the $3\times 3$ matrix $W^{\widehat a}{}_b = W^{\widehat a}{}_b(x^i) $ can arbitrarily depend on the local coordinates $t, x^a$. The total number of variables $\{V, \bm{K}, W^{\widehat a}{}_b\}$ is $1 + 3 + 3\times 3 = 13 = 16 - 3$, which obviously describes an arbitrary coframe, with the three of the sixteen components eliminated by the Schwinger gauge (\ref{Sgauge}).

The coframe (\ref{coframe}) gives rise to the general form of the spacetime line element (\ref{ds}) in the Arnowitt-Deser-Misner (ADM) parametrization \cite{ADM} 
\begin{equation}\label{LT}
ds^2 = V^2c^2dt^2 - \delta_{\widehat{a}\widehat{b}}W^{\widehat a}{}_c W^{\widehat b}{}_d
\,(dx^c - K^ccdt)\,(dx^d - K^dcdt).
\end{equation}
The off-diagonal components $g_{0a} = c\,\delta_{\widehat{c}\widehat{d}}W^{\widehat c}{}_a W^{\widehat d}{}_bK^b$ and $ g^{0a} = {\frac {K^a}{cV^2}}$ are related to the rotation effects.

The Riemannian (Levi-Civita) connection is uniquely determined by the metric and the coframe from the conditions of the absence of the nonmetricity (vanishing of the covariant derivative of the metric $D_ig_{\alpha\beta} = 0$) and the zero torsion assumption $D_i e^\alpha_j - D_j e^\alpha_i = 0$. Then for the ADM parametrization (\ref{LT}) of the general spacetime metric with the tetrad (\ref{coframe}), the components of the local Lorentz connection $\Gamma_{i\alpha\beta}$ have an explicit form: 
\begin{align}
\Gamma_{i\,\widehat{a}\widehat{0}} &= {\frac {c^2}V}\,W^b{}_{\widehat{a}}
\,\partial_bV\,e_i{}^{\widehat{0}} - {\frac cV}\,{\cal Q}_{(\widehat{a}
	\widehat{b})}\,e_i{}^{\widehat{b}},\label{connection1}\\
\Gamma_{i\,\widehat{a}\widehat{b}} &= {\frac cV}\,{\cal Q}_{[\widehat{a}
	\widehat{b}]}\,e_i{}^{\widehat{0}} + \left({\cal C}_{\widehat{a}\widehat{b}
	\widehat{c}} + {\cal C}_{\widehat{a}\widehat{c}\widehat{b}} + {\cal C}_{\widehat{c}
	\widehat{b}\widehat{a}}\right) e_i{}^{\widehat{c}},\label{connection2}
\end{align}
where we introduced (denoting by the dot $\dot{\,} = \partial_t$ the partial time derivative with respect to $t$) 
\begin{align}
{\cal Q}_{\widehat{a}\widehat{b}} &= g_{\widehat{a}\widehat{c}}W^d{}_{\widehat{b}}
\Bigl({\frac 1c}\dot{W}^{\widehat c}{}_d + K^e\partial_e{W}^{\widehat c}{}_d
+ {W}^{\widehat c}{}_e\partial_dK^e\Bigr),\label{Qab}\\
{\cal C}_{\widehat{a}\widehat{b}}{}^{\widehat{c}} &= W^d{}_{\widehat{a}}
W^e{}_{\widehat{b}}\,\partial_{[d}W^{\widehat{c}}{}_{e]},\qquad {\cal
	C}_{\widehat{a} \widehat{b}\widehat{c}} = g_{\widehat{c}\widehat{d}}
\,{\cal C}_{\widehat{a}\widehat{b}}{}^{\widehat{d}}.\label{Cabc}
\end{align}
As usual, the round brackets ${}_{(ab)}$ and the square brackets ${}_{[ab]}$ denote, respectively, the symmetrization and antisymmetrization of the marked indices. 

\subsection{Spin in gravitational and electromagnetic fields}\label{spingrav}

The general formalism of the Frenkel-Thomas model describes the motion of the spin in electromagnetic and gravitational (inertial) fields, as a particular case.

Let us consider a relativistic particle with mass $m$, electric charge $q$, anomalous magnetic moment (AMM) $\mu '$ and EDM $d$, 
\begin{equation}
\mu' = G\,\frac{q\hbar}{2m},\qquad  d = \eta^{\rm edm}\,\frac {q\hbar}{2mc}, \label{mude}
\end{equation}
where $G$ is the magnetic anomaly, $G = {\frac {g - 2}2}$, $g$ is the gyromagnetic factor, and $G$ and $\eta^{\rm edm}$ characterize the AMM and EDM values, respectively. The dynamics of a particle in the gravitational and electromagnetic fields is described by the system of equations \cite{NelsonEDM,FukuyamaEDM}: 
\begin{align}
{\frac {DU^\alpha}{d\tau}} = {\frac {dU^\alpha}{d\tau}} + U^i\Gamma_{i\beta}{}^\alpha U^\beta
= -\,{\frac qm}\,g^{\alpha\beta}F_{\beta\gamma}U^\gamma,&\label{DUG}\\
{\frac {DS^\alpha}{d\tau}} = {\frac {dS^\alpha}{d\tau}} + U^i\Gamma_{i\beta} {}^\alpha S^\beta
=  -\,{\frac qm}\,g^{\alpha\beta}F_{\beta\gamma}S^\gamma &\nonumber\\
-\,{\frac 2\hbar}\left[M^\alpha{}_\beta + {\frac {1}{c^2}}\left(M_{\beta\gamma}U^\alpha
U^\gamma - M^{\alpha\gamma}U_\beta U_\gamma\right)\right]S^\beta.&\label{DSG}
\end{align}
Here it was convenient to introduce the generalized polarization tensor 
\begin{equation}
M_{\alpha\beta} = \mu' F_{\alpha\beta} + c\,d\,\widetilde{F}{}_{\alpha\beta},\label{Mab}
\end{equation}
(where $\widetilde{F}{}_{\alpha\beta} = {\frac 12}\epsilon_{\alpha\beta\mu\nu}F^{\mu\nu}$) with components 
\begin{equation}
M_{\hat{0}\hat{a}} = c{\mathcal P}_a,\qquad
M_{\hat{a}\hat{b}} = \epsilon_{abc}{\mathcal M}^c,\label{MP}
\end{equation}
or in the 3-vector form 
\begin{equation}\label{MaPa}
\bm{\mathcal M} = \mu'\bm{\mathfrak{B}} +d\bm{\mathfrak{E}},\qquad
\bm{\mathcal P} = c\,d\bm{\mathfrak{B}} - \mu'\bm{\mathfrak{E}}/c.
\end{equation}
The components of the electromagnetic field strength tensor $F_{\alpha\beta}=e_\alpha^i e_\beta^j F_{ij}$ are calculated with respect to the anholonomic local Lorentz frame of reference:
\begin{equation}\label{EBun}
\bm{\mathfrak{E}}{}_a = \{ F_{\widehat{1}\widehat{0}}, F_{\widehat{2}\widehat{0}}, F_{\widehat{3}\widehat{0}} \},\quad
\bm{\mathfrak{B}}{}^a = \{ F_{\widehat{2}\widehat{3}}, F_{\widehat{3}\widehat{1}}, F_{\widehat{1}\widehat{2}} \},
\end{equation}
and are related to the holonomic components $\bm{E} = \{F_{10}, F_{20}, F_{30}\} = -\,\bm{\nabla}\Phi - \partial_t\bm{A} $ and $\bm{B} = \{F_{23}, F_{31}, F_{12}\} = \bm{\nabla}\times\bm{A}$ of the Maxwell tensor $F_{ij} = \partial_i A_j - \partial_j A_i$ by means of the tetrad fields
\begin{eqnarray}\label{EE}
\bm{\mathfrak{E}}{}_a &=& {\frac {1}{V}}\,W^b{}_{\hat{a}}\left(\bm{E} + c\bm{K}\times\bm{B}\right)_b,\\
\bm{\mathfrak{B}}{}^a &=& {\frac 1w}\,W^{\hat{a}}{}_b\,\bm{B}^b,\label{BB}
\end{eqnarray}
where $w = \det W^{\hat{a}}{}_b$.

In accordance with the general formalism of the model of a particle with spin, we can write down the explicit form of the force and the spin transfer matrix for the system (\ref{DUG})-(\ref{DSG}): 
\begin{eqnarray}
{\cal F}^\alpha &=& -\,U^i\Gamma_{i\beta}{}^\alpha U^\beta
- {\frac {q}{m}}\,F^\alpha{}_\beta\,U^\beta,\label{Fe}\\
\Phi^\alpha{}_\beta &=& -\,U^i\Gamma_{i\beta}{}^\alpha - {\frac {q}{m}}
\,F^\alpha{}_\beta - {\frac 2\hbar}\Bigl[M^\alpha{}_\beta\nonumber\\  \label{Pe}
&& + {\frac {1} {c^2}}\left(U^\alpha M_{\beta\gamma}U^\gamma - U_\beta M^{\alpha\gamma} U_\gamma\right)\Bigr].
\end{eqnarray}
One can check that the compatibility conditions (\ref{cc}) are satisfied.

Substituting (\ref{Fe}), (\ref{Pe}) and (\ref{U}) into (\ref {phi0a})-(\ref{phia}) and (\ref{Thomas1}), we derive
\begin{equation}\label{phiomsum}
\bm{\phi} = {\stackrel {(e)}{\bm{\phi}}} + {\stackrel {(g)}{\bm{\phi}}},\qquad
\bm{\omega} = {\stackrel {(e)}{\bm{\omega}}} + {\stackrel {(g)}{\bm{\omega}}}.
\end{equation}
The contribution of the electromagnetic field reads
\begin{align}
{\stackrel {(e)}{\bm{\phi}}} &= {\frac {q}{m}}\left[- \bm{\mathfrak B}
+ {\frac {\gamma} {\gamma + 1}}\,{\frac {\widehat{\bm v}(\widehat{\bm v}\cdot
		\bm{\mathfrak B})}{c^2}} + {\frac {\widehat{\bm v}\times \bm{\mathfrak E}}{c^2}}
\right]\nonumber\\
& + \,{\frac {2}\hbar}\left[- \bm{\mathcal M} + {\frac {\gamma} {\gamma + 1}}
\,{\frac {\widehat{\bm v}(\widehat{\bm v}\cdot\bm{\mathcal M})}{c^2}}
- {\frac {\widehat{\bm v}\times \bm{\mathcal P}}{c}}\right],\label{phiE}\\
{\stackrel {(e)}{\bm{\omega}}} &= {\frac {q}{m}}\,{\frac {\gamma  - 1}{\gamma}}\left[
\bm{\mathfrak B} - {\frac {\widehat{\bm v} (\widehat{\bm v}\cdot\bm{\mathfrak B})
		+ \widehat{\bm v}\times\bm{\mathfrak E}}{\widehat{v}^2}}\right],\label{omegaE}
\end{align}
and the contributions of the gravitational field are
\begin{align}\label{phiG}
{\stackrel {(g)}{\bm{\phi}}}{}_a &= U^i\epsilon_{abc}\left[{\frac 1{2\gamma}}
\Gamma_i{}^{cb} + {\frac {\gamma}{\gamma + 1}} \,{\frac {\widehat{v}_d}{c^2}}
\,\Gamma_{id}{}^b\widehat{v}^c + {\frac {1}{c^2}}\,\Gamma_{i\widehat{0}}{}^b
\widehat{v}^c\right],\\
{\stackrel {(g)}{\bm{\omega}}}{}_a &= -\,{\frac {\gamma}{\gamma + 1}}
U^i\epsilon_{abc}\left[{\frac {\widehat{v}_d} {c^2}}\,\Gamma_{id}{}^b\widehat{v}^c
+ {\frac 1 {c^2}}\,\Gamma_{i\widehat{0}}{}^b\widehat{v}^c\right].\label{omegaG}
\end{align}

The physical spin precession is the sum (\ref{Omab2}). The result reads explicitly: 
\begin{equation}\label{Omtot}
\bm{\Omega} = {\stackrel {(e)}{\bm{\Omega}}} + {\stackrel {(g)}{\bm{\Omega}}},
\end{equation}
where for the electromagnetic ${\stackrel {(e)}{\bm{\Omega}}} = {\stackrel {(e)}{\bm{\phi}}} + {\stackrel {(e)}{\bm{\omega}}}$ and for the gravitational ${\stackrel {(g)}{\bm{\Omega}}} = {\stackrel {(g)}{\bm{\phi}}} + {\stackrel {(g)}{\bm{\omega}}}$ parts we find, respectively: 
\begin{align}
{\stackrel {(e)}{\bm{\Omega}}} = {\frac {q}{m}}\left[-\,{\frac {1}{\gamma}}\,\bm{\mathfrak B}
+ {\frac {1}{\gamma + 1}}{\frac {\widehat{\bm v}\times\bm{\mathfrak E}}{c^2}}\right]\nonumber\\
+ \,{\frac {2}\hbar}\left[ - \bm{\mathcal M} + {\frac {\gamma}
{\gamma + 1}}\,{\frac {\widehat{\bm v}(\widehat{\bm v}\cdot\bm{\mathcal M})}{c^2}}
- {\frac {\widehat{\bm v}\times \bm{\mathcal P}}{c}}\right],\label{OmegaE}\\
{\stackrel {(g)}{\bm{\Omega}}}{}_a = \epsilon_{abc}\,U^i\left[{\frac 1{2\gamma}}
\Gamma_i{}^{cb} + {\frac {1}{\gamma + 1}}\,\Gamma_{i\widehat{0}}{}^b
\widehat{v}^c/c^2\right].\label{OmegaG}
\end{align}
The exact formula (\ref{OmegaG}) can also be used in the flat spacetime for the non-inertial frames of reference and curvilinear coordinates, since the $\Gamma_{i\beta}{}^\alpha$ connection contains the information about both the gravitational and inertial effects. 

\section{Quantum Dirac fermion dynamics in external classical fields}\label{SpinQuantum}

\subsection{Generally covariant Dirac equation}\label{Dirac}

The study of quantum systems in a gravitational field, and in particular, the study of the generally relativistic dynamics of fermions on a curved manifold has a long history, which began almost immediately after the establishment of the Dirac spinor equation \cite{Tetrode,Weyl,Bade,Oliveira,diracS,Kobzarev:1966,Hehl,Kiefer}. A special mention deserves the work of Kobzarev and Okun \cite{KobzarevOkun}, who demonstrated that, unlike the electric dipole moment, there should be no anomalous ``gravitational dipole moments'' even if there are $CP$-noninvariant fermion interactions.

The most general description of the electromagnetic interactions should take into account the possible non-minimal coupling with the AMM and EDM of the particle, and the corresponding covariant Dirac equation for the spinor field $\Psi$ with the rest mass $m$, AMM $\mu'$ and EDM $d$ has the form \cite{ostor} 
\begin{equation}
\left(i\hbar\gamma^\alpha D_\alpha - mc + {\frac{\mu'}{2c}}\sigma^{\alpha\beta}F_{\alpha\beta}
+ {\frac{d}{2}}\sigma^{\alpha\beta}\widetilde{F}{}_{\alpha\beta}\right)\Psi=0.\label{Diracgen}
\end{equation}
The spinor covariant derivative 
\begin{equation}
D_\alpha = e_\alpha^i D_i,\qquad D_i = \partial _i - {\frac {iq}{\hbar}}
\,A_i + {\frac i4}\sigma^{\alpha\beta}\Gamma_{i\,\alpha\beta},\label{eqin2}
\end{equation}
describes the minimal interaction of a fermion particle with the external classical fields: the electromagnetic 4-potential $A_i = (-\,\Phi, \bm{A})$ (coupled to the electric charge $q$ of the fermion) and the potentials of the gravitational field $(e^\alpha_i, \Gamma_i{}^{\alpha\beta})$. The tetrad indices of the Dirac matrices manifest the definition of the three-component physical spin (pseudo) vector in the local Lorentz rest frame of a particle. In the limit of the flat Minkowski spacetime, the equation (\ref{Diracgen}) reduces to the Dirac-Pauli equation for a particle with AMM and EDM \cite{Com}.

We can recast the Dirac equation (\ref{Diracgen}) into the Schr\"odinger form, but the corresponding ``naive'' Hamiltonian is non-Hermitian (see e.g. \cite{BLP}). This problem is solved by rescaling the spinor wave function $\psi = (\sqrt{-g} e^0_{\widehat 0})^\frac{1}{2}\Psi$, and the resulting Schr\"odinger equation 
\begin{equation}
i\hbar\frac{\partial \psi} {\partial t}= {\cal H}\psi\label{sch}
\end{equation}
then contains the Hermitian (and self-adjoint) Hamiltonian 
\begin{eqnarray}
{\cal H} &=& \beta mc^2V + q\Phi + {\frac c 2}\left(\pi_b\,{\cal F}^b{}_a \alpha^a
+ \alpha^a{\cal F}^b{}_a\pi_b\right)\nonumber\\ && + \,{\frac c2}\left(\bm{K}\!\cdot\bm{\pi}
+ \bm{\pi}\!\cdot\!\bm{K}\right) +\,{\frac {\hbar c}4}\left(\bm{\Xi}\!\cdot\!\bm{\Sigma}
- \Upsilon\gamma_5\right)\nonumber\\ \label{HamiltonDP}
&& - \,\beta V\left(\bm{\Sigma}\cdot\bm{\mathcal M} + i\bm{\alpha}\cdot\bm{\mathcal P}\right).
\label{eq54}
\end{eqnarray}
Here, as usual, $\alpha^a = \beta\gamma^a$ ($a,b,c,\dots = 1,2,3$) and the spin matrices $\Sigma^1 = i\gamma^{\hat 2}\gamma^{\hat 3}, \Sigma^2 = i\gamma^{\hat 3}\gamma^{\hat 1}, \Sigma^3 = i\gamma^{\hat 1}\gamma ^{\hat 2}$ and $\gamma_5=i\alpha^{\hat{1}}\alpha^{\hat{2}} \alpha^{\hat{3}}$. The boldface font is used to denote the 3-vectors ${\bm K} = \{K^a\},\,{\bm\alpha} = \{\alpha^a\}, \,{\bm\Sigma} = \{\Sigma^a\},\,{\bm\pi} = \{\pi_a\}$. The latter object denotes the kinetic momentum operator, $\bm\pi=-i\hbar\bm{\nabla} - q\bm A$. The minimal coupling gives rise to the terms in (\ref{HamiltonDP}) with the objects 
\begin{eqnarray}\label{AB1}
{\cal F}^b{}_a &=& VW^b{}_{\widehat a},\\
\Upsilon &=& V\epsilon^{\widehat{a}\widehat{b}\widehat{c}}\Gamma_{\widehat{a}\widehat{b}\widehat{c}} =
- V\epsilon^{\widehat{a}\widehat{b}\widehat{c}}{\cal C}_{\widehat{a}\widehat{b}\widehat{c}},\label{AB2}\\
\Xi^a &=& {\frac Vc}\,\epsilon^{\widehat{a}\widehat{b}\widehat{c}}\Gamma_{\widehat{0}\widehat{b}\widehat{c}}
= \epsilon_{\widehat{a}\widehat{b}\widehat{c}}\,{\cal Q}^{\widehat{b}\widehat{c}},\label{AB3}
\end{eqnarray}
while the terms with $\bm{\mathcal M}^a$ and $\bm{\mathcal P}_a$, which we defined in (\ref{MaPa}), are responsible for the non-minimal interaction. 

\subsection{Foldy-Wouthuysen representation}\label{FWrep}

The spin of particles is a purely quantum quantity. Accordingly, one needs to be careful in the treatment of interaction of the quantum spin with external classical fields, and the formalism described above requires an adequate physical interpretation. In standard textbooks on the quantum field theory, an accurate definition of the spin and angular momentum operators of a relativistic particle is missing. For example, in the excellent and popular textbook of Ryder the author refers a reader interested in this question to the special literature (\cite{Ryder}, reference [15] in chapter 2).  At the same time, the ``proton spin puzzle'' remains a topical issue in the physics of hard hadron processes for about 35 years: the experiments on the scattering of polarized electrons and muons on polarized protons have shown that quarks in protons carry only a relatively small part ($\leq 30\%$) of the proton spin, and the orbital moments of quarks and gluons play a significant role (see reviews \cite{Anselmino:1994gn,Teramond, JiNature, Efremov:1989sn} and references therein). Accordingly, the deep inelastic scattering on polarized protons and deuterons occupies an important place \cite{Accardi} in the experimental program of the new electron-ion collider eIC to be built at Brookhaven. The question of the correct interpretation of the orbital momentum is also crucial in the recently booming physics of the so-called twisted states, \cite{BliokhSOI,BliokhDN,Bliokh2017,SmirnovaBliokh,silenko2019relativistic}. Finally, it may seem quite surprising to see the very existence of such a subject as the relativistic quantum chemistry of heavy atoms, in which the relativistic description of an electron spin is of a fundamental importance (see Refs. \cite{Dyall,ReiherWolfBook,PengReiher,Autschbach,ReiherTCA,Liu,local,NakajimaH,ReiherArXivBook,ReiherRev} and references therein).

The key point in describing this variety of phenomena is the relativistic quantum mechanics, based on the pioneering work of Foldy and Wouthuysen (FW) \cite{FW} and its further development. The simple and trivial relation between quantum-mechanical operators and the corresponding classical variables is a distinctive feature of Schr\"{o}dinger's non-relativistic quantum mechanics (QM). This relation between the quantum operators of the energy, momentum and angular momentum is distorted in the naive Hamiltonian interpretation of the Dirac equation. The correct transformation of the Dirac equation to the Schr\"{o}dinger representation has been given in 1950 by Foldy and Wouthuysen \cite{FW}. In this approach, understanding the role of the particle position operator \cite{Pryce,NewtonWigner,FG,KhriplovichPomeranskyJETP,PomeranskyUFN} was very essential. In particular, a determination of the position operator is important for the construction of the Berry curvature \cite{BliokhDN,Bliokh2017,XiaoDuval,CDuval,QNiu,Mohrbach,Gosselin1,Gosselin2,Gosselin3,BliokhBerry1,BliokhBerry2}. Unfortunately, the corresponding achievements are not properly reflected in the books on QM and are not taken into account by many authors (see, for example, the monograph \cite{ReiherWolfBook}, articles \cite{electrondensity,QChem1990,QChem1998,BliokhSOI,Lloyd,BliokhDN,Bliokh2017,SmirnovaBliokh,BBBarnett,BBPRL2017,BB}, and the related criticism and literature in \cite{Reply2019, PRAFW}). There is also quite a number of misleading statements about the existence of a spin-orbital interaction for the free Dirac particles \cite{BliokhSOI,BliokhDN,Bliokh2017,SmirnovaBliokh}. We will focus here only on the fundamental behavior of the spin of relativistic particles in external fields and on the role of the gravitational field and the rotation of the Earth in the high-precision spin experiments.
        
Below, we mostly follow Ref. \cite{PRAFW}. At the very foundation, there is the well-known 10-dimensional Poincar\'{e} algebra with the 4-momentum $p _\mu$ and the angular momentum $j_{\mu\nu} = -\, j_{\nu\mu} $ generators, where $\mu ,\nu = 0,1,2,3 $ \cite{Pryce,Dirac,CurrieRevModPhys,JordanMuku}. Let us split the 4-momentum and the angular momentum into the temporal and spatial components: $H = p_0$, $\bm {p} =\{ p_a\} $, $\bm {j} =\{\epsilon ^ {abc} j_{bc }/2\} $, and $\bm {\kappa} =\{ j _ {0a}\} $. In the resulting set, $\bm {p}$, $ H $ are the generators of the infinitesimal spatial and temporal translations, and $\bm j$ and $\bm\kappa$ generate the infinitesimal rotations and the Lorentz transformations (boosts) that satisfy the known commutation relations \cite{Pryce,CurrieRevModPhys,JordanMuku,BakamjianThomas,Foldy56,Foldy61}. They should be complemented by operators of the orbital angular momentum (OAM) and the spin part of the total angular momentum:
\begin{equation}
\bm j=\bm l+\bm s,\qquad \bm l\equiv\bm{q}\times\bm{p}.
\label{spinOAM}
\end{equation}
The coordinates $q_a$ should satisfy the commutation relations \cite{Pryce,CurrieRevModPhys,JordanMuku}
\begin{equation}
\begin{split}
[q_a,p_b]&=i\hbar\delta_{ab},\quad [q_a,j_b]=i\hbar\epsilon_{abc}q_c,\\
[q_a,\kappa_b]&={\frac{1}{2c^2}}\left(q_b[q_a,H] + [q_a,H]q_b\right) - i\hbar t\delta_{ab}\, , \label{position}
\end{split}
\end{equation}
and thus
\begin{equation}
\begin{array}{c}
[l_b,p_b]=i\hbar\epsilon_{abc}p_c,\qquad
[s_a,p_b]=0.
\end{array}
\label{addeq}
\end{equation}
The commutativity of the particle coordinates
\begin{equation}
[q_a,q_b]=0\,, \label{kcommfq}
\end{equation}
turns out to be a very nontrivial condition \cite{Pryce,JordanMuku}. The separation of the spin and the orbital angular momentum is fixed by the commutation relations \cite{Pryce,JordanMuku,AcharyaSudarshan}
\begin{equation}
\begin{split}
[q_a,l_b]&=i\hbar\epsilon_{abc}q_c,\quad [q_a,s_b]=0, \quad [p_a,s_b]=0,\\
[l_a,l_b]&=i\hbar\epsilon_{abc}l_c,\quad [s_a,s_b]=i\hbar\epsilon_{abc}s_c, \quad [l_a,s_b]=0\, .
\label{spinOAMope}
\end{split}
\end{equation}
	
For a free spinning particle, we have \cite{JordanMuku}
\begin{equation}
\begin{split}
\bm j&=\bm l+\bm s, \quad \bm l\equiv\bm q\times\bm p,\\
\bm \kappa& = {\frac {1}{2c^2}}(\bm q{\cal H}+{\cal H}\bm q)
-\frac{\bm s\times \bm{p}}{\beta mc^2 + {\cal H}} - t\bm p,\\
{\cal H}&=\beta\epsilon,\quad \epsilon = \sqrt{m^2c^4 + c^2\bm p^2},
\end{split}\label{tomFW}
\end{equation}
where $\bm{q}$ is the position operator. The last term in the formula for $\bm{\kappa}$ is missing in Refs. \cite{JordanMuku,BakamjianThomas,SuttorpDeGroot,Bacry88}. The spin is a three-component (pseudo) vector defined here in particle's rest frame \cite{PRAFW}, whereas the OAM is always defined in the laboratory frame. Evidently, the Hamiltonian (\ref{tomFW}) commutes with the OAM and spin operators. However, in the set of the Dirac operators $\bm p, {\cal H}_D, \bm j, \bm\kappa, \bm q, \bm s_D$, where $\bm s_D=\hbar\bm\Sigma/2$ and the Dirac radius vector $\bm r$ is the position operator, we find that all the Poincar\'{e} algebra relations are satisfied, except for the commutators containing $\bm\kappa$.
	
One can construct the total angular momentum from the spatial parts of the two antisymmetric tensors $L^{\mu\nu} $ and $S^{\mu\nu}$:
\begin{equation}
J^{\mu\nu}=L^{\mu\nu}+S^{\mu\nu}=x^{\mu}p^{\nu}-x^{\nu}p^{\mu}+S^{\mu\nu}.\label{JLS}
\end{equation}
Let us turn to the widely spread description of the spin by the four-component operator \cite{FG,BLP}
\begin{equation}\begin{array}{c}
a^\mu=(a^0,\bm a)=\left(\frac{\bm p\cdot\bm s}{m},\,
\bm s+\frac{\bm p(\bm s\cdot\bm p)}{m(\epsilon + mc^2)}\right)\, ,
\end{array}\label{fourspintens}
\end{equation}
obtained from $\bm s$ by a Lorentz transformation. Then one can define the antisymmetric tensor
\begin{equation}
S^{\mu\nu}={\frac 1{mc}}\,\varepsilon^{\mu\nu\alpha\beta}a_\alpha p_\beta\, ,\label{spintas}
\end{equation}
in terms of which
\begin{equation}
a^\mu ={\frac{1}{2mc}}\varepsilon^{\alpha\beta\nu\mu}p_\alpha S_{\beta\nu}\, . \label{4-spin}
\end{equation}
The orbital angular momentum operator $\bm l$ is the spatial part of the antisymmetric tensor $L^{\mu\nu} = (-\bm\kappa,-\bm l)$, where $\bm\kappa=(\bm q{\cal H}+{\cal H}\bm q)/2c^2 - t\bm p$. The explicit form of $\bm{l}$ is invariant with respect to Lorentz transformations. Similarly, the spatial part of $S^{\mu\nu}$ is \cite{PRAFW}
\begin{equation}
\bm\zeta=\bm s-\frac{\bm p\times(\bm p\times\bm s)}{m(\epsilon + mc^2)},\label{zetatwo}
\end{equation}
and it would be logical to use it as a definition of the spin operator.

In the FW representation, the unnecessary components of the tensor (\ref {spintas}) can be removed by a redefinition of the particle position operator \cite{Pryce,SuttorpDeGroot,DeKerfBauerle},
\begin{equation}
\bm{\mathcal{X}}=\bm x +\frac{\bm s\times\bm p}{m(\epsilon + mc^2)}\,,\label{newposo}
\end{equation}
where $\bm{x}$ is the usual position operator of the center of charges. The angular momentum operator in the FW representation has the usual form \cite{Pryce,SuttorpDeGroot,DeKerfBauerle}
\begin{equation}
\bm{\mathcal{L}}=\bm{\mathcal{X}}\times\bm p\, ,\label{newAM}
\end{equation}
and the total angular momentum operator retains the standard form
\begin{equation}
{\bm{j}}=\bm l+\bm s=\bm{\mathcal{L}}+{\bm\zeta}\,.\label{podeGSp}
\end{equation}
We emphasize that $\bm {\mathcal {X}}$ and ${\bm\zeta}$ correspond to the case (d) in Pryce's classification \cite {Pryce} and are defined in the laboratory frame. Additional arguments for this choice of operators of $\bm {\mathcal {X}}$ and ${\bm\zeta}$ are presented in Refs. \cite{KhPom,KhriplovichPomeranskyJETP,PomeranskyUFN,Bauke,CKT,Deriglazov1,Deriglazov2}.

It is easy to understand why just the operators $\bm x$, $\bm l$ and $\bm s$ are the generally accepted operators of the position, OAM and spin. The commutativity of the coordinates $x_a\, (a = 1,2,3)$ allows one to use the ordinary geometry. The operators $\bm l $ and $\bm s $, despite the fundamental difference of their definitions, satisfy the familiar commutation relations (see Eq. (\ref{spinOAMope})), providing their consistent quantization. In contrast, in the alternative set of the basic operators $\bm {\mathcal {X}},\bm{\mathcal {L}}$, and $\bm\zeta$, the coordinates $\mathcal {X}_a\, (a = 1,2,3)$ do not commute, and one should use the noncommutative geometry. In addition, the OAM and spin operators, $\bm {\mathcal {L}}$ and $\bm\zeta$, do not satisfy the commutation relations (\ref{spinOAMope}) and are not quantized.

The problem of the correspondence between the nonrelativistic and relativistic quantum mechanics in the FW representation is rigorously solved in Ref. \cite{JINRLett12}. By making use of the FW representation, the transition to the classical limit in the relativistic QM for particles with an arbitrary spin corresponds to the Wentzel-Kramers-Brillouin approximation in the zeroth order of $\hbar$, which is similar to the nonrelativistic QM. As a result,  when the conditions of this approximation are satisfied, the use of the FW representation allows one to reduce the construction of the classical limit in the relativistic QM to a replacement of operators in the Hamiltonian and in the quantum-mechanical equations of motion by the corresponding classical quantities, \cite {JINRLett12}. 
	
Let us now return to the analysis of the quantum dynamics of a fermion particle in external fields. To determine the physical content of the Schr\"{o}dinger equation (\ref{sch}), one should pass to the FW representation. A purely gravitational case without the electromagnetic field was studied in Refs. \cite{OST,OSTRONG,ostgrav}, and here we consider the general case with an account of both the gravity and the electromagnetism.

The exact FW transformation is discussed in detail in Refs. \cite{E,erik,FW}. One can construct the FW transformation for the Dirac Hamiltonian (\ref{HamiltonDP}) with the general method developed in Refs. \cite{PRA,PRAnonstat,PRA2}. It allows one to obtain the FW Hamiltonian which is exact in all terms of the zeroth and the first orders in the Planck constant $\hbar$ and it also includes the second-order terms that describe the contact interactions. Here, it is sufficient for us to take into account only the first-order terms in $\bm{\Xi}, \Upsilon, \bm{\mathcal P}, \bm{\mathcal M}$. Omitting nonessential technical details (see Refs. \cite{OST,OSTRONG,ostgrav,ostor} for the computational methods), we find the FW Hamiltonian:
\begin{eqnarray}
{\cal H}_{FW} &=& \beta\epsilon' + q\Phi + \frac c2\left(\bm K\!\cdot\!\bm\pi
+ \bm \pi\cdot\!\bm K\right)\nonumber\\ && +\frac\hbar2\bm\Pi\cdot\bm\Omega_{(1)}+
\frac\hbar2\bm\Sigma\cdot\bm\Omega_{(2)}.\label{Hamlt}
\end{eqnarray}
Here $\bm{\Pi} = \beta\bm{\Sigma}$ and we have in the semiclassical limit
\begin{equation}
\epsilon' = \sqrt{m^2c^4V^2 + c^2\delta^{cd}{\cal F}^a{}_c\,{\cal F}^b{}_d\,\pi_a\,\pi_b\,},\label{eQ}
\end{equation}
\begin{eqnarray}
\Omega^a_{(1)} &=& {\frac {c^2}{\epsilon'}}{\cal F}^d{}_c \pi_d\biggl[{\frac 12}
{\Upsilon}\delta^{ac} - \epsilon^{abe}V{\cal C}_{be}{}^c\nonumber\\ && + {\frac {\epsilon'}
	{\epsilon' + mc^2V}}\epsilon^{abc}W^e{}_{\widehat{b}}\,\partial_eV\nonumber\\
&& +\,{\frac{eV^2}{\epsilon' + mc^2V}}\,\epsilon^{acb}{\mathfrak E}_b
- {\frac{2V}{c\hbar}}\epsilon^{acb}{\mathcal{P}}_b\biggr],\label{FO}\\
\Omega^a_{(2)} &=& {\frac c2}\,\Xi^a - {\frac {c^3}{\epsilon'(\epsilon'+mc^2V)}}
\,\epsilon^{abc}{\cal Q}_{(bd)}\delta^{dn}{\cal F}^k{}_n\pi_k{\cal F}^l{}_c\pi_l \nonumber\\
&& -\,{\frac{ec^2V^2}{\epsilon'}}{\mathfrak B}^a + {\frac {2V}\hbar}\biggl[- {\mathcal{M}}^a\nonumber\\
&& + {\frac{c^2}{\epsilon'(\epsilon' + mc^2V)}}\delta^{an}{\cal F}^c{}_n\pi_c{\cal F}^d{}_b
\pi_d{\mathcal{M}}^b\biggr].\label{finalOmegase}
\end{eqnarray}

\subsection{Quantum spin dynamics vs classical theory of spin}\label{spinCQ}

To analyze the spin dynamics, it is necessary to evaluate the commutator of the polarization operator $\bm\Pi =\beta\bm\Sigma$ with the FW Hamiltonian (\ref{Hamlt}). This yields the dynamic equation describing the spin precession in external gravitational and electromagnetic fields:
\begin{equation}
\frac{d\bm \Pi}{dt}=\frac{i}{\hbar}[{\cal H}_{FW},\bm \Pi]=\bm\Omega_{(1)}
\times\bm \Sigma+\bm\Omega_{(2)}\times\bm\Pi.\label{spinmeq}
\end{equation}
For the practical problems in the high-energy particle physics in accelerators and storage rings, it is sufficient to work with quasiclassical quantities and equations. This results in the following explicit quasiclassical equation describing the precession of the mean spin 3-vector ${\bm s}$:
\begin{eqnarray}
{\frac {d{\bm s}}{dt}} &=& \bm \Omega\times{\bm s}
= (\bm \Omega_{(1)}+\bm \Omega_{(2)})\times{\bm s},\label{dots}
\end{eqnarray}

Using Eq. (\ref{Hamlt}), we also obtain the velocity operator in the quasiclassical approximation:
\begin{eqnarray}
{\frac {dx^a}{dt}} &=& \frac{i}{\hbar}[{\cal H}_{FW},x^a] = \beta\,{\frac
	{\partial\epsilon'}{\partial \pi_a}} + cK^a \nonumber\\
&=& \beta\,{\frac {c^2}{\epsilon'}}\,{\cal F}^a{}_b\delta^{bc}{\cal F}^d{}_c
\pi_d + cK^a.\label{velocity}
\end{eqnarray}
Let us compare this with the relation between the holonomic and anholonomic components of the particle velocity. It is convenient to parametrize the anholonomic 4-velocity components by the spatial 3-velocity $\widehat{v}^a$ ($a = 1,2,3 $), as shown in (\ref{U}). Then we have
\begin{eqnarray}
U^a &=& {\frac {dx^a}{d\tau}} = e^a_\alpha U^\alpha = {\frac \gamma V}
(cK^a + VW^a{}_{\widehat b}\,\widehat{v}^{b}),\label{Ua}\\
U^0 &=& {\frac {dt}{d\tau}} = e^0_\alpha U^\alpha = {\frac \gamma V},\label{U0}
\end{eqnarray}
and, therefore, we derive for the components of the holonomic velocity
\begin{equation}
{\frac {dx^a}{dt}} = {\frac {U^a}{U^0}} = {\cal F}^a{}_b\,\widehat{v}^b + cK^a.\label{va}
\end{equation}
Comparing this equation with Eq. (\ref{velocity}), we thus can identify the velocity operator in the Schwinger gauge (\ref{coframe}) with
\begin{equation}
\beta\,{\frac {c^2}{\epsilon'}}\,{\cal F}^b{}_a\pi_b = \widehat{v}_a.\label{Fpv}
\end{equation}
From this we find $\delta^{cd}{\cal F}^a{}_c\,{\cal F}^b{}_d \pi_a\pi_b = (\epsilon')^2\widehat{v}^2/c^2$, and by making use of this in Eq. (\ref{eQ}), we obtain $(\epsilon')^2 = m^2c^4V^2 + (\epsilon')^2\widehat{v}^2/c^2$, and, consequently,
\begin{equation}
\epsilon' = \gamma\,mc^2\,V.\label{eV}
\end{equation}
The two equations, (\ref{Fpv}) and (\ref{eV}), are crucial for demonstrating the full consistency between the quantum and the classical spin dynamics. Namely, from Eqs. (\ref {Fpv}) and (\ref{eV}) we derive
\begin{align}
{\frac {\epsilon'}{\epsilon' + mc^2V}} &= {\frac \gamma {1 + \gamma}},\label{factors}\\
{\frac {c^3}{\epsilon'(\epsilon' + mc^2V)}}\,{\cal F}^b{}_a\pi_b{\cal F}^d{}_c\pi_d
&= {\frac \gamma {1 + \gamma}}\,{\frac {\widehat{v}_a\widehat{v}_c}{c}},\label{factors1}
\end{align}
and that makes it possible to finally establish the quantum-classical correspondence and to compare the classical model of a spinning particle in external fields with the quantum dynamics of a Dirac fermion.

Substituting Eqs. (\ref{connection1}) and (\ref{connection2}), we can recast Eq. (\ref{OmegaG}) into
\begin{equation}
{\stackrel {(g)}{\bm{\Omega}}} = -\,{\frac 1{\gamma}}\,\bm{\mathcal{B}} +
{\frac {1}{\gamma + 1}}\,{\frac {\widehat{\bm{v}} \times\bm{\mathcal{E}}} {c^2}},\label{omgem}
\end{equation}
where the generalized gravitoelectric and gravitomagnetic fields are defined by 
\begin{eqnarray}\label{ge}
{\mathcal E}^a &=& {\frac {\gamma}V}\delta^{ac}\left(c{\cal Q}_{(\widehat{c}\widehat{b})}
\widehat{v}^b - c^2\,W^b{}_{\widehat{c}}\,\partial_bV\right),\\
{\mathcal B}^a &=& {\frac {\gamma}V}\left(-\,{\frac c2}\,{\Xi}^a - {\frac 12}\Upsilon
\,\widehat{v}^a + \epsilon^{abc}V{\cal C}_{bc}{}^d\widehat{v}_d\right).\label{gm}
\end{eqnarray}
The remarkable similarity of Eq. (\ref{omgem}) and the first term in Eq. (\ref{OmegaE}) suggests an introduction of the effective magnetic and electric fields 
\begin{eqnarray}
\bm{\mathfrak B}_{\rm eff} &=& \bm{\mathfrak B} + {\frac mq}\,\bm{\mathcal B},\label{Beff}\\
\bm{\mathfrak E}_{\rm eff} &=& \bm{\mathfrak E} + {\frac mq}\,\bm{\mathcal E}.\label{Eeff}
\end{eqnarray}
One can easily understand the presence of the factor $m/q$ (the ratio of the gravitational ``charge'' to the electric charge) from the dimensional arguments. Accordingly, the general precession velocity (\ref{Omtot}) is rewritten as follows:
\begin{align}
\bm{\Omega} &= {\frac {q}{m}}\left[- \,{\frac 1{\gamma}}\,\bm{\mathfrak B}_{\rm eff} +
{\frac {1}{\gamma + 1}}{\frac {\widehat{\bm v}\times\bm{\mathfrak E}_{\rm eff}}{c^2}}
\right]\nonumber\\ &+ \,{\frac {2}\hbar}\left[ - \bm{\mathcal M} + {\frac {\gamma}{\gamma + 1}}
\,{\frac {\widehat{\bm v}(\widehat{\bm v}\cdot\bm{\mathcal M})}{c^2}}
- {\frac {\widehat{\bm v}\times \bm{\mathcal P}}{c}}\right].\label{OmegaT}
\end{align}

Using equations (\ref{Fpv})--(\ref{factors}), we thereby finally demonstrate the full agreement between the classical limit of the quantum-mechanical dynamics (\ref{dots}), (\ref{FO})-(\ref{finalOmegase}) and the corresponding equation of motion of the classical spin (\ref{ds1}) and (\ref{OmegaT}) in the most general case of an arbitrary gravitational (inertial) and electromagnetic field acting on the particle. Note that it is necessary to use Eq. (\ref{U0}) to relate the derivatives with respect to the proper and coordinate time, ${\frac {d}{d\tau}} = {\frac {\gamma}{V}}{\frac {d} {dt}}$. This result, that manifests the validity of the {\it equivalence principle} (EP) for the spin as a substantially quantum object \cite{Teryaev:2016edw}, has been established in the pioneering work of Kobzarev and Zakharov \cite{Kobzarev:1966} for the case of the weak gravitational field. The later results for the arbitrary strong fields \cite{OST,OSTRONG} can be therefore considered as a consistent generalization of EP for spin.

\subsection{Gravitoelectromagnetism and precession of spin in the gravitational field}\label{GEM}

In the absence of the electromagnetic field, the dynamics of spin is determined by the angular velocity (\ref{omgem}) of its precession under the action of the gravitoelectric (\ref{ge}) and gravitomagnetic (\ref{gm}) fields. Depending on the form of these fields, one can study in the framework of the developed formalism the spin effects in any physical and astrophysical situations, including the cases of the strong fields in the vicinity of binary systems of ultracompact objects such as neutron stars and black holes. However, in experiments and observations in terrestrial laboratories and in the solar system, one can confine oneself to the consideration of the linear approximation in the framework of the so-called gravitoelectromagnetism (GEM), when the metric (\ref{LT}) is described by 
\begin{eqnarray}
V &=& 1 - {\frac {\mathit \Phi}{c^2}},\quad W = 1 + {\frac {\mathit \Phi}{c^2}},\quad
W^{\widehat a}{}_b = W\,\delta^a_b,\label{gemvw}\\
{\bm K} &=& {\frac {2{\bm{\mathcal A}}}{c^2}}.\label{gemvwk}
\end{eqnarray}
Although the quantities $({\mathit\Phi}, {\bm{\mathcal A}})$ do not form a 4-vector, they are in many ways a formal analogue of the electromagnetic 4-potential $(\Phi, \bm{A})$. In particular, substituting (\ref{gemvw}) and (\ref{gemvwk}) into (\ref{Qab}), (\ref{Cabc}) and (\ref{AB1})-(\ref{AB3}), we find that the gravitoelectric (\ref{ge}) and gravitomagnetic (\ref{gm}) fields have an ``almost Maxwellian'' form: 
\begin{equation}\label{gemEB}
\bm{\mathcal E} = \gamma\,\bm{\nabla}{\mathit\Phi},\quad \bm{\mathcal B} = -\,{\frac {\gamma}{c}}
\,\bm{\nabla}\times \bm{\mathcal A} - {\frac {\gamma}{c^2}}\,\widehat{\bm{v}}\times\bm{\nabla}{\mathit\Phi}.
\end{equation}
As a consequence, for the case of the gravitoelectromagnetism, the precession angular velocity (\ref{omgem}) reduces to 
\begin{equation}\label{Ogem}
{\stackrel {(g)}{\bm{\Omega}}} = {\frac {1}{c}}\,\bm{\nabla}\times \bm{\mathcal A}
+ {\frac {(2\gamma + 1)}{(\gamma + 1)c^2}}\,\widehat{\bm{v}}\times\bm{\nabla}{\mathit\Phi}.
\end{equation}
 
It is interesting to note that although the spin of a Dirac particle precesses in a magnetic field twice as fast as compared to the classical orbital momentum (resulting in the same precession of spin and velocity and in the conservation of helicity, that is used, for example, in the search of an anomalous magnetic moment of muons \cite{MuonAMM}), for the case of the gravitomagnetic field the precession values coincide, which manifests the validity of the equivalence principle and yields the non-conservation of the helicity \cite{Teryaev:1999su,Teryaev:2016edw}. This, in particular, can lead to the transformation of a neutrino into a (Majorana) antineutrino or a sterile (Dirac) neutrino \cite{Teryaev:1999su,Teryaev:2016edw, Kamenshchik:2016tut, Kamenshchik:2015iua}.

From the practical point of view of the physics on the Earth and in the solar system,  of greatest interest is the case of the gravitational field created by a body with the mass $M$ and the angular momentum $\bm{J}$. The exact solution of the Einstein equations for such a source, that for the case of a non-rotating body $\bm{J} = 0$ reduces to the Schwarzschild metric, was obtained by Kerr \cite{Kerr}. The Kerr metric has a fairly complicated structure, but it reduces to a special case of the gravitoelectromagnetic field (\ref{gemvw})-(\ref{gemvwk}) far from the massive source, with
\begin{equation}\label{geLT}
{\mathit \Phi} = {\frac {G_NM}{r}},\qquad \bm{\mathcal A} = {\frac {G_N\,\bm{J}\times\bm{r}}{c\,r^3}}.
\end{equation}
Curiously, long before the discovery of Kerr's exact solution, this configuration was derived by Lense and Thirring in 1918 \cite{LenseThirring,LenseThirringHehl,LenseThirringHistory} as a weak gravitational field of a slowly rotating massive body.

Substituting (\ref{geLT}) into (\ref{Ogem}), we recast the spin precession in the gravitational field of a rotating massive body into a sum of the two terms 
\begin{eqnarray}\label{Osum}
{\stackrel {(g)}{\bm{\Omega}}} &=& {\stackrel {(dS)}{\bm{\Omega}}} + {\stackrel {(LT)}{\bm{\Omega}}},\\
{\stackrel {(dS)}{\bm{\Omega}}} &=& {\frac {(2\gamma + 1)}{(\gamma + 1)}}\,
{\frac {G_NM\,\bm{r}\times\widehat{\bm{v}}}{c^2\,r^3}},\label{omDS}\\
{\stackrel {(LT)}{\bm{\Omega}}} &=& {\frac {G_N}{c^2\,r^3}}\left[
{\frac {3(\bm{J}\cdot\bm{r})\,\bm{r}}{r^2}} - \bm{J}\right].\label{omLT}
\end{eqnarray}
Sometimes one can read in the literature that the first and the second terms arise as a gravitoelectric and as a gravitomagnetic effect, respectively. This is not quite true though, since, as we see from (\ref{Ogem}), both the gravitoelectric and gravitomagnetic fields (\ref{gemEB}) contribute to the expression (\ref{omDS}).

The first term (\ref{omDS}), known as the ``{\it de Sitter precession}'' or the geodetic precession,  persists also for the static Schwarzschild field, while the second term (\ref{omLT}) is nontrivial only for stationary fields generated by a rotating source. The latter is often found in the literature under the name of the ``{\it Lense-Thirring precession}'', although to avoid a misunderstanding, it should be clarified that these physicists have nothing to do with the derivation of the precession formula (\ref{omLT}). In their work, the spacetime metric was found as an approximate solution of the Einstein equations, which is a special case of the gravitoelectromagnetic field in the form (\ref{geLT}), and much later, the actual spin dynamics in the Lense-Thirring metric was investigated by Schiff \cite{Schiff,Schiff:1960}, Kobzarev and Zakharov \cite{Kobzarev:1966}, and Schwinger \cite{Schwinger:1974} (a collection of the original and review papers \cite{Ruffini} is a useful resource on this topic).

At the same time, it is fair to attribute the result (\ref{omDS}) to de Sitter, who was the first to obtain it by analyzing the celestial dynamics of the Sun-Earth-Moon system \cite{deSitter2,Desitter:1917}. Soon, however, de Sitter's original derivation was substantially generalized and improved by Schouten \cite{Schouten:1919,Schouten:1922}, Kramers \cite{Kramers} and, in particular, by Fokker \cite{Fokker}, justifying an alternative name ``de Sitter-Fokker precession'' which is also encountered in the literature. This effect is reliably established, as a confirmation of Einstein's general relativity theory, in observations for the Sun-Earth-Moon system, in which we can  treat the Earth-Moon pair as a gyroscope (or as a particle with the spin angular momentum that arises from the Moon's rotation around the Earth) moving on the orbit of the Earth in the gravitational field of the Sun. Taking into account that the mass of the Sun is $M_\odot = 1.9\times 10^{30}\,$kg, the radius of the Earth's orbit is equal to the astronomical unit $R = 1\,$a.u.$=1.49\times 10^{ 11}\,$m, and the speed of motion can be found as $v = \sqrt{\frac {G_NM_\odot}{R}}$ (obviously, the motion is non-relativistic, so the Lorentz factor is $\gamma = 1$), we find for the magnitude of the precession of the Moon's orbit in its motion around the Earth 
\begin{equation}\label{omDSex}
{\stackrel {(dS)}{\Omega}} = {\frac {3}{2}}\,{\frac {G_NM_\odot}{c^2\,R^2}}\,\sqrt{\frac {G_NM_\odot}{R}}
= 2.8 \times 10^{-15}\,{\frac 1{\rm s}},
\end{equation}
which is approximately $1.9$ arcseconds per century. This value, found back in the works of de Sitter, was confirmed experimentally \cite{Shapiro:1988} based on the analysis of the laser ranging data for the Moon with an accuracy of about 2{\%}.

In contrast to the geodetic precession effect, the experimental verification of the Lense-Thirring precession had to await for the Gravity Probe B space mission, which was based on Schiff's theoretical work \cite{Schiff,Schiff:1960} ah's independent proposal outlined in the memorandum \cite{Pugh}, \text{color}{magenta}{first made public in \cite{Ruffini}. }

In the course of the Gravity Probe experiment \cite{GravityProbeB, Everitt:2015}, a spacecraft was launched on April 20, 2004 into a near-Earth polar orbit at an altitude of 642 km, with the four classical gyroscopes on board, the dynamics of which was observed from August 28, 2004 to 14 August 2005. The task was to test both the de Sitter and the Lense-Thirring effects of the spin precession of a particle moving at a nonrelativistic velocity $\bm{v}$ in an orbit with the radius $R$ in the Earth's gravitational field. The net effect, in accordance with (\ref{Osum})-(\ref{omLT}), is described by
\begin{equation}
{\stackrel {(g)}{\bm{\Omega}}} = \frac{3G_NM_\oplus}{2c^2 R^3}\,\bm{R}\times\bm{v}
+\frac{G_NI_{\oplus}}{c^2 R^3}\Bigl[3\frac{(\bm{\omega}_\oplus\cdot\bm{R})\,\bm{R}}{R^2}
- \bm{\omega}_\oplus\Bigr]. \label{eq:Schiff}
\end{equation}
The Earth, as a source of the gravitational field, is characterized here by the mass $M_\oplus$, the angular velocity $\bm{\omega}_\oplus$ and the moment of inertia $I_\oplus$ (so that $\bm{J} = I_\oplus \,\bm{\omega}_\oplus$), where, taking into account the nonsphericity of the Earth, $I_\oplus = C_\oplus M_{\oplus}R_{\oplus}^2$ relative to the polar axis with the coefficient $C_\oplus = $0.3307 \cite{Williams}.

For the conditions of the Gravity Probe B experiment, the theoretical calculation predicts the angular velocity of the de Sitter precession and the angular velocity of the Lense-Thirring precession, respectively (in units of milliarcseconds (mas) per year (y)): 
\begin{equation}
{\stackrel {(dS)}{\Omega}} = -\,6606.1\,{\frac {\rm mas}{\rm y}},\qquad
{\stackrel {(LT)}{\Omega}} = -\,39.2\,{\frac {\rm mas}{\rm y}}.\label{GPBT}
\end{equation}
The measurements confirmed both values: 
\begin{equation}
{\stackrel {(dS)}{\Omega}} = -\,6601.8 \pm 18.3\,{\frac {\rm mas}{\rm y}},\quad
{\stackrel {(LT)}{\Omega}} = -\,37.2\pm 7.2\,{\frac {\rm mas}{\rm y}}.\label{GPBE}
\end{equation}
The accuracy of the Gravity Probe B result corresponds, in terms of the angular velocity of the Earth's rotation, to the notable sensitivity 
\begin{equation}\label{GPBA}
{\frac {\Delta {\stackrel {(dS)}{\Omega}}}{\omega_\oplus}} \approx 4\times 10^{-11},\quad
{\frac {\Delta {\stackrel {(LT)}{\Omega}}}{\omega_\oplus}} \approx 1.5\times 10^{-11}\,.
\end{equation}

\section{Gravitational corrections to spin dynamics in storage rings}\label{spinSR}

The following discussion will be primarily linked to the plans of the ultrasensitive searches for the EDM of protons \cite{srEDM,AbusaifCYR,YannisHybrid} and neutrons \cite{ChuppRMPEDM,n2EDM}. In the former case, we are talking about dedicated storage rings in which spin rotation due to the magnetic moment of the proton is eliminated. In the latter case, we discuss the UCNs in  storage cells. The terrestrial laboratories rest in a non-inertial frame of reference attached to the gravitating and rotating Earth. We will analyse the subtle effects, the consistent description of which requires the tools of the general relativity theory (GR). 

\subsection{Spin in cyclotron: flat space}\label{spinCF}

The dynamics of spin in the orbit of a cyclic accelerator in the flat Minkowski space is well known \cite{Thomas,Frenkel,BMT,NelsonEDM,FukuyamaEDM}. In accelerator experiments, the angle of rotation of the spin with respect to the momentum of the particle is measured, that is, the difference $\bm{\Omega}_s = \bm{\Omega}_L - \bm{\Omega}_c$ between the angular velocities of the Larmor precession $\bm{\Omega}_L$ of spin and the cyclotron angular velocity $\bm{\Omega}_c$.

The Minkowski spacetime geometry is defined by $V = 1$, $W^{\widehat a}{}_b = \delta^a_b$, $\bm{K} = 0$, and hence the coframe and connection are trivial: $e^ \alpha_i = \delta^\alpha_i$, $\Gamma_{i\beta}{}^\alpha = 0$. Then the equation of motion of a charged particle in an electromagnetic field (\ref{DUG}) reduces to 
\begin{equation}
m\,\frac{d U^{\alpha}}{d\tau} = -\,q\,F^{\alpha\beta}U_{\beta}\label{eq:Motion}
\end{equation}
and gives the familiar cyclotron angular velocity 
\begin{gather}\label{Cyclotron}
\bm{\Omega}_c^{(e)} = {\frac {q}{m\gamma}}\left( -\,\bm{B}
+ {\frac{(\bm{B}\cdot\bm{v})\,\bm{v} + \bm{v}\times\bm{E}}{v^2}}\right)
\end{gather}
of rotation ${\frac {d\bm{N}}{d\tau}} = \gamma\bm{\Omega}_c^{(e)}\times\bm{N}$ of a unit vector that determines the direction of particle's motion, $\bm{N} := {\frac {\bm{p}}{p}} = {\frac {\bm{v}}{v}}$, \cite{ObukhovSilenkoTeryaev}.

The spin of a particle with the 4-velocity $U^\alpha$ in the laboratory frame is described by the 4-vector $S^\alpha = (S^0, \bm{S})$ (orthogonal to the velocity $U_\alpha S^\alpha = 0$), and inverting the relations (\ref{S0}), (\ref{sa}), 
\begin{gather}
S^0=\frac{\gamma}{c^2}(\bm{s}\cdot\bm{v}),\quad
\bm{S} = \bm{s} + \frac{\gamma^2}{c^2(\gamma + 1)}(\bm{s}\cdot\bm{v})\,\bm{v},
\label{SpinVector}
\end{gather}
where $\bm{s}$ is the spin vector in the comoving frame. Taking into account the EDM contribution, the 4-vector of spin $S^\alpha$ of a particle with the charge $q$, magnetic dipole moment (MDM) $\mu_0$, anomalous magnetic $\mu'$ moment, and electric dipole $d_{\rm edm }$ moment (\ref{mude}) satisfies the generalized Frenkel-Thomas-Bargmann-Michel-Telegdi (FT-BMT) equation (\ref{DSG}) in the laboratory system. According to this dynamical equation, the physical 3-spin vector $\bm{s}$ precesses (\ref{ds1}) in the electromagnetic field with an angular velocity (\ref{OmegaE}). Combining (\ref{OmegaE}) with (\ref{Cyclotron}), we get the precession angular velocity $\bm{\Omega}_s = {\stackrel {(e)}{\bm{\Omega}}} - \bm{\Omega}^{(e)}_c$ with respect to the detectors, which is conveniently represented as the sum of the MDM and EDM contributions: 
\begin{eqnarray}
\bm{\Omega}_s &=& \bm{\Omega}_s^{\rm mdm} + \bm{\Omega}_s^{\rm edm}\,,\label{Oesum}\\
\bm{\Omega}_s^{\rm mdm} &=& {\frac{q}{m}}\left\{-\,G\,\bm{B} + \left(G - {\frac{1}{\gamma^2-1}}
\right){\frac {\bm{v}\times\bm{E}}{c^2}}\right.\nonumber\\
&&\left. +\,{\frac{\gamma}{\gamma+1}}\left(G - {\frac{1}{\gamma-1}}\right)
{\frac {(\bm{v}\cdot\bm{B})\,\bm{v}}{c^2}}\right\}\,,\label{Osmdm}\\
\bm{\Omega}_s^{\rm edm} &=& -\,\frac{\eta^{\rm edm}\,q}{mc}\left\{ \bm{E} + \bm{v}\times\bm{B}
-\,{\frac{\gamma}{\gamma+1}}\,{\frac {(\bm{v}\cdot\bm{E})\,\bm{v}}{c^2}}\right\}\,.\nonumber\\
{}\label{Fukuyama1}
\end{eqnarray}

Optimization for the ultra-small EDM signal requires the suppression of the spin rotation due to the magnetic moment, that is, it is necessary to provide 
\begin{equation}
\bm{\Omega}_s^{\rm mdm}=0\,. \label{eq:Freezing1}
\end{equation}
Then, in the absence of an EDM, the spin would retain its orientation with respect to the momentum, i.e.,  be tangential to the orbit in the storage ring. This is the {\it ``frozen spin''} mode, when the  spin would rotate only due to the EDM, and the storage rings with the frozen proton spin are in the focus of attention \cite{srEDM,AbusaifCYR,YannisHybrid}. We will discuss different frozen spin options below. 

\subsection{Gravitational corrections to cyclotron and Larmor rotation}\label{Larmor}

For the first time, the influence of the terrestrial gravity on the spin dynamics in accelerator searches for the EDM was considered for the special case of the magnetic focusing of the beam by the authors of the review \cite{SilenkoTeryaev2005,ObukhovSilenkoTeryaev,SilenkoTeryaev2006}.

Before analyzing the possible effects of external fields in the particle physics, it is worthwhile to recall the physical conditions on the Earth's surface where accelerators and storage rings are located: The Earth rotates at an angular velocity (assuming the sidereal day $T_\oplus = $23 hours 56 minutes 4.1 seconds = 86164.1 s) 
\begin{eqnarray}
\omega_\oplus = {\frac {2\pi}{T_\oplus}} = 7.29\times 10^{-5} {\rm s}^{-1},\label{oE}
\end{eqnarray}
and thus the experiments are carried out in a non-inertial frame of reference. In addition, the Earth is a massive source of the gravitational field with the mass $M_\oplus = 5.97 \times 10^{24}\,$kg. Despite such a mass, on the surface of the Earth, which has an average radius $R_\oplus = 6.378 \times 10^6\,$m, the gravitational field is rather weak: the corresponding value of the gravitational potential is equal to 
\begin{equation}
{\frac {G_NM_\oplus}{c^2R_\oplus}} = 6.95\times 10^{-10}.\label{GMC2R}
\end{equation}

Accordingly, for the spacetime geometry in which the particle physics experiments take place, an approximate description of the Earth's gravitational field as a gravitoelectromagnetic field (\ref{gemvw})-(\ref{gemvwk}) is valid to a very good accuracy, where, however, instead of (\ref {geLT}) we have 
\begin{equation}\label{LTter}
{\mathit \Phi} = {\frac {G_NM_\oplus}{r}},\quad \bm{\mathcal A} = \left({\frac {G_N\,\bm{J}_\oplus}
	{c\,r^3}} - {\frac {c\,\bm{\omega}_\oplus}{2}}\right)\times\bm{r}.
\end{equation}
The spatial local coordinates are chosen as a Cartesian system $(x, y, z)$ with the origin at the center of the Earth, where, without the loss of generality, $z$ is the axis of rotation, so that the angular velocity vectors $\bm{\omega}_\oplus = \left(0, 0, \omega_\oplus\right)$ and the angular momentum $\bm{J}_\oplus = \left(0, 0, J_\oplus\right)$, respectively. From the formal point of view, (\ref{LTter}) describes the Lense-Thirring metric of the gravitational field of a slowly rotating massive body, with an account of the non-inertiality of the local spatial system that rotates relative to the distant fixed stars.
From the potentials (\ref{LTter}) we obtain the gravitoelectric and the gravitomagnetic (\ref{gemEB}) fields of the Earth: 
\begin{eqnarray}
\bm{\mathcal E} &=& \gamma\,\bm{g}_\oplus,\label{gEter}\\
\bm{\mathcal B} &=& \gamma\left(\bm{\omega}_\oplus + {\frac {\bm{g}_\oplus\times\widehat{\bm{v}}}{c^2}}\right)\nonumber\\
&& + \,{\frac {\gamma\,G_N}{c^2\,r^3}}\left(\bm{J}_\oplus - {\frac {3(\bm{J}_\oplus\cdot\bm{r})
		\,\bm{r}}{r^2}}\right),\label{gBter}
\end{eqnarray}
where, as usual, the 3-vector of the Newtonian acceleration is given by
\begin{equation}
\bm{g}_\oplus = -\,{\frac {G_NM_\oplus}{r^3}}\,\bm{r}.\label{N}
\end{equation}

Since the angular momentum of the Earth is $J_\oplus = I_\oplus\omega_\oplus = 5.85\times 10^{33}$\,kg\,m${}^2$s$^{-1}$ (where we estimate the moment of inertia as $I_\oplus = C_\oplus M_\oplus R^2_\oplus$ with the empirical coefficient $C_\oplus = 0.3307$), in the laboratory on its surface we have 
\begin{equation}\label{oJLT}
{\frac {2G_NJ_\oplus}{c^2R_\oplus^3}} = 3.31\times 10^{-14}\,{\rm s}^{-1}.
\end{equation}
Comparing this with (\ref{oE}), we find that the first term in the gravitomagnetic potential (\ref{LTter}) is smaller than the second one by 9 orders of magnitude. Therefore, for all practical purposes, one can assume in all calculations that the gravitomagnetic potential of the Earth's field is $\bm{\mathcal A} = -\,c\,\bm{\omega}_\oplus\times\bm{r}/2$.

The general-relativistic system of equations (\ref{DUG})-(\ref{DSG}) of particle's motion with the 4-velocity $U^\alpha = \{\gamma, \gamma\widehat{\bm{v}}\}$, spin $S^\alpha = \{S^0, \bm{S}\}$ and the dipole moments in terms of components has the explicit form 
\begin{eqnarray}\label{dg0}
{\frac {d\gamma}{d\tau}} &=& {\frac q{mc^2}}\,\gamma\,\bm{\mathfrak{E}}{}_{\rm eff}
\cdot\widehat{\bm{v}},\\ \label{dgv0}
{\frac {d(\gamma\widehat{\bm{v}})}{d\tau}} &=& {\frac q{m}}\gamma\left(
\bm{\mathfrak{E}}{}_{\rm eff} + \widehat{\bm{v}}\times\bm{\mathfrak{B}}{}_{\rm eff}\right),
\end{eqnarray}
\begin{eqnarray}
{\frac {dS^{\widehat{0}}}{d\tau}} &=& \bm{S}\cdot\left\{{\frac q{mc^2}}
\,\bm{\mathfrak{E}}{}_{\rm eff} - {\frac {2\gamma^2}{c^2\hbar}}\,\widehat{\bm{v}}
\times\bm{\Delta}\right\},\label{ds0}\\
{\frac {d\bm{S}}{d\tau}} &=& S^{\widehat{0}}\left\{{\frac q{m}}\bm{\mathfrak{E}}{}_{\rm eff}
- {\frac {2\gamma^2}{\hbar}}\,\widehat{\bm{v}}\times\bm{\Delta}\right\}\nonumber\\
+ \bm{S}&\times&\left\{{\frac q{m}}\bm{\mathfrak{B}}{}_{\rm eff} + {\frac {2\gamma^2}
	{\hbar}}\left(\bm{\Delta} - \widehat{\bm{v}}\,{\frac {\bm{\Delta}\cdot\widehat{\bm{v}}}
	{c^2}}\right)\right\}.\label{dgsa}
\end{eqnarray}
The two effective objects, $\bm{\mathfrak{B}}{}_{\rm eff}$ and $\bm{\mathfrak{E}}{}_{\rm eff}$, which were introduced in (\ref{Beff}) and (\ref{Eeff}), encode in a compact form the electromagnetic, inertial, and gravitational fields acting on the particle (where (\ref{gEter}) and (\ref{gBter}) should be used for terrestrial conditions), and the effects of the magnetic and electric dipole moments are encoded in the yet another effective entity 
\begin{equation}\label{Delta}
\bm{\Delta} := \bm{\mathcal M} + {\frac 1c}\,\widehat{\bm{v}}\times\bm{\mathcal P}.
\end{equation}

The Larmor angular velocity of the spin precession encompasses the electromagnetic and the gravitational contributions 
\begin{equation}
\bm{\Omega}_L = {\stackrel {(e)}{\bm{\Omega}}} + {\stackrel {(g)}{\bm{\Omega}}},\label{omter}
\end{equation}
where the electromagnetic part ${\stackrel {(e)}{\bm{\Omega}}}$ is given by (\ref{OmegaE}) and the gravitational part ${\stackrel {(g)}{\bm{\Omega}}}$ is obtained by substituting the gravitoelectromagnetic potential (\ref{LTter}) into (\ref{Ogem}): 
\begin{eqnarray}\label{Osumter}
{\stackrel {(g)}{\bm{\Omega}}} &=& -\,\bm{\omega}_\oplus + {\stackrel {(dS)}{\bm{\Omega}}}
+ {\stackrel {(LT)}{\bm{\Omega}}},\\
{\stackrel {(dS)}{\bm{\Omega}}} &=& {\frac {(2\gamma + 1)}{(\gamma + 1)}}\,
{\frac {\widehat{\bm{v}}\times\bm{g}_\oplus}{c^2}},\label{omDSter}\\
{\stackrel {(LT)}{\bm{\Omega}}} &=& {\frac {G_N}{c^2\,r^3}}\left[{\frac
	{3(\bm{J}_\oplus\cdot\bm{r})\,\bm{r}}{r^2}} - \bm{J}_\oplus\right].\label{omLTter}
\end{eqnarray}

The cyclotron angular velocity $\widehat{\bm{\Omega}}{}_c$ describes the rotation ${\frac{d\widehat{\bm{N}}}{d\tau}} = \gamma\widehat{\bm{\Omega}}{}_c\times\widehat{\bm{N}}$ of the unit vector that determines the direction of particle's motion, $\widehat{\bm{N}} = {\frac {\widehat{\bm{v}}}{\widehat{v}}}$. Using the equations of motion (\ref{dgv0}) and (\ref{dg0}), we find \cite{ObukhovSilenkoTeryaev} the cyclotron velocity, that includes the corrections from the Earth's gravity and rotation: 
\begin{eqnarray}
\widehat{\bm{\Omega}}{}_c &=& {\frac {q}{m\gamma}}\left\{ -\,\bm{\mathfrak B} +
{\frac {\widehat{\bm{v}}}{\widehat{v}{}^2}}\times\bm{\mathfrak E}\right\}\nonumber\\
&& -\,\bm{\omega}_\oplus + {\frac {2\gamma^2 - 1}{\gamma^2}}\,{\frac
	{\widehat{\bm{v}}\times\bm{g}_\oplus}{\widehat{v}{}^2}}.\label{Oa}
\end{eqnarray}

Combining (\ref{omter}) with (\ref{Oa}), we obtain the precession angular velocity $\bm{\Omega}_s = \bm{\Omega}_L - \bm{\Omega}_c$ with respect to the detectors: 
\begin{eqnarray}
\bm{\Omega}_s &=& -\,{\frac {1}{\gamma^2 - 1}}\,{\frac qm}\,{\frac {\widehat{\bm{v}}\times
		\bm{\mathfrak{E}}{}_{\rm eff}}{c^2}} + {\stackrel {(LT)}{\bm{\Omega}}}\nonumber\\
&& - {\frac {2}{\hbar}}\left\{\bm{\Delta} - {\frac {\gamma}{\gamma + 1}}
\,\widehat{\bm{v}}\,{\frac {\bm{\Delta}\cdot\widehat{\bm{v}}}{c^2}}\right\}.\label{OFS}
\end{eqnarray}
Or, in expanded form: 
\begin{eqnarray}
\bm{\Omega}_s &=& -\,\frac{\gamma}{\gamma^2-1}{\frac {\widehat{\bm{v}}\times\bm{g}_\oplus}{c^2}}
+ {\frac {G_N}{c^2\,r^3}}\left[{\frac {3(\bm{J}_\oplus\cdot\bm{r})\,\bm{r}}{r^2}}
- \bm{J}_\oplus\right]\nonumber\\
&& + \,{\frac {q}{m}}\left(G - \frac{1}{\gamma^2-1}
\right){\frac {\widehat{\bm{v}} \times\bm{\mathfrak{E}}}{c^2}} \nonumber\\
&& -\,{\frac {qG}{m}}\left(\bm{\mathfrak{B}} - {\frac {\gamma}{\gamma + 1}}
\,\widehat{\bm{v}}\,{\frac {\bm{\mathfrak{B}}\cdot\widehat{\bm{v}}}{c^2}}\right)\nonumber\\
&& -\,{\frac {q\eta^{\rm edm}}{mc}}\left\{\bm{\mathfrak{E}} + \widehat{\bm{v}}
\times\bm{\mathfrak{B}} - {\frac {\gamma}{\gamma + 1}}\,\widehat{\bm{v}}
\,{\frac {\bm{\mathfrak{E}}\cdot\widehat{\bm{v}}}{c^2}}\right\}.\label{OMFSblue}
\end{eqnarray}
Remarkably, the explicit contribution of the Earth's rotation $\bm{\omega}_\oplus$ cancels out. However, one should keep in mind the non-inertial contributions to the anholonomic velocity $\widehat{\bm{v}} = \bm{v} + \bm{\omega}_\oplus\times\bm{r}$ and the anholonomic electric and magnetic fields (\ref{EE}), (\ref{BB}), which for terrestrial conditions are reduced to $\bm{\mathfrak{E}} = \bm{E} - (\bm{\omega}_\oplus\times\bm{r})\times\bm{B}$, $\bm{\mathfrak{B}} = \bm{B}$. Here the term proportional to the cross product  of the anholonomic velocity and the electric field decribes rotation of the spin of the particle moving in the electric field, while the term proportional to the magnetic anomaly describes precession of the magnetic moment in the magnetic field in the ring. It is instructive to compare Eq. (\ref{OMFSblue})  with Eqs. (\ref{Osmdm}) and (\ref{Fukuyama1}) for the flat space.

\subsection{Gravitational shift of a beam orbit in a storage ring}\label{beam}

The explicit gravitational corrections obtained above should be supplemented with an indirect correction to the Larmor precession due to the focusing fields. Before proceeding to the discussion, we recall that in the accelerator physics it is common to introduce a local basis associated with the motion of a particle beam along a circular orbit, defining $\bm{e}_x = \bm{e}_\rho$ as a unit vector along the radius $\bm{\rho}$ from the center of the accelerator to a point in the orbit, $\bm{e}_y$ as a vertical unit vector, and finally $\bm{e}_z = \bm{e}_x\times\bm{e }_y$ as a tangential unit vector. From the equations of particle's motion (\ref{dg0}), (\ref{dgv0}) we find the gravitational force  
\begin{equation}\label{Fg}
\bm{F}_g = {\frac {2\gamma^2 - 1}{\gamma}}\,m\bm{g}_\oplus,\qquad \bm{g}_\oplus = -\,|g_\oplus|\bm{e}_y,
\end{equation}
and in order to prevent the accumulated beam from falling to the Earth, this force must be compensated \cite{SilenkoTeryaev2006,OrlovGravity,ObukhovSilenkoTeryaev} by the focusing fields, which are created by magnetic or electric quadrupoles. 

Beam particles are making the radial and vertical betatron oscillations in the storage ring with an angular velocity $\omega_{x,y}= Q_{x,y}\,\Omega_c$, where the dimensionless $Q_{x,y}$ are called the betatron tunes. The cyclotron angular velocity $\Omega_c$ depends only on particle's velocity and the storage ring circumference. In the first approximation, we can assume that the particle moves in an oscillatory potential with a ``spring constant'' $\langle k\rangle$, which can be expressed in terms of the betatron frequency as 
\begin{equation}
\langle k \rangle \,\approx \gamma m \omega_y^2\,.
\end{equation}
Then the gravitational vertical coherent displacement of the beam trajectory will be equal to \cite{KolyaSpringConstant,QuantumOscillations} 
\begin{equation}
\Delta y \approx \frac{(2\gamma^2-1)\,|g_\oplus|}{\gamma\,Q_y^2\,\Omega_c^2}\,.\label{eq:BeamShift1}
\end{equation}
For the search of the EDM of protons, a storage ring with the electric bending of protons with the kinetic energy of $T_p=233$ MeV is of interest. Historically, the first proposed device was a storage ring with electric focusing and the circumference of 500 m, $\Omega_c = 2.26\times 10^6\,$s$^{-1}$ and $Q_y =0.45$ \cite{srEDM,AbusaifCYR}. In this case, the estimate (\ref{eq:BeamShift1}) gives the gravitational coherent shift of a beam in the storage ring 
\begin{equation}
\Delta y_E \approx 1.3 \times 10^{-11}\ \rm{m}\,.\label{eq:yE}
\end{equation}
Recently \cite{YannisHybrid}, a hybrid version with magnetic focusing, 800 m circumference with $\Omega_c =1.4 \times 10^6\,$s$^{-1}$ and $Q_y =2.3$ was proposed, for which 
\begin{equation}
\Delta y_B \approx 1.3 \times 10^{-12}\ \rm{m}\, . \label{eq:yH}
\end{equation}
Such a microscopic gravitational displacement of a beam in a storage ring has rightly been of no concern to  accelerator physicists, still it leads to observable effects in the precision spin dynamics.

\subsection{Electric focusing effect on spin precession}\label{focus}

We now discuss in more detail the most interesting case of the electric focusing in a purely electric storage ring \cite{srEDM,AbusaifCYR}. Let us split the electric field into the two terms $\bm{E} = \bm{E}_\rho + \bm{E}_f$, where the purely radial field $\bm{E}_\rho$ keeps the particle in orbit in the storage ring plane. The particle velocity is orthogonal to $\bm{E}_\rho$. The focusing field $\bm{E}_f$ is required to compensate the gravitational contribution to the particle acceleration in order for the orbit to be closed in the storage ring plane. Using the explicit form of the equations of particle's motion (\ref{dg0}), (\ref{dgv0}) 
\begin{equation}
{\frac {d\widehat{\bm{v}}}{dt}} = {\frac {q}{\gamma m}}\,\bm{E}_\rho 
+\left\{{\frac {q}{\gamma m}}\,\bm{E}_f + {\frac {2\gamma^2 - 1}{\gamma^2}}\,\bm{g}_\oplus
- \bm{\omega}_\oplus\times\widehat{\bm{v}}\right\},
\end{equation}
we find the focusing field from the condition that the expression in the curly brackets vanishes: 
\begin{equation}
\bm{E}_f = -\,{\frac {m}{q}}\,{\frac {2\gamma^2 - 1}{\gamma}}\,\bm{g}_\oplus
+ {\frac {\gamma\,m\,\bm{\omega}_\oplus\times\widehat{\bm{v}}}{q}}.\label{Ef}
\end{equation}
The contour integral for a single revolution in the storage ring  
\begin{equation}
\oint \bm{E}_f d\bm{r}=0\,,
\end{equation}
and the field $\bm{E}_f$ can be realized electrostatically \cite{VergelesJETP}.

The focusing makes an indirect gravitational contribution to the precession angular velocity of the spin, 
\begin{equation}
\bm{\Omega}^{f}_s = {\frac{q}{m}}\left(G - {\frac{1}{\gamma^2 - 1}}\right)
{\frac {\widehat{\bm{v}}\times\bm{E}_f}{c^2}}\,,\label{omfoc}
\end{equation}
see the second line in (\ref{OMFSblue}). Importantly, this precession corresponds to the rotation of the magnetic moment, just like the de Sitter geodetic precession.

Taken together, with an account of the closedness of the beam trajectory, the sum of (\ref{omfoc}) with the first term in (\ref{OMFSblue}) gives the total gravitational correction 
\begin{eqnarray}
\bm{\Omega}_s &=& {\frac {[1 - G\,(2\gamma^2-1)]}{\gamma}}\,{\frac {\widehat{\bm{v}}\times\bm{g}_\oplus}{c^2}}
\nonumber\\
&& +\,{\frac {G_N}{c^2\,r^3}}\left[{\frac {3(\bm{J}_\oplus\cdot\bm{r})\,\bm{r}}{r^2}}
- \bm{J}_\oplus\right] + \gamma\,{\frac {\widehat{\bm{v}}\times(\bm{\omega}_\oplus
		\times\widehat{\bm{v}})}{c^2}}.\nonumber\\ {}\label{FakeEDMelectric1}
\end{eqnarray}
Only the first term, which combines the geodetic effect and the focusing effect, is of practical importance for the accelerator experiments on the proton EDM: 
\begin{equation}\label{FakeEDMelectric2}
\bm{\Omega}^{\rm rel}_g = {\frac {[1 - G\,(2\gamma^2-1)]}{\gamma}}\,{\frac {\widehat{\bm{v}}\times\bm{g}_\oplus}{c^2}}.
\end{equation}
This result was obtained in 2016 by Obukhov et al. \cite{ObukhovSilenkoTeryaev} for arbitrary energies. The remaining contributions in (\ref{FakeEDMelectric1}) are merely of methodological value. It is worthwhile to note that this answer is universal for the electric focusing, regardless of the nature of particle's bending in the orbit. 

\section{Gravity and search for the EDM of protons with frozen spin}\label{frozen}

\subsection{Electric bending and electric focusing: $G > 0$}\label{elec}

A convenient quantity is the spin tune $\nu_s = \Omega_s/\Omega_c$. The angle of rotation of the spin relative to the momentum per revolution is equal to $\theta_s = 2\pi\nu_s$. When the Larmor frequency is a multiple of the cyclotron frequency, i.e., 
\begin{equation}
\nu_s = k = 0,  \pm 1, \ \pm 2, ... , \label{InetgerResonance}
\end{equation}
the spin orientation with respect to the momentum at the detection point, at the local spin rotator, or at the collision point of colliding beams is retained after each revolution of the beam, i.e., the local spin freezing is realized. For $\nu_s=0$ the spin is frozen globally. Only in this case, the EDM signal is accumulated coherently, i.e., the bending radial electric field acts as an EDM rotator, along the entire storage ring.

On the other hand, the condition (\ref{InetgerResonance}) corresponds to the integer spin resonance, when the beam polarization becomes unstable due to an imperfection magnetic fields in a storage ring. This is a real problem, and special methods have been developed for the rapid crossing of the integer resonances during the acceleration of stored particles \cite{lee1997spin,mane2005spin}. In 2014 a novel method of compensating of the polarization instabilities at the integer resonance by solenoidal spin navigators was proposed \cite{Filatov2014} and is being actively developed \cite{Filatov2020PRL,Filatov2020EPJ,filatov2021hadron} for the NICA collider at JINR. The practical implementation of this method is important for precision spin experiments, including the search for the EDM of protons and deuterons, within the framework of the SPD spin program at the NICA \cite{SPD:NICA} complex. The local freezing is especially important for deuterons, for which maintaining the longitudinal polarization by the standard approach with the Siberian snakes is impractical because the unrealistically large snake field integrals are required. In the physical program of the electron-nuclei collider eIC, which is under construction at the Brookhaven National Laboratory, the deep inelastic scattering on longitudinally polarized deuterons is one of the key items, and precisely the integer spin resonance is being discussed as a working point for the deuteron beams \cite{Vadim2020,filatov2021hadron}.

Let us consider protons in a storage ring with the purely electrostatic bending, $\bm{B} = 0$. According to (\ref{Osmdm}), the global frozen spin condition, $\nu_s=0$, corresponds to 
\begin{equation}
\bm{\Omega}_s^{\rm mdm} = {\frac {q}{m}}\left(G-\frac{1}{\gamma^2-1}\right)
{\frac {\widehat{\bm{v}}\times\bm{E}}{c^2}} = 0\label{fro1}
\end{equation}
which requires 
\begin{equation}
\gamma^2 = {\frac{1+G}{G}}\,. \label{FrozenCondition2}
\end{equation}
The solution exists only for $G > 0$, for protons it corresponds to the so-called magic kinetic energy $T_p = 233$ MeV \cite{srEDM,AbusaifCYR}. Under the condition (\ref{FrozenCondition2}), the result (\ref{FakeEDMelectric2}) becomes 
\begin{eqnarray}
\bm{\Omega}^{\rm GR}_{\it fake} = \bm{\Omega}^{\rm rel}_g &=& {\frac {\left[1 - G\,(2\gamma^2-1)\right]}
{\gamma}}\,{\frac {\widehat{\bm{v}}\times\bm{g}_\oplus}{c^2}}\nonumber\\
&=& -\,{\frac {|g_\oplus|}{c}}\,\sqrt{G}\,\bm{e}_\rho\,,\label{OrlovEDM1}
\end{eqnarray}
and reproduces \cite{NikolaevFerrara} the result of Orlov et al. \cite{OrlovGravity} obtained in 2012 by solving the spin equations of motion in the Schwarzschild metric with the  magic energy condition (\ref{FrozenCondition2}) imposed.

As mentioned above, the formula (\ref{OrlovEDM1}) corresponds to the spin precession in a radial ``magnetic field'' of the gravitational origin and generates a false EDM signal. This should be compared to 
\begin{equation}
\bm{\Omega}_s^{\rm edm} = -\,{\frac{q\eta^{\rm edm}}{mc}}\,\bm{E}.
\end{equation}
For the radial bending electric field $E = 8\times 10^6$ V/m, the resulting gravity-induced false EDM yields \cite{AbusaifCYR} 
\begin{equation}
\begin{split}
d_p^{\it fake} \approx 1.8\times 10^{-28}\ e\cdot{\rm cm}\,, \\
\eta_{p}^{\it fake} \approx 1.7 \times 10^{-14} \, .
\end{split}
\end{equation}

A unique feature of the electrostatic storage ring is that it can be run with concurrent clockwise and counterclockwise beams on the same orbit. In contrast to the EDM signal that does not depend on the sign of the beam velocity, the gravity-induced false EDM is of the opposite sign for the CW and CCW beams. As a result, the two effects can be unambiguously separated, and an interesting opportunity arises to use the measured gravitational effect as a reference signal for checking the presence or absence of the poorly understood systematic effects in the operation of the storage ring.

The requirements for the accuracy of checking the identity of the trajectories of the counter-rotating beams at the level of $\sim\,$5 picometers are discussed in \cite{AbusaifCYR,KawallSQUID}. In the recent experiment by the JEDI collaboration at the COSY accelerator, equipped with a far from the optimal beam position monitors, it was demonstrated that the position of a deuteron beam centroid with a diameter of the order of 1 mm can be controlled with an accuracy of 1 micrometer and even better \cite{JamalQuantum}. It is noteworthy that this is only one order of magnitude higher than the amplitude of the zero quantum oscillations of the beam particles in an approximately harmonic profile, in which individual particles in COSY perform the betatron oscillations. A theory of collective oscillations with amplitudes below the one-particle zero quantum oscillations was developed in \cite{QuantumOscillations}. In a static ring, the quantum limit for monitoring  the centroid of a bunch of $N$ particles by static beam position monitors scales with $\propto 1/\sqrt{N}$.

\subsection{Electrostatic bending and magnetic focusing: $G > 0$}\label{mag}

A new version of an electrostatic proton storage ring with the focusing by magnetic quadrupoles was recently proposed by Semertzidis et al. \cite{YannisHybrid,HybridEDMring}. It is claimed that, as compared to the electric focusing, the magnetic focusing makes it possible to increase the sensitivity for the search of the EDM of the proton. It is important for us that in this case the gravitational attraction of the Earth will be compensated by the Lorentz force due to the focusing radial magnetic field $\bm{B}_f$: 
\begin{equation}
q\,\widehat{\bm{v}}\times\bm{B}_f = -\,{\frac {2\gamma^2-1}{\gamma}}\,m\bm{g}_\oplus\,,
\end{equation}
from where we find 
\begin{equation}
\bm{B}_f = {\frac {m}{q}}\,{\frac {2\gamma^2 - 1}{\gamma}}\,{\frac {\widehat{\bm{v}}\times\bm{g}_\oplus}
{\widehat{v}{}^2}}.\label{Bf}
\end{equation}
The rotation of the magnetic moment of the proton in this focusing magnetic field gives rise to a false EDM effect.

Since the focusing effect does not depend on the mechanism of the bending in orbit, the problem completely reduces to the one solved in \cite{ObukhovSilenkoTeryaev}. Substituting (\ref{Bf}) into the third line (\ref{OMFSblue}) in combination with the first term that includes the geodetic de Sitter effect, we find 
\begin{equation}
\bm{\Omega}^{\rm GR}_{\it fake} = -\,{\frac {\gamma\left[1 + G\,(2\gamma^2-1)\right]}{\gamma^2 - 1}}
\,{\frac {\widehat{\bm{v}}\times\bm{g}_\oplus}{c^2}}.\label{FakeEDMmagnetic1}
\end{equation}
The frozen spin condition (\ref{FrozenCondition2}) yields \cite{AbusaifCYR} 
\begin{equation}\label{ObukhovEDM}
\bm{\Omega}^{\rm GR}_{\it fake} = -\,(3 + G)\,{\frac {|g_\oplus|}{c}}\,\sqrt{G}\,\bm{e}_\rho\,.
\end{equation}
As compared to the case of the electric focusing (\ref{OrlovEDM1}), for the magnetic focusing the proton false EDM is thus enhanced by the factor $(3+G) \approx 4.8$.

The magnetic focusing in an electric storage ring does not prevent a concurrent accumulation of the counter-rotating  beams with the same energy. However, in the quadrupole magnets, the focusing of a clockwise beam changes to the defocusing of a counterclockwise beam, and vice versa. In contrast to the purely electric storage rings, the trajectories of the two beams will not be locally the same, with the irradicable gravitational separation of the trajectories at the level of $2\Delta y_B \approx 3$ picometer (\ref{eq:yH}). The issue of the separation of signals of the true EDM and quite an enhanced gravity-induced false EDM, endangering the targeted proton EDM $d_p < 10^{-29}\,e\cdot$cm in the regime of the magnetic focusing, remains open. 

\subsection{Frozen spin for hybrid bending}\label{hybrid}

For particles with a negative magnetic anomaly, $G<0$, for example, for a deuteron with $G_d = -\,0.1416$, the frozen spin condition (\ref{eq:Freezing1}) can only be satisfied under the hybrid bending with the crossed vertical magnetic and radial electric fields. In this case, according to the equation (\ref{Osmdm}), the two fields are related via
\begin{equation} \label{eq:HybridFreezing}
\bm{B} = \frac{\left[G(\gamma^2 - 1) - 1\right]}{G(\gamma^2-1)}
\,{\frac {\widehat{\bm{v}}\times\bm{E}}{c^2}}\,.
\end{equation}
The hybrid confinement in orbit is also required for protons with a non-magic energy. 

An example is provided by the prototype ring PTR proposed by the CPEDM collaboration \cite{AbusaifCYR}. A purely electrostatic version with the radial electric field $E_r = 7\times 10^6$ V/m is designed for the accumulation of protons with the energy $T_p =30$ MeV. An additional vertical magnetic field $B_y = 0.0327\ {\rm T}$ in the same ring increases the rigidity of the ring and provides the frozen spin of protons with the energy  45 MeV \cite{AbusaifCYR}. In the hybrid storage ring, the angular velocity of rotation of the EDM is equal to  
\begin{equation}\label{OmegaHybrid}
\bm{\Omega}_s^{\rm  edm} = -\,{\frac{\eta^{\rm  edm}q}{mc}}\,{\frac{1+G}{G\gamma^2}}\,\bm{E}\,.
\end{equation}
The result for the false EDM gravitational signal depends only on the kind of the focusing. For the focusing with the electric quadrupoles at a generic energy, it is given by the equation (\ref{FakeEDMelectric2}) \cite{ObukhovSilenkoTeryaev}, and for the magnetic focusing by the equation (\ref{FakeEDMmagnetic1}), \cite{SilenkoTeryaev2006}.

With the hybrid bending, the accumulation of the counter-rotating  beams of the same energy is impossible. The long-term control of the drift of electric and magnetic fields and the constancy of their ratio (\ref{eq:HybridFreezing}) become quite non-trivial tasks. An additional problem will be an elimination  of the rotation of the magnetic moment in the parasitic radial component of the magnetic field \cite{SpinTuneMapping,RathmannNikolaevSlimPRAB}. One of the key objectives of experiments on the prototype ring PTR will be to study precisely these systematic effects.

As a possible solution to these problems, without necessarily being attached to a specific PTR ring, Koop proposed the simultaneous accumulation in a hybrid ring on the same orbit either counter- or co-rotating beams with either different masses or energies \cite{KoopAsymmetricCollider,KoopPhysScripta}, see also the interesting development of these ideas by Talman \cite{TalmanHybrid}. These beams with a  rational ratio of velocities can be either counter-propagating  or rotating in the same direction. One beam will be polarized with a globally frozen spin and will measure the EDM of these particles. The second beam will be a comagnetometer and can be either polarized with a locally frozen spin or unpolarized. The parasitic radial magnetic field can be controlled by the vertical separation of the beams. For the co-magnetometry purposes, i.e., for  tracking the bending fields in the storage ring, it may be sufficient to control the cyclotron frequency of the second beam. Koop found possible solutions for the counterotating pairs $(p,p)$ and ($p,{}^3$He), and the co-rotating ones $(p,d)$ and ($p,{}^6$Li), see also  \cite{TalmanHybrid} for a discussion of the time reversal invariance tests in such rings in the collider mode when both beams are polarized. It is worthwhile to mention the existence of solutions with the table-top storage rings of the radii of several meters \cite{KoopPhysScripta}. 

The case of the hybrid focusing by a combination of the electric and magnetic quadrupoles is discussed in \cite{NikolaevFerrara,AbusaifCYR}. 

\subsection{Main results of JEDI collaboration and proton EDM in electric storage ring PTR}\label{JEDI}

The main task of the PTR will be a study of systematic effects as an imperative for the subsequent transition to an ultimate all electric frozen-spin proton storage ring operating at the magic energy of 233 MeV. Such a dedicated  machine with  an extremely high sensitivity  to the proton EDM (\ref{pEDMtarget}) is viewed as a part of the CERN post-LHC program. In the all electric mode with the proton energy of 30 MeV, the work will focus on the study of suppression of systematic effects running the storage ring with concurrent counter-rotating beams. In the hybrid bending mode with the proton energy of 45 MeV, the central topic will be the spin dynamics in the first ever implementation of the frozen spin mode. A detailed presentation of this program and the technical details can be found in the PTR conceptual design study \cite{AbusaifCYR} prepared by the CPEDM collaboration. In addition to solving the above problems, it is possible to conduct a direct search for the proton EDM on the PTR with an already interesting sensitivity $d_p \sim 10^{-24}\ e\cdot$cm. But to begin with, we summarize here the main achievements of the JEDI collaboration.

Before the start of the spin program at the NICA collider, the COSY Synchrotron in J\"ulich will have been the only one in the world machine fully equipped for the precision polarization experiments \cite{MaierCOSY,FeldenCOSY,WilkinCOSYlegacy}. In a series of experiments at COSY, the JEDI collaboration obtained the record-breaking results: 
\begin{itemize}
\item {A demonstration of measuring the polarization of deuterons to an accuracy of $10^{-6}$ \cite{BrantjesPolarimetry}.}
\item {A technique has been developed for measuring the deuteron spin precession frequency with an accuracy of $10^{-10}$ \cite{SpinTuneContinuousJEDI,PrecessingSpinJEDI}.}
\item {A feedback technique for a continuous control of the spin phase to an accuracy of 0.15 rad has been developed \cite{PhaseLockingJEDI,PhaseMeasurementJEDI}.}  
\item {The coherence time exceeding 1000~s for deuteron spins idly precessing in the horizontal plane has been achieved \cite{SCT1000sJEDI,SCTChromaticityJEDI}. The previous record result of 0.5~s for electrons and positrons was obtained at the Budker INP \cite{VassermanSCT}.}
\item {A new method of the spin tune mapping was developed, allowing for the first time to evaluate experimentally the integral systematic impact on the spin precession of the unwanted magnetic fields due to the imperfection of the magnetic system in the COSY plane \cite{SpinTuneMapping}.}
\item {The beam based alignment  to ensure a passage of the orbit  through the center of the quadrupole magnets has been realized \cite{BBAJEDI}.}
\item {The radiofrequency Wien filter as a novel spin rotator has been proposed, commissioned and is in operation at COSY  \cite{SlimRFWFsimulation,SlimDrivingCircuit,SlimComissioning}.}

\end{itemize}
This is a very incomplete list of important JEDI results that were recognized as a sufficient foundation for  preparation of the PTR project \cite{AbusaifCYR}.

A novel approach to a search for the proton EDM at the all electric PTR, based on running the  RF WF  operating at the cyclotron frequency, was proposed by the CPEDM collaboration (see Appendix H in \cite{AbusaifCYR}). Here the RF WF acts as a static device. When a clockwise (CW) bunch passes through the Wien filter, the  counterclockwise (CCW) bunch must be at the diametrically opposite point of the ring. Since the magnetic field in the Wien filter changes its sign by the time of arrival of the CCW bunch with the opposite velocity sign, the Wien filter condition imposed on the CW bunch is also satisfied for the CCW bunch: 
\begin{equation}
\bm{E}_{\rm wf} + \bm{v}\times\bm{B}_{\rm wf} = \bm{E}_{\rm wf} + (-\bm{v})\times(-\bm{B})_{\rm wf} = 0\, .
\end{equation}
It follows from the equation (\ref{Fukuyama1}) that, due to the contribution of the EDM, the axis of the stable spin $\bm{c} = \bm{\Omega}_c/\Omega_c$ is tilted by an angle $\xi_{\rm edm}$, 
\begin{equation}
\tan \xi_{\rm edm} = \frac{\eta_{\rm edm} \beta}{2\left[1 - \beta^2(1+G)\right]}\, .
\end{equation}

The spin rotation in the Wien filter changes the spin tune \cite{SpinTuneMapping}, 
\begin{equation}
\begin{split}
\cos\pi(\nu_s+\Delta\nu_s) &= \cos\pi\nu_s \cdot \cos\frac{1}{2}\psi \\
&- (\bm{c}\cdot\bm{w}) \sin\pi\nu_s \cdot\sin\frac{1}{2}\psi\, ,
\end{split}
\end{equation}
where $\psi$ is the angle of the spin rotation in the Wien filter. With the vertically aligned magnetic axis $\bm{w}$ of the  WF, the scalar product $(\bm{c} \cdot\bm{w}) = \pm \sin \xi_{\rm edm}$ is of opposite  sign for the two bunches. Thus the difference between the spin precession frequencies of the two bunches would yield the EDM signal 
\begin{equation}
\nu_s^{\rm cw} - \nu_s^{\rm ccw} = {\frac{1}{2}}\,\xi_{\rm edm}\psi\,.
\end{equation}
The described experiment at the all electric PTR run for one year would enable to set the upper limit on the proton EDM $d_p < 2\times 10^{-24}\ e\cdot$cm  \cite{AbusaifCYR}.

\section{Spin as antenna for axion-type particles in the Universe}\label{axion}

\subsection{Axions beyond QCD}\label{axionbeyond}

Quite paradoxically, the axion physics is surprisingly diverse. The existence of the axion phenomena was first theoretically predicted in electrodynamics by Tellegen \cite{Tellegen:1948,Tellegen:1956}, who proposed the concept of a gyrator in the electric network theory as an element of a physical system that has the property of ``rotating'' the field strengths into the field excitations\footnote{{{We follow the terminology of Mie \cite{Mie} and Sommerfeld \cite{Sommerfeld} to distinguish the electric and magnetic field strengths $\bm{E}, \bm{B}$ from the electric and magnetic excitations $\bm{D}, \bm{H}$, see \cite{Birk}.}}}. The corresponding constitutive relation has the form $\bm{E} = -\,s\,\bm{H}$, $\bm{B} = s\,\bm{D}$, where $s$ is the gyrator, or, in the modern terminology, its reciprocal is called the axion $a = 1/s$. In such a system, the spatial parity is obviously violated, and the object $a$ itself is a pseudoscalar from a geometric point of view. Although there are no material substances in nature with such a constitutive law, Lindell and Sihvola \cite{Lindell} suggested that such a system could be artificially constructed as a metamaterial, which they called a perfect electromagnetic conductor (PEMC), since it can be considered as a natural generalization of an ideal electric conductor. Tretyakov's group \cite{Tretyakov} demonstrated the possibility of manufacturing such an artificial metamaterial and investigated its properties. 

The unusual constitutive law of the Tellegen/PEMC metamaterial, however, is not something completely exotic, if we notice that in fact this is a very special case of a real material medium with the magnetoelectric properties. A characteristic property of such a substance (as a rule, of a crystalline nature) is the occurrence of an electric polarization in it in an external magnetic field, or its magnetization in an external electric field. Such crystals are anisotropic media, which are characterized by the non-trivial electric permittivity and magnetic permeability tensors, as well as by the magnetoelectric susceptibility tensor (as a specific example, we can mention the magnetoelectric Cr$_2$O$_3$). The isotropic part of the magnetoelectric susceptibility tensor can be naturally identified with the axion, the existence and the magnitude of which was established already in Astrov's classic experiments \cite{Astrov} with the uniaxial Cr$_2$O$_3$ crystals (proposed by Dzyaloshinskii \cite{Dzyal} from the magnetic symmetry analysis), and later confirmed in the more accurate studies by Rado and Folen \cite{Rado} and Wiegelmann et al. \cite{Wiegelmann}. In this sense, the axion was reliably measured in condensed matter physics \cite{HORS1,HORS2}.

Speaking of axions in the condensed matter physics, one should not forget the topological materials, first of all the so-called topological insulators \cite{TarasenkoUFN,PankratovUFN,KvonUFN}, whose theoretical and experimental studies have been developing in an avalanche-like manner recently. Such materials are the three-dimensional dielectric crystalline structures which have the conducting states localized on the surface of the crystal. The existence of such non-dissipative metallic surface states stems from the nontrivial topological properties of the band structure of the crystal, and their topological nature determines the stability of such states to defects and inhomogeneities of the conducting boundary of the material. The electromagnetic response of a three-dimensional topological insulator is described by the axion electrodynamics of an isotropic medium, the polarization and magnetic properties of which are given by the effective electric permittivity and magnetic permeability, as well as by the pseudoscalar magnetoelectric susceptibility parameter $a$. They are determined by the microscopic model of a topological insulator \cite{Qi:2008,Qi:2009,Qi:2011,Karch:2009,Nenno:2020,Sekine:2021}. In particular, the nontrivial axion $a$ is calculated as an integral in the momentum space of the topological Chern-Simons 3-form constructed from the Berry connection on the space of periodic Bloch functions of a crystal \cite{Malashevich:2010,Li:2010}.

The axion electrodynamics \cite{Ni:1977,Wilczek,CarrollFielJackiw,Jackiw,Itin,Kostelecky} is a remarkable theoretical laboratory for studying systems with the broken fundamental ($P$, $C$, $T$) symmetries, which throws a kind of a bridge from the condensed matter physics (where axions have already been discovered) to the high-energy physics and cosmology, where axions still have the status of hypothetical fields and particles. In this regard, it is worthwhile to quote Wilczek, who in one of his pioneebring papers \cite{Wilczek} astutely noted that ``\dots it is\dots not beyond the realm of possibility that fields whose properties partially mimic those of axion fields can be realized in condensed-matter systems'', thereby emphasizing the unity of the physical science in the apparently rather distant areas. A detailed discussion of the fundamental connections of the particle physics and cosmology with the condensed matter physics can be found in Volovik's book \cite{Volovik:2009} (see also \cite{volovik1998axial} for realization of the axion in superfluid $^3He-A$ by sound waves in the fermionic system).

In continuation of Sec.~\ref{CP}, we now turn to the practical issue of observation of the so far elusive cosmic axions.

\subsection{Detecting axions in flat spacetime}\label{axionflat}

The axion-matter coupling constant $f_{(a)}$, the axion mass and the contribution of cold axions to the dark matter depend on the time when the axion phase transition occurred in the expanding inflationary Universe. According to \cite{AxionPreskill,AxionSikivie,AxionDineFischler}, when the angular frequency of the axion field is about three times the expansion rate of the Universe, coherent oscillations of the cold cosmological axion field start. This determines the possibilities and prospects for active experimental searches for axions; for extensive literature on the subject, we refer to the reviews \cite{BudkerAxion,DiLuzioAxion,SikivieAxion,AxionSnowmass}. Attributing the local energy density of the dark matter $\rho_{\rm DM} \approx 400$ MeV/cm$^3$ \cite{Read_2014}  to axions in the invisible halo of our Galaxy, the amplitude of the classical axion field $a(x) = a_0\cos\left(\omega_{(a)} t- {\bm k}_{(a)} \cdot {\bm x}\right) $ can be evaluated as \cite{GrahamAxion}
\begin{equation}
a_0 = {\frac {1}{m_{(a)}}}\,\sqrt{\frac{2\rho_{\rm DM}\hbar}{c^3}}\,. \label{AxionAmplitude}
\end{equation}
Recent searches for interaction of solar axions with matter in the Baksan underground laboratory  yielded the upper bounds $m_{(a)} \leq 320$ eV/$c^2$ and $m_{(a)} \leq 4.6$ eV/$c^2$ for the KSVZ and DFSZ axion axions,  respectively \cite{AxionBaksan2022}.
 
The astrophysical upper bounds on the axion mass, $m_{(a)} < 10^{-2}$\,eV/$c^2$, are based on the particle physics methods. One estimates contributions to the fluxes of gamma rays due to the decay of axions produced via the bremsstrahlung mechanism \cite{TurnerSN1987A1,TurnerSN1987A2} in nucleon-nucleon collisions in pulsars \cite{AxionPulsar} and in the explosion of the supernova SN1987A \cite{amassobs}. The question of the lower limit on the axion mass remains open. In the focus of our discussion will be the minimal axion model with the Weinberg relation (\ref{AxionMass}) between the axion mass and the nucleon coupling constant, and we do not dwell on  more speculative axion-like particles, for example, in supersymmetric models, see reviews \cite{Grahamannurev,RevModPhysaxion,DiLuzioAxion,SikivieAxion,YannisYoun2021,AxionSnowmass}.

Quite naturally, in the $U(1)_{PQ}$ current, in addition to the chromodynamic anomaly, there is also an electromagnetic anomaly which generates, by analogy with the axion-gluon $L_a$ (\ref{Laxion}), the axion-photon interaction 
\begin{equation}\label{AxionPhoton} 
L_{a\gamma\gamma} = -\,g_{a\gamma\gamma}{\frac{1}{f_{(a)}c}}\,{\frac{\alpha}{\pi}}\,a(x)\,\bm{E}(x)\cdot\bm{B}(x)\,,
\end{equation}
with the constant $g_{a\gamma\gamma} \sim 1$ \cite{SVZAxion,EricAxion,KimAxion,DineAxion}. The most remarkable manifestation of this interaction is the inverse Primakoff effect -- the conversion of axions in a static external magnetic field into a photon with the energy equal to the mass of the axion, i.e., with the angular frequency \cite{SikivieHaloscope,AnselmAxion} 
\begin{equation}
\omega_{(a)} = {\frac{m_{(a)}c^2}{\hbar}}\,.\label{AxionFrefuency}
\end{equation}
The count of single microwave photons excited in the magnetic field of the superconducting resonator of the Sikivie haloscope depends on the the so-called axion wind -- the flux of galactic relic axions in terrestrial laboratory, caused by the motion of the Earth in the Galaxy. Sometimes it is more convenient to talk about the motion of the detector through the field of cold axions with the nonrelativistic velocity $v_{(a)} \approx 10^{-3}c$. The inverse Primakoff effect was and remains the basis of numerous searches in experiments with haloscopes. In recent experiments of the CAPP \cite{CAPPaxion} and ADMX \cite{ADMX2018} collaboration, the sensitivity of cryogenic axion haloscopes was already close to the level sufficient to begin the critical test of the existence of the dark matter, consisting of the KSVZ axions \cite{SVZAxion,KimAxion}, and it can exceed the observational threshold with the continuous progress in the superconducting resonator technique in the gigahertz region corresponding to the axion masses $m_{(a)} \sim 10\,\mu$eV$/c^2$. Unfortunately, the axion mass is unknown and one is bound to resort to a frequency scanning  and a sufficiently high sensitivity is possible only under the slow scanning, which limits the covered mass interval \cite{YannisAxionScan}. A detailed coverage of the extensive program of experiments ADMX \cite{ADMX}, ADMX-HF \cite{ADMX-HF}, HAYSTAC \cite{HAYSTAC}, CAPP \cite{CAPPaxion,CAPP1,CAPP2}, and RADES \cite{RADES} is beyond the scope of this article, and we refer readers to the detailed discussion of the issue and planned new experiments in the reviews \cite{Grahamannurev,RevModPhysaxion,ICHEP2018Axion,YannisYoun2021,BudkerAxion,DiLuzioAxion,SikivieAxion,AxionSnowmass}.

Still another application of the direct and inverse Primakoff effect is to the ``shining the laser light through the wall'' approach \cite{AnselmAxion}, when an intermediate axion is produced in a magnetic field by the Primakoff mechanism, then penetrates through a wall opaque to the light, and subsequently regenerates back into a photon in the magnetic field \cite{LSW}. A number of experiments were carried out with helioscopes, which make it possible to detect ultrarelativistic axions emitted by the Sun in the X-ray range (see, for example, \cite{CASTNature,BabyIAXO,AxionBaksan2022}).

The interaction of axions with fermions leads to a rich variety of phenomena. At first, we will discuss this by neglecting the rotation of the Earth. Inverting (\ref{ThetaToAxion}), we obtain an estimate of the oscillating contribution to the EDM of nucleons \cite{GrahamAxion2011,GrahamAxion} 
\begin{equation}
\begin{split}
d_n^{\rm ax}(x) = \eta^{\rm ax}\,{\frac{\mu_N}{c}} = {\frac{a(x)}{f_{(a)}}} \kappa_{(a)} {\frac{\mu_N}{c}}\,, \label{OscillatingAxionEDM}
\end{split}
\end{equation}
where the chiral suppression of the EDM  \cite{BaluniEDM,CrewtherEDM} is shown explicitly:
\begin{equation}
\begin{split}
\kappa_{(a)} \sim \frac{m^*}{\Lambda_{QCD}}\approx 10^{-2}\,. \label{ChiralSuppression}
\end{split}
\end{equation}
The Weinberg interaction (\ref{AxionFermion}) gives a new contribution to the non-minimal dipole terms in the generalized Dirac equation (\ref{Diracgen}): 
\begin{equation}
{\frac{\mu'}{2c}}\overline{\Psi}\sigma^{\alpha\beta}\Psi F_{\alpha\beta} 
+ {\frac{d}{2}}\overline{\Psi}\sigma^{\alpha\beta}\Psi\widetilde{F}{}_{\alpha\beta}
-\,{\frac{\hbar}{2f_{(a)}}}g_f\partial_{\mu}a(x)\overline{\Psi}\gamma^\mu\gamma_5\Psi.\label{DiracAxion}
\end{equation}

The role of the EDM in the spin precession was discussed in Sec.~\ref{spinCF}. The oscillating axion contribution (\ref{OscillatingAxionEDM}) must be included in $d(x) = d^{\rm edm}+d^{\rm ax}(x)$. The EDM enters (\ref{DiracAxion}) with the dual electromagnetic field strength $\widetilde{F}{}_{\alpha\beta}$. From the point of view of the spin dynamics, the interaction of the oscillating axion contribution $d^{\rm ax}(x)$ in the EDM with an external electric field is equivalent to the action of a radio-frequency spin rotator. In the NMR-type experiments, when the frequencies do coincide, such a rotator obviously induces the detectable rotation and depolarization of a spin  precessing in an external magnetic field \cite{Froissart,SpinTuneMapping}. This underlies, for example, the program of the CASPEr \cite{BudkerAxion} experiment, see also \cite{GrahamSpin2018,Grahamannurev,GramolinNature,Graham2021,BudkerGyroscope}.

Now we focus on the Weinberg interaction (\ref{AxionFermion}). For nonrelativistic fermions $f$, the corresponding Hamiltonian of a direct interaction with the axion field reads
\begin{equation}
{\mathcal H}_{a\bar{f}f} = {\frac{\hbar c}{2}} {\frac{g_f}{f_{(a)}}}\,\bm{\sigma}
\cdot\Bigl( \bm{\nabla}a(t,{\bm x}) + \frac{{\bm p}_f}{m_f c^2} 
\,\dot{a}(t,{\bm x})\Bigr)\, , \label{NonRelHamilt}
\end{equation}
where ${\bm \sigma}$ are the Pauli matrices, ${\bm p}_f = m_f{\bm v}_{(a)}$ is the momentum of the fermion in its motion relative to the axion field (recall that the constant $f_ {(a)}$ has the dimension of the axion field $a(t,{\bm x})$, see (\ref{ThetaToAxion})).  In typical laboratory experiments, the cold axion field can be treated as homogeneous one, and only the second term in the brackets is of practical importance. As first proposed by Kolokolov et al. in  \cite{Kolokolov1989pseudomagnetic,IgorMagnon,IgorAxionWind1}, it can be reinterptreted as  interaction with (pseudo)magnetic field, referred to also as an axion wind. Later on it was reintroduced in \cite{PospelovAxion,StadnikAxion} and became the generally accepted one. It can be derived from the classical considerations, but a correct quantum-mechanical derivation is possible only on the basis of the FW transformation, see Sec.~\ref{SpinQuantum}.

An oscillating $P$-odd interaction of axions with electrons, proportional to $\bm \sigma\cdot {\bm v}_e$, leads to various phenomena in atomic and molecular physics \cite{StadnikAxion}. Note that the speed of atomic electrons $v_e$ can significantly exceed $v_{(a)}$. The relevant experiments on this topic are discussed in the detailed review \cite{BudkerRMP}. On the other hand, for static spins the same Hamiltonian has a meaning of interaction with an oscillating external pseudomagnetic field proportional to the velocity $v_{(a)}$ of the spin motion with respect to the galactic axion field. One can search for manifestations of this pseudomagnetic field in magnetic media, for example, via magnon excitations (see \cite{IgorMagnon,IgorAxionWind1,IgorAxionWind2,vorob1995detectors,GramolinNature,Graham2021} and the cited literature), and via the resonant spin rotation using the NMR methods mentioned above, when, with a proper correction for the rotation of the Earth \cite{GrahamSpin2018}, the frequency of the pseudomagnetic field oscillations coincides with the frequency of the spin precession in the laboratory \cite{BudkerAxion,Grahamannurev,GramolinNature,Graham2021,BudkerGyroscope}. The search for the axion signal by analyzing the neutron EDM data, accumulated over a decade, on the spin precession of ultracold neutrons in a storage cell with a mercury, $^{199}$Hg, comagnetometer, see Sec.~\ref{fakeEDM}, was carried out in \cite{AbelAxion}. Here one looked for a temporal variation of the ratio of the precession frequencies of the neutron spin and the $^{199}$Hg comagnetometer,  expected in view of the strong Schiff suppression of the axion contribution to the EDM of an atom as compared to the contribution to the EDM of a neutron. Then the observed signal must be treated as the contribution of the axion field directly to the neutron EDM. The achieved sensitivity is at least six orders of magnitude lower than one needs to observe the QCD axion (the sensitivity estimates were criticized for an insufficient account for to the randomness of the phase of the axion field, see \cite{Roussy:2020ily}). Still  this analysis for the first time demonstrated a possibility of a laboratory study in the interval of the record-low axion masses $10^{-24}$ eV/$c^2 \leq m_{(a)} \leq 10^{-17}$ eV/$c^2$. A similar analysis of the 2016-2017 data on the search for the EDM of the $^{180}$Hf$^{19}$F$^+$ ion \cite{CairncrossHfF}, discussed above, was carried out in \cite{Roussy:2020ily}. In this case, the theory predicts that the oscillation of the $CP$-odd electron-nucleon coupling constant dominates in the axion signal in the low-mass region \cite{FlambaumPospelov,FlambaumSamsonovJHEP}. With due attention to the phase uncertainty of the axion field, the axion mass region $10^{-22}$ eV/$c^2\leq m_{(a)} \leq 10^{-15}$ eV/$c^2$ has been studied. In the covered mass interval, the sensitivity is even lower than that in the neutron experiment \cite{AbelAxion}. 
 		
However, one must bear in mind that the common axion \cite{SVZAxion,KimAxion,EricAxion,DineAxion} is only one of the possible candidates for the dark matter, and any new look at the manifestations of the dark matter is of interest in as broad as possible range of observables. Thus, in connection with axions, a new interest arose in revisiting  the Schiff shielding in oscillating external fields \cite{FlambaumResonant,TranTan,FlambaumTranTan}. Among new topics that grew out of the axion physics, are fresh looks at the possible time dependence of fundamental constants and masses due to interaction with the dark matter \cite{StadnikTimeVariation}, see also \cite{BudkerScalar2019,BudkerScalar2021} for the discussion of oscillating fundamental constants due to the galactic halo of scalar fields (relaxions). Let us also mention a possible detection of  the electron-axion interaction by observing the recoil electrons in the liquid xenon as a part of a XENON collaboration program for the search of the weakly interacting dark matter \cite{XENONaxion}.
 
\subsection{Search for axions in the storage ring experiments}\label{axionring}
 
Using the spin of particles in storage rings as an axion antenna belongs to the NMR class of experiments, although in a completely specific setup. The first such an experimental search for axions has been proposed by JEDI collaboration at the synchrotron COSY \cite{StephensonAxion,PretzAxion,JEDIaxion} and the results have been released in 2022 \cite{JEDIaxion2022}. The novelty of this approach is that spins in the accelerator move with ultra-relativistic velocities, so that one can expect an enhancement of the pseudomagnetic field in the accelerator orbit by the factor $c/v_{(a)} \approx 10^{3}$ as compared to the fields acting on static spins. This was pointed out in \cite{FrozEDMPRD}, but without elaborating implications for the search of axions. The first complete solution of the problem of using the spin as an axion antenna in this mode was given in \cite{SilenkoAxionAntenna}, with the FW transformation playing a crucial role.

The Hamiltonian corresponding to the equations (\ref{Diracgen}) and (\ref{DiracAxion}) in the Dirac representation has the form 
\begin{eqnarray}
{\mathcal H} = \beta mc^2 + c\bm{\alpha}\cdot\bm{\pi} + q\Phi + \mu'(i\bm\gamma\cdot\bm E
- \bm{\Pi}\cdot\bm B)&&\nonumber\\ -\,d(\bm \Pi\cdot\bm E+i\bm\gamma\cdot\bm B)
+ {\frac{\hbar\,g_f}{2f_{(a)}}}(c\bm{\Sigma}\cdot\bm{\nabla}a - \gamma_5\,\dot{a})\,,&& \label{eqelect}
\end{eqnarray}
where the notation is the same as in the equations (\ref{eq54}), (\ref{Hamlt}). We have already discussed the role of the axion contribution to the EDM. A new element is the contribution of the Weinberg interaction describing the axion wind. After a relativistic transformation to the FW representation, following the method \cite{JMP,Silenko2008,PRA2}, we find in the semiclassical approximation the corresponding contribution to the Larmor angular velocity of the spin rotation \cite{SilenkoAxionAntenna} 
\begin{equation}
\bm\Omega^{\rm ax} = {\frac{g_f}{f_{(a)}}}\left[{\frac{c\bm{\nabla}a}{\gamma}} + {\frac {\bm{v}}{c}}\left(
\dot{a} + {\frac{\gamma}{\gamma+1}}\,\bm{v}\cdot\bm{\nabla}a\right)\right].\label{eq15}
\end{equation}
 
As far as the cold axion field in the galactic halo is concerned, the terms $\propto {\bm \nabla} a(x)$ can be omitted. Then the total axion contribution, including the effects of the oscillating EDM and the pseudomagnetic field, to the instantaneous angular velocity of the spin rotation with respect to the momentum of the particle in a purely magnetic storage ring takes the form 
\begin{eqnarray}
\bm{\Omega}^{\rm ax} &=& {\frac{a_0}{f_{(a)}}}\,\Bigl[g_f\omega_{(a)}\sin(\omega_{(a)}t){\frac{\bm{v}}{c}}\nonumber\\
&& - \,\kappa_{(a)} \gamma \cos(\omega_{(a)} t)\frac{\bm v}{c}\times\bm{\Omega}_c\Bigr]\,.\label{AxionOmega}
\end{eqnarray}
Here, we have rewritten the EDM contribution $\propto {\bm v}\times {\bm B}$ in  the FT-BMT equation (\ref{Fukuyama1}) by substituting ${\bm B}$ in terms of the cyclotron angular velocity (\ref{Cyclotron}). The unambiguous relation of the two contributions in (\ref{AxionOmega}) is explained by a simple kinematic relationship, evident from (\ref{Fukuyama1}), between the electric field in the comoving particle's system and the magnetic field in a purely magnetic storage ring.

The sum (\ref{AxionOmega}) is tantamount to   endowing the static storage ring by the two radio-frequency spin rotators, which do not affect the orbital motion of the particle. The pseudomagnetic field rotates the spin from the vertical orientation into the horizontal one about particle's momentum, and as a spin rotator, it imitates a radio-frequency solenoid operating at a frequency of $\omega_{(a)}/2\pi$. The spin rotation rate is proportional to the field integral in the rotator \cite{Froissart,SpinTuneMapping,RathmannNikolaevSlimPRAB}. In this case, the longitudinal pseudomagnetic field acts along the entire circumference of the ring. The EDM contribution, which is expressed in terms of the cyclotron angular velocity ${\bm \Omega}_c$, also rotates the spin from the vertical to the horizontal position and vice versa, but  about the radial axis. It is equivalent to an RF Wien filter with a radial magnetic field.
 
Consider first the simplest example of the axion spin resonance for the purely magnetic bending of protons or deuterons. The axion signal during the slow energy, i.e., the cyclotron frequency $\Omega_c/2\pi$, scan  will be a spontaneous spin rotation in the vertical plane when the resonance condition is satisfied 
\begin{equation}
\Omega_s = G\gamma \Omega_c = \omega_{(a)}\, . \label{AxionResonance}
\end{equation}
In the scheme adopted in the JEDI experiment, the spin of the bunch of deuterons lies in the ring plane and the linear in time accumulation of the vertical polarization serves as an axion signal \cite{StephensonAxion,PretzAxion,JEDIaxion,AxionEDMPRD2019}. All the other RF spin rotators in the ring, including the RF Wien filter discussed in \cite{KimSemertzidis}, are better switched off.

In the spin resonance mode (\ref{AxionResonance}), the ratio of the two frequencies in the square brackets (\ref{AxionOmega}) reads
\begin{equation}
 \frac{g_f\omega_{(a)}}{\kappa_{(a)} \gamma \Omega_c} = {\frac{g_f G}{\kappa_{(a)}}}  \sim 10^2 G \gg  1\, .
\end{equation}
The coupling of the spin with the axion pseudomagnetic field turns out to be much more significant than the coupling of the axion contribution to particle's EDM with the comoving electric field. To this end we note that this impact of the pseudomagnetic field was overlooked in the early simulations of the sensitivity of spin in storage rings as an axion antenna \cite{PretzAxion}, so that the sensitivity to axions in such experiments was substantially underestimated. 
 
As we see from (\ref{AxionOmega}), the phases of the radial and longitudinal axion spin rotators differ by $\pi/2$. The description of the spin evolution induced by the axion pseudomagnetic field is simplified in a precession-linked reference system, which is rotating with the angular velocity $\Omega_s=G\gamma \Omega_c$. The spin precession is frozen in this rotating system; a description of the resonance evolution of the spin envelope is found in \cite{SpinTuneMapping,Silenko:2017a,silenko2017general,PretzAxion}. The amplitude of the axion signal would depend on the difference $\Delta$ of the axion field oscillation and the spin precession phases, while the angular velocity of the resonance rotation of the spin in the vertical plane is equal to
\begin{equation}
\Omega_{res} = {\frac{a_0}{2f_{(a)}}}\,{\frac{v}{c}}\,\gamma \,|g_f\,G - \kappa_{(a)}\,|\,\Omega_c\, \label{OmegaRes}
\end{equation}
and does not depend on the phase $\Delta$.

In the usually discussed scheme with the in-plane initial polarization \cite{StephensonAxion,PretzAxion,JEDIaxion,AxionEDMPRD2019}, the axion signal will be proportional to $\sin \Delta$ \cite{SpinTuneMapping,RathmannNikolaevSlimPRAB}. There is no way to control the phase $\Delta$ and a buildup of the vertical polarization by interaction with  the axion field would be irreproducible from one beam  fill to another. As a practical remedy, the JEDI collaboration invoked filling the ring with four bunches with different polarizations \cite{PretzAxion,AxionEDMPRD2019,JEDIaxion2022}. Ramping the magnetic field of the ring while keeping constant beam orbit, the JEDI experiment covered the spin precession frequency range from 119.997 kHz to 121.457 kHz, or an axion mass range of 4.95--5.02~neV$/c^2$. No signal of the axion-induced spin rotation has been observed. When interpreted in terms of the oscillating EDM of the deuteron, the JEDI has set the upper bound $d^\text{ax} < 5\cdot 10^{-23}\,e\cdot$cm \cite{JEDIaxion2022}.
  
Here we note that the problem of the phase $\Delta$ would not arise at all if one tracks the rotation of the initial vertical polarization caused by the axion in the direction of the ring plane \cite{KolyaNICA}. In this case, the axion field signal will be a linear in time growth of the amplitude of the precessing horizontal polarization, which can be measured by the method developed in \cite{SpinTuneContinuousJEDI}. Because of the short proton spin coherence time $\tau_{SCT}$  \cite{LLMNR}, the JEDI scheme can not be used for protons, while in the scheme with initial vertical polarization a signal of the oscillating horizontal polarization accumulated  for the time $\sim \tau_{SCT}$ can still be detected experimentally.
 
Only a limited range of spin precession frequencies above 110 KHz is accessible at the  magnetic storage ring COSY. To this end, the hybrid version of the PTR will be a unique broadband axion antenna in the low frequency domain. Specifically, here one starts with the zero proton spin precession frequency at the frozen spin point. Beyond this point, the electric and magnetic fields must be varied synchronously to  preserve the orbit radius and the injection energy. The angular velocity of the spin precession will be proportional to the change of the magnetic field from the frozen spin value, 
\begin{equation}
\Omega_s = \,{\frac{q}{mc}} {\frac{1+G}{\gamma^2}}\,|\Delta B|\,. \label{Scanning PTR}
\end{equation}
The attainable band of frequencies will depend on the range of magnetic fields tolerated  by the air magnetic winding of PTR and electric fields in the electrostatic deflectors \cite{AbusaifCYR}. The same mechanism of variation of the spin precession frequency with orbit retention shall work in all magnetic storage rings if a straight section is converted into static Wien filter. A practical solution for NICA, suggested recently in \cite{Senichev2022ide}, would be complementing the ring with long, $\sim 100$ m,  bypasses which can be operated without affecting the equipment in main rings. Here one can approximate the Wien filter with alternating magnetic dipoles and electrostatic deflectors suggested in the quasi-frozen spin approach \cite{ValetovQFS}. The crucial point is that this way the attainable band of spin precession frequencies can be expanded by  more than one order in magnitude compared to what was achieved by JEDI at COSY.
 
Active experiments on the subject have not yet resulted in a direct observation of axions and axion-like particles. But the intertwining of the most fundamental problems, from the $CP$ nonconservation in the quantum chromodynamics to the nature of the dark matter, justifies all efforts. Only the combined search for axions in astrophysical observations, in electromagnetic interactions, in experiments with static spins, and in the storage ring physics makes it possible to cover the entire interesting spectrum of axion masses. It is gratifying that the sensitivity of a number of experimental techniques approaches the threshold for testing the basic theoretical concepts and may exceed this threshold in the foreseeable future, albeit in a narrow frequency intervals, for the time being. 

\subsection{Axion effects in dynamics of spin in the gravitational field}\label{axionspin}
 
Since the gravitational interaction is universal, it should be taken into account in the context of the spin-axion problems. In Sec.~\ref{Spinclassic} and \ref{SpinQuantum} above, we considered the quantum and classical spin dynamics in external electromagnetic, gravitational, and inertial fields. The corresponding results can be generalized by adding an axion field to this list, with a minimal extension of the geometric formalism. Let us write down the generalization of the Dirac equation (\ref{Diracgen}) for a fermionic particle in external fields 
\begin{eqnarray}
\Bigl(i\hbar\gamma^\alpha D_\alpha - mc + {\frac{\mu'}{2c}}\sigma^{\alpha\beta}F_{\alpha\beta}
+ {\frac{d}{2}}\sigma^{\alpha\beta}\widetilde{F}{}_{\alpha\beta}&&\nonumber\\
-\,{\frac{\hbar}{2f_{(a)}}}g_f e^i_\alpha\,\partial_ia(t,{\bm x})\gamma^\alpha\gamma_5\Bigr)\Psi = 0,&&
\label{DiracAxion1}
\end{eqnarray}
taking into account the direct interaction of the particle with the axion field, which is described by the last term.

Quite remarkably, one can reformulate (\ref{DiracAxion1}) as the Dirac equation in the Riemann-Cartan space-time, in which the gravitational field is described by the two independent geometric structures, the curvature and the torsion, if we identify the torsion pseudovector with the covariant gradient of the axion field \cite{ObukhovAxion}: 
\begin{equation}\label{axitor}
\check{T}{}_\alpha = {\frac{2g_f}{f_{(a)}}} e^i_\alpha\,\partial_ia.
\end{equation}
The analysis of the general-relativistic Dirac theory in such a geometry \cite{OSTgen,OSTbaldin} shows that the equation (\ref{DiracAxion1}) reduces to the Schr\"odinger equation (\ref{sch}) with the Hamilton operator (\ref{HamiltonDP}), in which the two key objects (\ref{AB2})-(\ref{AB3}) are redefined as $\bm{\Xi}\rightarrow\bm{\Xi} + \bm{\Xi}^{\rm ax}$ and $\Upsilon\rightarrow\Upsilon + \Upsilon^{\rm ax}$, getting additional contributions from the axion field: 
\begin{eqnarray}
\bm{\Xi}_a^{\rm ax} &=& {\frac{2g_f}{f_{(a)}}}\,{\mathcal F}^b{}_a\partial_b a,\label{axiXi}\\
\Upsilon^{\rm ax} &=& {\frac{2g_f}{cf_{(a)}}}\left(\partial_ta + cK^b\partial_b a\right).\label{axiUps}
\end{eqnarray}
The gravitational field manifests its presence here through the components of the Schwinger tetrad, also encoded in the form of the object (\ref{AB1}).

In turn, this leads to the generalization of both the gravitoelectric $\bm{\mathcal E}\rightarrow\bm{\mathcal E} + \bm{\mathcal E}^{\rm ax}$ and the gravitomagnetic $\bm{\mathcal B }\rightarrow\bm{\mathcal B} + \bm{\mathcal B}^{\rm ax}$ fields in which the axion field adds to the usual expressions (\ref{ge}) and (\ref{gm}) contributions 
\begin{eqnarray}\label{axige}
{\mathcal E}_a^{\rm ax} \!&=&\! {\frac{\gamma cg_f}{f_{(a)}V}}\,\epsilon^{abc}\widehat{v}_b{\mathcal F}^d{}_c\partial_d a, \\
{\mathcal B}_a^{\rm ax} \!&=&\! -\,{\frac{\gamma cg_f}{f_{(a)}V}}\Bigl\{\!{\mathcal F}^b{}_a\partial_b a 
+ {\frac {\widehat{v}_a}{c^2}}(\partial_t a + cK^b\partial_b a)\!\Bigr\}.\label{axigm}
\end{eqnarray}

As a consequence, after the FW transformation, the spin precession angular velocity (\ref{OmegaT}) is modified $\bm{\Omega}\rightarrow\bm{\Omega} + \bm{\Omega}^{\rm ax}$ by the specific axion term \cite{ObukhovAxion}
\begin{eqnarray}
\bm{\Omega}^{\rm ax} = -\,{\frac {1}{\gamma}}\,\bm{\mathcal B}^{\rm ax} + {\frac {1}{1 + \gamma}}
\,{\frac {\widehat{\bm{v}}\times\bm{\mathcal E}^{\rm ax}}{c^2}}\,.\label{axiS}
\end{eqnarray}
It is important to emphasize that all these results are valid for any configurations of the electromagnetic, gravitational, and axion fields, which makes them applicable to any physical and astrophysical problems, including the case of the strong fields (for example, in the vicinity of compact massive objects).

The results obtained admit an equivalent formulation in the framework of the classical Frenkel-Thomas-BMT spin theory, in which the action of an axion field on the spin vector in the curved spacetime is described by the general-relativistic covariant equation 
\begin{equation}\label{DSax}
{\frac {DS^\alpha}{d\tau}} =  {\frac{g_f}{f_{(a)}}} (e^i_\gamma\,\partial_ia)\,\varepsilon^{\alpha\beta\gamma}\,S_\beta,
\end{equation}
where $\varepsilon^{\alpha\beta\gamma} = \varepsilon^{\alpha\beta\gamma\delta}U_\delta$, cf. Balakin-Popov \cite{Balakin} and Dvornikov \cite{Dvornikov:2019}. In the presence of the electromagnetic field, one should also add the usual terms from the right-hand side of Eq. (\ref{DSG}).

In the context of experiments in high-energy physics at accelerator laboratories located on the Earth, one needs to specialize from the general formalism to the conditions of the terrestrial gravity and rotation. In this case, the gravitational field is adequately described in the gravitoelectromagnetism approximation (\ref{gemvw}), (\ref{gemvwk}), and (\ref{LTter}). Using this approximation (and also taking into account the fact that for the Earth's gravity we can put $V = W = 1$ with very good accuracy), we derive
\begin{equation}
\bm{\Xi}^{\rm ax} = {\frac{2g_f}{f_{(a)}}}\,\bm{\nabla}a,\quad
\Upsilon^{\rm ax} = {\frac{2g_f}{cf_{(a)}}}\left\{\dot{a} + \bm{\omega}_\oplus\!\cdot\!(\bm{r}\times\bm{\nabla}a)
\right\},\label{axiUps1}
\end{equation}
and simplify (\ref{axige}) and (\ref{axigm}) to 
\begin{eqnarray}\label{axige1}
\bm{\mathcal E}^{\rm ax} \!&=&\! {\frac{\gamma cg_f}{f_{(a)}}}\,\widehat{\bm{v}}\times\bm{\nabla}a\,, \\
\bm{\mathcal B}^{\rm ax} \!&=&\! {\frac{-\,\gamma cg_f}{f_{(a)}}}\!\left[\bm{\nabla}a + {\frac
{\widehat{\bm{v}}}{c^2}}\left\{\dot{a} + \bm{\omega}_\oplus\!\cdot\!(\bm{r}\times\bm{\nabla}a)
\right\}\right]\!.\label{axigm1}
\end{eqnarray}

As a result, we find that the axion field in the terrestrial conditions leads to a correction in the spin motion \cite{ObukhovAxion}
\begin{eqnarray}
\bm{\Omega}^{\rm ax} &=& {\frac{g_f}{f_{(a)}}}\,\left[\,{\frac {c}{\gamma}}
\,\bm{\nabla}a + {\frac {\widehat{\bm{v}}}c}
\,\Bigl\{\,\dot{a} + \bm{\omega}_\oplus\cdot(\bm{r}\times\bm{\nabla}a) \right.\nonumber\\
&&\left. \qquad\ +\,{\frac {\gamma}{\gamma + 1}}\,{\widehat{\bm{v}}\cdot\bm{\nabla}a}
\Bigr\}\right].\label{axiO}
\end{eqnarray}
Noteworthy is the peculiar ``mixing'' of axion effects with inertial/gravitational ones.

In general, the conclusions for flat space are confirmed, with a correction due to the rotation of the Earth. In particular, the nonrelativistic Hamiltonian for the axion contribution, which we get in the FW picture for (\ref{DiracAxion1}), 
\begin{eqnarray}
{\mathcal H}{}_{FW}^{\rm ax} &=& -\,{\frac {\hbar}{2}}\,\bm{\Pi}\cdot\bm{\mathcal B}^{\rm ax} \nonumber\\
&=& {\frac {\hbar cg_f}{2f_{(a)}}}\,\beta\bm{\Sigma}\cdot\Bigl[\bm{\nabla}a + {\frac {\bm{p}_f}{m_fc^2}}
\,{\stackrel {(m)} {\partial_t}}a\Bigr]\,,\label{axiH}
\end{eqnarray}
agrees with (\ref{NonRelHamilt}), however, the rate of the change of the axion field is given by the material derivative ${\stackrel {(m)} {\partial_t}}a = \partial_ta + \bm{v}^{\rm rot}\cdot \bm{\nabla}a$, where $\bm{v}^{\rm rot} = \bm{\omega}_\oplus\times\bm{r}$ is the dragging velocity due to the motion of the frame, located on the rotating Earth. Thus, a longitudinal pseudomagnetic field acting on a spin can be generated not only by a time-dependent axion configuration, but also by a static inhomogeneous axion field.

Note that in addition to the direct influence of the Earth's gravity and rotation through the spacetime metric and the coframe components in the structure of the gravitomagnetic and gravitoelectric fields, the gravitational field implicitly manifests its presence also through the form of the axion field obtained as a solution of the scalar wave equation in the curved spacetime. The corresponding analysis of such effects was carried out by Stadnik and Flambaum \cite{StadnikAxion}, however, without taking into account the Earth's rotation. 

\section{Geometric magnetic field in electrostatic laboratory on a rotating Earth}\label{geo}

\subsection{Magnetic and electric fields in noninertial laboratory}\label{max}

So far, we have considered the spin dynamics in the prearranged external electric and magnetic fields. We now turn to the discussion of these fields as such in non-inertial reference frames under special boundary conditions. From the point of view of searches for the EDM, we are interested in the case of an electrostatic storage ring in the coordinate system $K$, attached to the rotating Earth, with a static distribution of electric charges and zero currents. It is clear that these charges move and create a magnetic field in the reference frame $K'$ of fixed distant stars. One needs to find out whether the magnetic field, measured in the terrestrial physical laboratory, would be non-zero.

Maxwell's theory on an arbitrary curved manifold in the most compact form is formulated in terms of differential forms \cite{Birk}: 2-forms of the field strength $F = F_{ij}dx^i\wedge dx^j/2$, 2-forms of the field excitations $H = H_{ij}dx^i\wedge dx^j/2$, and 3-form of the electric current $J = J_{ijk}dx^i\wedge dx^j\wedge dx^k/6$: 
\begin{eqnarray}\label{Max0}
dF = 0,\qquad dH = J,\qquad H = \sqrt{\frac {\varepsilon_0}{\mu_0}}\,{}^*F. 
\end{eqnarray}
This formulation goes back to Gustav Mie \cite{Mie} and Arnold Sommerfeld \cite{Sommerfeld}, and its great advantage is its universality: Maxwell's theory has the same form in all coordinates and frames of reference, and the general covariance is obvious. The system (\ref{Max0}) encompasses the homogeneous equation (the first equality), the inhomogeneous equation (the second equality), and finally the constitutive law (the last equality), which establishes the relationship between the components of the field strength tensor $F_{ij}$ and the excitation tensor $H_ {ij}$, where the star ${}^*$ denotes the Hodge dualization operation. In components, the constitutive relation reads 
\begin{equation}\label{MaxHF}
H^{ij} = \sqrt{\frac {\varepsilon_0}{\mu_0}}\,{\frac {\sqrt{-g}}{2}}\varepsilon^{ijkl}\,F_{kl}.
\end{equation}
As usual, $\varepsilon_0, \mu_0$ are the electric and magnetic constants of the vacuum, and the quantity $\sqrt{\mu_0/\varepsilon_0} \approx 377\,$ohm characterizes the so-called vacuum impedance. Note also that $c = 1/\sqrt{\varepsilon_0\mu_0}$.

In practice, the use of the local coordinates $x^i = (t, \bm{x})$ leads to the identification of the components of the field strength tensor $F_{ij} = (\bm{E}, \bm{B})$ and the excitation tensor $H_{ij} = (\bm{D}, \bm{H})$ as the electric and magnetic fields, and the current component $J_{ijk} = (\rho, \bm{J})$ as the charge density and the electric current density, which turns the system (\ref{Max0}) into the familiar Maxwell equations
\begin{eqnarray}
\bm{\nabla}\times \bm{E} + \dot{\bm{B}} = 0,\qquad \bm{\nabla}\cdot\bm{B} = 0,\label{Max1}\\
\bm{\nabla}\times \bm{H} - \dot{\bm{D}} = \bm{J},\qquad \bm{\nabla}\cdot\bm{D} = \rho.\label{Max2}
\end{eqnarray}
Here the dot denotes the partial derivative with respect to the coordinate time $t$, and the components of the vector operator $\bm{\nabla}$ have the usual meaning of partial derivatives with respect to the spatial coordinates $\bm{x}$.

Remarkably, Maxwell's equations in the gravitational field have the form (\ref{Max1})-(\ref{Max2}) of electrodynamics in a medium\footnote{See the corresponding discussion in \S\,90 of the book \cite{LLvol2}.}, however, the properties of this inhomogeneous and anisotropic ``medium'' are determined not by the physical matter, but by the spacetime geometry. In particular, the electric permittivity and magnetic permeability tensors are constructed from the metric components (\ref{LT}), and their explicit form is encoded in the constitutive relation (\ref{MaxHF}). In addition, the off-diagonal components of the metric (that is, $\bm{K}\neq 0$) play a special role, being responsible for the {\it magnetoelectric} phenomena which are described by the emergence of the electric polarization in response to an applied magnetic field, and the magnetization caused by an applied electric field. The possibility of magnetoelectric effects was first predicted by Landau and Lifshitz (see \S\,51 in \cite{LLvol8}) and experimentally confirmed in experiments by Astrov \cite{Astrov} for the class of material substances identified by Dzyaloshinskii \cite{Dzyal} from the magnetic symmetry analysis.

Here we will focus, based on the results of \cite{VergelesJHEP,VergelesJETP}, on the theoretical discussion of specific magnetoelectric effects under the conditions of terrestrial gravity and rotation, when the spacetime geometry is described by the metric (\ref{LT}) in the gravitoelectromagnetic approximation (\ref {gemvw}), (\ref{gemvwk}).

Local coordinates, as such, do not have a direct physical meaning, and, as a consequence, the components of the electric and magnetic fields $\bm{E}, \bm{B}$ with respect to the coordinate basis are not observable quantities. In the laboratory, only the field components (\ref{EBun}) are measurable with respect to the (anholonomic, in general) basis of the local Lorentz frame of reference, which is determined by the corresponding tetrad. To distinguish the coordinate objects from the physical objects, we use a different font $\bm{\mathfrak E}, \bm{\mathfrak B}$ for the anholonomic components of the electromagnetic field strength tensor $F_{\alpha\beta}=e_\alpha^i e_ \beta^j F_{ij}$. For the class of problems under consideration, it is more convenient to switch from the Schwinger gauge (\ref{Sgauge}) to a tetrad in the Landau-Lifshitz gauge (\ref{Lgauge}). They differ by a Lorentz transformation, and a direct calculation yields
\begin{equation}
\begin{split}
e^{\widehat 0}_0 &= \widetilde{V},\quad e^{\widehat 0}_a = \widetilde{W}\,{\frac {2{\mathcal A}_a}{c^3}},\quad
e^{\widehat b}_a = W\,\delta^b_a + 2\,{\frac {{\mathcal A}^b{\mathcal A}_a}{c^4}},\\
\widetilde{V} &= 1 - {\frac {{\mathit \Phi}}{c^2}} + {\frac {2\,|{\mathcal A}|^2}{c^4}},\quad
\widetilde{W} = 1 + {\frac {3{\mathit \Phi}}{c^2}} - {\frac {2\,|{\mathcal A}|^2}{c^4}}.
\end{split}\label{cofLL}
\end{equation}
Here $|{\mathcal A}|^2 = \delta^{ab}{\mathcal A}_a{\mathcal A}_b$, and the gravitoelectromagnetic potentials for the terrestrial conditions are given by the expressions (\ref{LTter}).

Using this coframe, we obtain a relation between the anholonomic physical fields and the coordinate ones 
\begin{eqnarray}
\bm{\mathfrak E}_a &=& {\frac {1}{\widetilde{V}}}\,e^b_{\widehat a}\,\bm{E}_b,\label{EELL}\\
\bm{\mathfrak B}^a &=& {\frac {1}{\det e}}\,e^{\widehat a}_b\biggl(\bm{B}^b
- {\frac {2\widetilde{W}}{c^3\widetilde{V}}}\,[\bm{E}\times \bm{\mathcal A}]^b\biggr),\label{BBLL}
\end{eqnarray}
where $\det e = \det e^{\widehat a}_b$, and $e_{\widehat a}^b$ is the inverse 3-frame to $e^{\widehat a}_b$.

The constitutive relation (\ref{MaxHF}) has the most transparent form in terms of the physical fields: 
\begin{eqnarray}\label{DE} 
\bm{D}^a &=& \varepsilon_0\biggl(\det e\,e^a_{\widehat b}\,\bm{\mathfrak E}{}^b - {\frac
{\widetilde{W}}{c}}\,\epsilon^{abc}\,e^{\widehat d}_b\,\bm{\mathfrak B}_d\,\bm{\mathcal A}_c\biggr),\\
\bm{H}_a &=& {\frac {1}{\mu_0}}\,\widetilde{V}\det e\,e^{\widehat b}_a\bm{\mathfrak B}_b.\label{HB}
\end{eqnarray}
The last terms in (\ref{DE}) and in (\ref{BBLL}) as well are responsible for the magnetoelectric effect induced by the non-inertiality (due to the rotation of the Earth) of the laboratory reference frame: the electric field generates a magnetic field \cite{VergelesJHEP,VergelesJETP}. If there is no rotation ($\bm{\mathcal A} = 0$), the effect disappears. The corresponding magnetic field in the terrestrial laboratory (\ref{BBLL}) we will call the geometric magnetic field $\bm{\mathfrak B}_\omega$ (where the symbol ``${}_\omega$'' is not an index, but shows the origin of the field).

The qualitative conclusion is confirmed by the analysis of Maxwell's equations. Let us consider the case of static configurations for zero currents $\bm{J} = 0$. Then the first of the equations of the inhomogeneous system (\ref{Max2}) can be easily integrated by the ansatz
\begin{equation}
\bm{H} = \sqrt{\frac {\varepsilon_0}{\mu_0}}\,\bm{\nabla}\psi\,,\label{Hpsi}
\end{equation}
where the constant factor is introduced for the convenience. Substituting this into (\ref{HB}), we then use (\ref{BBLL}) to find the coordinate magnetic field 
\begin{equation}
\bm{B} = {\frac 1{c\widetilde{V}}}\biggl(\bm{\nabla}\psi + {\frac {2\widetilde{W}}{c^2}}
\,\bm{E}\times \bm{\mathcal A}\biggr),\label{Bsol}
\end{equation}
and then from the homogeneous system (\ref{Max1}) we derive the equation for the scalar function 
\begin{equation}
\bm{\nabla}\cdot\left[(\bm{\nabla}\psi)/\widetilde{V}\right] - {\frac {2}{c^2}}
\,\bm{E}\cdot\bm{\nabla}\times(\bm{\mathcal A}\widetilde{W}/\widetilde{V}) = 0.\label{Epsi1}
\end{equation}
Substituting here the gravitomagnetic potential (\ref{LTter}), with an account of the smallness of the gravitational radius of the Earth, see (\ref{eq:GravRad}), the last equation can be simplified to the Poisson equation 
\begin{equation}
\Delta\psi = -\,{\frac{2}{c}}\,\bm{E}\cdot\bm{\omega}_\oplus.\label{Poisson_Magn_Geom}
\end{equation}
This result was obtained in \cite{VergelesJHEP,VergelesJETP}, and the above general approach was developed in \cite{ObukhovEfield}. The electric field $\bm{E}$ entering the right-hand side can be calculated in the first approximation neglecting the rotation of the Earth. The complete system of equations should be solved perturbatively, using the Earth's angular velocity as a small parameter. In the lowest order, from (\ref{EELL}) and (\ref{HB}), (\ref{Hpsi}) we have $\bm{\mathfrak E} = \bm{E}$ and $\bm{ \mathfrak B}_\omega = \bm{\nabla}\psi/c$.

For what follows, the symmetry properties of the geometric magnetic field are of primary importance. As an axial vector, it changes the sign upon an inversion of another axial vector, the angular velocity $\bm{\omega}_\oplus$. Of course, we cannot force the Earth to rotate in the opposite direction. But the geometric field $\bm{\mathfrak B}_\omega = \bm{\nabla}\psi/c$ also changes its sign when the polar vector of the electric field in the terrestrial laboratory changes sign, which can have practical consequences. 

\subsection{Geometric magnetic field and experimental searches for the EDM}\label{geoEDM}

\subsubsection{Charged sphere on rotating Earth}\label{sphere}

As an illustration, let us consider the case of a charged conducting sphere with the radius $a$ and the total electric charge $Q$ resting on the rotating Earth. Using the electric field $\bm{E}(\bm{r})$ of such a sphere, the equation (\ref{Poisson_Magn_Geom}) is solved exactly. The corresponding geometric magnetic field is 
\begin{eqnarray}
\bm{\mathfrak B}_\omega(\bm{r}) &=& \bm{\mathfrak B}_{\rm dip} 
- {\frac {\bm{E}({\bm r})\times (\bm{\omega}_\oplus\times\bm{r})}{c^2}},\label{BQ}\\
\bm{\mathfrak B}_{\rm dip} &=& {\frac {\mu_0}{4\pi}}\times\left\{
\begin{split}
3\,{\frac {(\bm{\mathfrak m}\cdot\bm{r})\,\bm{r}}{r^5}} &- {\frac {\bm{\mathfrak m}}{r^3}}, &  r>a,\\
{\frac {2\,\bm{\mathfrak m}}{a^3}}, & &  r<a.
\end{split} \right.\label{IE40}
\end{eqnarray}
Here the first term in (\ref{BQ}) describes the configuration (\ref{IE40}) created by the magnetic dipole moment 
\begin{equation}
\bm{\mathfrak m} = {\frac {Q\,a^2\,\bm{\omega}_\oplus}{3}} \label{momsphere}
\end{equation}
of the charged sphere rotating with the Earth. The second term in (\ref{BQ}) manifests the result of the Lorentz transformation with the local velocity $\bm{\omega}_\oplus\times\bm{r}$ from the inertial frame $K'$ to the rotating Earth frame $K$. Here the smallness parameter of the geometric field is $\omega_\oplus r/c$, cf. with (\ref{eq:EtaOmega}). 

\subsubsection{Fake EDM in the search for the EDM of ultracold neutrons}\label{fakeEDM}

In searches for of the EDM of neutrons, UCNs are contained in a storage cell in the uniform and parallel electric and magnetic fields. In this case, the EDM is extracted from the frequency shift $\Delta f$ of the neutron spin precession (\ref{PrecessFreq}) after the inversion of the electric field  $\bm{E}_0$, 
\begin{equation}
d_n = {\frac {\pi\hbar\Delta f}{2|\bm{E}_0|}} \,,\label{eq:EDM extraction}
\end{equation}
under the assumption that the inversion of the electric field does not affect the magnetic field. But this is clearly violated by the geometric magnetic field.

The UCN storage cell can be considered as a flat capacitor. The one-dimensional problem has a simple solution (the $z$ axis is chosen along the electric field inside the cell, in the median plane $z=0$) 
\begin{equation} 
\bm{\mathfrak B}_\omega = \left(0,\,0,\,-\,{\frac{2\omega^z_\oplus E_0}{c^2}}\,z\right)
= -\,{\frac{2\omega_\oplus^z\,z}{c^2}}\,\bm{E}_0 \label{IIiE30}
\end{equation}
with the constant gradient inside the cell 
\begin{equation}\label{eq:Gradient}
{\frac{d\,\bm{\mathfrak B}_\omega}{d\,z}} = -\,{\frac{2\omega_\oplus^z}{c^2}}\,\bm{E}_0\,.
\end{equation}
Here $\omega_\oplus^z$ is the projection of the angular velocity vector onto the $z$ axis (= direction of the electric field).

In searches for the EDM of neutrons, the frequency of the neutron spin precession is measured relative to that of the mercury atoms serving as a comagnetometer. The mercury Hg atoms are uniformly distributed in the cell volume and for the mercury comagnetometer the mean geometric magnetic field is equal to zero: $\langle \bm{\mathfrak B}_\omega^{({\rm Hg})}\rangle =0$. On the other hand, the center of mass of the neutron gas is displaced with respect to the center of mass of the mercury by $\langle z \rangle$, which leads to a nonvanishing mean geometric magnetic field acting on the magnetic moments of neutrons, 
\begin{gather}\label{eq:HgeomNeutron}
\langle \bm{\mathfrak B}_\omega^{(n)}\rangle = -\,{\frac{2\langle z\rangle
\omega_\oplus^z}{c^2}}\,\bm{E}_0 .
\end{gather}
It changes the sign when the electric field in the cell is inverted, and generates a false EDM signal \cite{VergelesJHEP}
\begin{gather}
d_{\rm fake} = -\,{\frac{2\langle z \rangle \omega_\oplus^z}{c^2}}\,\mu_n\, .\label{eq:dFalse}
\end{gather}
In the experiment \cite{pendlebury2015revised}, the displacement of the neutron center of mass was $\langle z \rangle \simeq 2.8$ mm, in the recent experiment \cite{PSIEDM} it was $\langle z \rangle \simeq 3.9$ mm. Using the last value, we find $d_{\rm fake}\approx 2.5\times 10^{-28}$ $e\cdot$cm. This is still small as compared to the most accurate experimental result $d_n = (0.0\pm 1.1_{\rm stat} \pm 0.2_{\rm sys})\times 10^{-26}$ $e\cdot$cm \cite{PSIEDM}, but it will become significant in the next-generation experiments under discussion with a sensitivity up to $d_n \sim 10^{-28}$ $e\cdot$cm \cite{n2EDM}. Note also that in neutron cells with a typical height $h \sim 15$ cm, the contribution of the false EDM along the cell height varies in a broad interval
\begin{equation}\label{eq:dFalseSpread}
\Delta d_{\rm fake} = \pm\,{\frac{h\omega_\oplus^z}{c^2}}\mu_n \simeq \pm\,6\times 10^{-27} e\cdot\rm{cm}.
\end{equation}

\subsubsection{Geometric magnetic field in electrostatic proton storage ring}\label{geoproton}

The electrostatic proton storage ring is a cylindrical capacitor-deflector with a narrow gap $\delta \ll h$, where $h$ is the height of the electrodes \cite{srEDM,AbusaifCYR}. The radius of the storage ring is negligibly small compared to the radius of the Earth. The solution of the two-dimensional electrostatic problem in the gap between the electrodes is well known, 
\begin{equation}
\bm{E}_0 = -\,\bm{\nabla}\Phi(r) = -\,{\mathcal E}_0\,{\frac{\rho\bm{r}}{r^2}},
\qquad \Phi(r) = {\mathcal E}_0\,\rho \ln{\frac{r}{\rho}}\,,\label{IIE10}
\end{equation}
where $\rho$ is the median radius. Outside the gap between the electrodes, the electric field disappears. From the point of view of an observer in the $K'$ system of distant stars, static charges in the laboratory create opposite currents in the $K'$ system and generate the magnetic field in the gap between the electrodes. The speed of charge's motion determines the small parameter (\ref{eq:EtaOmega}), which is four orders of magnitude greater than $\eta_{p}^{\rm edm}\sim 10^{-15}$ in the planned proton storage rings \cite{srEDM, AbusaifCYR,YannisHybrid}.

In a storage ring located at the north or south poles of the Earth in a system of distant stars, the geometric magnetic field is equal to 
\begin{equation}
\bm{\mathfrak B}'_\omega(\bm{r}) = {\frac{\bm{v}(\bm{r})\times\bm{E}_0(\bm{r})}{c^2}},
\label{IIE20}
\end{equation}
where $\bm{v}(\bm{r}) = \bm{\omega}\times\bm{r}$. For an experimentalist in the terrestrial laboratory, it is compensated by the Lorentz transformation to the laboratory system. But such a complete compensation is absent at an arbitrary latitude. 

Referring to \cite{VergelesJETP,VergelesJHEP} for the complete solution, we write out the final result for the geometric magnetic field between the storage ring electrodes: 
\begin{eqnarray}
\bm{\mathfrak B}_\omega &=& {\frac{{\cal E}_0\rho}{c^2}}\Bigg\{\bm{\omega}_{\rm T}\,\ln
\left(\frac{r}{\rho}\right) + {\frac 12}\,\bm{\omega}_{\rm T} - {\frac {(\bm{r}\cdot\bm{\omega}_{\rm T})
		\,\bm{r}}{r^2}}\Bigg\}\nonumber \\
&\simeq& {\frac {{\mathcal E}_0\rho}{2c^2}}\left(\bm{\omega}_{\rm T}
- {\frac {2(\bm{r}\cdot\bm{\omega}_{\rm T})\,\bm{r}}{r^2}}\right),\label{IIE1300}
\end{eqnarray}
where $\bm{\omega}_{\rm T}$ is the projection of the angular velocity of rotation of the Earth onto the plane of the storage ring. Here, at the last step, we neglected the value of $|\log(r/\rho)| < \delta/(2\rho) \ll 1$.

The background magnetic fields are a main headache in the planned experiments to search for the EDM of protons in all-electric storage rings with the frozen proton spin \cite{srEDM,AbusaifCYR}. Modern technologies allow for the very radical shielding of the Earth's magnetic field $\bm{B}_\oplus$, which is directed along the magnetic meridian and on the scale of the storage ring can be considered as homogeneous with the constant projection onto the ring plane.

The Earth's magnetic field $\bm{B}_\oplus$ and the geometric magnetic field $\bm{\mathfrak B}_\omega$ differ significantly in that the geometric magnetic field cannot be shielded by magnetic shields. We take the magnetic meridian as the $y$ axis, so that the projection of the field onto the plane of the accelerator ring $\bm{B}_\oplus^{\rm T} = (0,\, B_\oplus^{\rm T})$ . Unlike the Earth's magnetic field, the geometric magnetic field is quadrupole along the particle's orbit, $\bm{\mathfrak B}_\omega = {\mathfrak B}_\omega\,(\sin 2\theta, \cos 2\theta )$. The position of the particle in the orbit is determined by the angle $\theta$, so that $\bm{r} = r\,(\cos\theta,\, -\sin \theta)$.

In all-electric rings, the most dangerous are the radial magnetic fields in the comoving system. In the above two cases, they are $B_\oplus^{(r)} = (\bm{r}\cdot\bm{B}_\oplus)/r = -\,B_\oplus\sin\theta$ and $ {\mathfrak B}_\omega^{(r)} = (\bm{r}\cdot\bm{\mathfrak B}_\omega)/r = {\mathfrak B}_\omega\,\sin\theta$. According to \cite{AbusaifCYR}, the rotation of proton's spin per revolution in the first approximation is proportional to the integral $\oint d\theta\, B_\oplus^{(r)}$. Both the Earth's magnetic field and the geometric magnetic field have the property 
\begin{equation}
\oint d\theta\,B_\oplus^{(r)}  = \oint d\theta\, {\mathfrak B}_\omega^{(r)} = 0.
\label{eq:EDMlike}
\end{equation}
To the first approximation, the geometric magnetic field does not give rise to a false EDM signal, but the question of a possible geometric Berry phase, discussed in \cite{AbusaifCYR}, calls for a scrutiny.

\section{Gravitational quantum anomalies and dynamics of dense hadron matter}\label{anomaly}
		
Non-central heavy ion collisions are the source of the quark-gluon and hadron matter, which has a huge angular momentum \cite{Kharzeev:2004ey} and moves with huge accelerations \cite{Kharzeev:2005iz}.

To estimate the scale of the quantities corresponding to these phenomena, it is useful to compare them with the macroscopic inertial effects, which are in the focus of this review. It turns out that the (local) angular velocity of rotation of the strongly interacting matter is 25 orders of magnitude higher than the angular velocity of rotation of the Earth, and the acceleration is several orders of magnitude higher than the acceleration of free fall \cite{Teryaev:2020syj}.

Indeed, one can estimate the local angular velocity $\Omega$ by assuming that velocity's change is about the speed of light $c$ on the scales of the size of a nucleus $R_A$. Its ratio to the angular velocity of the Earth's rotation (\ref{oE}) can be conveniently represented
\begin{equation}
\eta_{\rm rot} = {\frac {\Omega}{\omega_\oplus}} = {\frac{c}{R_A}}\cdot{\frac {T_\oplus}{2\pi}}
= {\frac{1}{2\pi}}\cdot{\frac{c T_\oplus}{R_A}}\approx 10^{27}
\end{equation}
as the ratio of light day (the distance traveled by the light during the Earth's revolution around its axis $T_\oplus$, and approximately 150 times greater than its distance from the Sun) to the size of the nucleus.

As noted above, the non-central collision of heavy ions leads to the generation of a large (on a microscopic scale) angular momentum of the order of $10^4\hbar$ \cite{Kharzeev:2004ey}, about $10\%$ of which is accumulated in the resulting quark-gluon medium (or in the hadronic one, depending on the collision energy) \cite{Baznat:2013zx}. At the same time, it is precisely the differential rotation described by the local angular velocity (vorticity) that is essential for the    transfer of the angular orbital momentum into the spin one. One can relate the estimate of acceleration to the angular velocity estimate by multiplying and dividing the obvious expression for it by $T_\oplus/2\pi$: 
\begin{equation}
\eta_{\rm acc} = {\frac{c}{R_A}}\cdot {\frac {c}{g_\oplus}} = \eta_{\rm rot}
\,{\frac{2\pi c}{T_\oplus g_\oplus}} \approx 10^{30}.
\end{equation}
An additional factor $\sim 2000$ is proportional to the ratio of the speed of light to the speed acquired during the day when moving with the acceleration $g_\oplus$.

Another representation of the quantity $\eta_{\rm acc}$ can be obtained by using the relation between the free fall acceleration and the first cosmic velocity $v_\oplus$: 
\begin{equation}
g_\oplus = \frac{v_\oplus^2}{R_\oplus},\nonumber
\end{equation}
so that
\begin{equation}
\eta_{\rm acc} = {\frac{c}{R_A}}\cdot {\frac {c}{g_\oplus}} =\frac{c^2}{v_\oplus^2} \cdot \frac{R_\oplus}{R_A}.
\end{equation}

According to the equivalence principle, the presence of large accelerations and angular velocities corresponds to the study of huge gravitational fields, while it can be added that the above estimate for the acceleration \cite{Teryaev:2020syj} corresponds to an ``effective'' gravitational energy of the order of the rest energy, analogous to the black hole or to the (flat) universe born from the vacuum. In this sense, the effects of gravity and cosmology can be studied not only in condensed matter physics \cite{Volovik:2000ua}, but also in the physics of heavy ion collisions. In other words, in order to obtain comparable accelerations due to the gravity, the Planck mass should become of order of a hadron mass.

Of significant interest is the key question of how a rotating and accelerated medium that exists on small distances during small periods of time can manifest itself in the measurements carried out by a detector at rest. The first possibility is related to the quantum measurement which plays quite a practical role in the case of the motion of the spin. Indeed, if we consider the spin simply as a directed segment in a rotating frame of reference, then it obviously rotates in this frame with the angular velocity equal in magnitude to the angular velocity of rotation of the frame, so that the equivalence principle (one of the formulations of which is the identity of the classical and quantum rotators) is trivially satisfied. The situation changes if one takes into account the quantum nature of spin and its measurement by a device located in a rotating reference frame \cite{Teryaev:2020syj}. In this case, the dynamics of the quantum spin, which coincides with the classical one, is a nontrivial consequence of the equivalence principle. Thus, the consideration of the spin dynamics in a noninertial frame of reference, besides the practical value, connects such fundamental areas of physics as the gravity and the theory of quantum measurements.

Can the spin of a particle be considered as ``measured'' in a rotating quark-gluon medium? A typical process, used in polarization measurements in a detector, is the weak decay resulting in a characteristic anisotropy of its products. The most important in this case is the decay of $\Lambda \to p \pi^-$, since the yield of $\Lambda$ hyperons in heavy ion collisions is quite large.

A possible influence of the rotation of the medium on this decay, in which this rotation is manifest, can be viewed similarly to the influence of an external, in particular, magnetic field. At the moment, there is no reason to regard this influence, if it exists, as a significant one. At the same time, the relation of the rotation of the medium, treated as a classical system, with the quantum spin, controlled by the angular momentum conservation law, can apparently be considered as a measurement.

The establishment of a thermodynamic equilibrium of the spin with the rotating medium is one of the main methods for calculating the polarization \cite{Becattini:2020ngo}. In this case,  the treatment in terms of the 4-velocity $v_\mu$ of the local equilibrium and the relativistic invariance underlies the interpretation of the four-dimensional tensor $\partial_\mu v_\nu - \partial_\nu v_\mu$ as the relativistic angular velocity, which, in addition to the vorticity $\bm{\omega} = \bm{\nabla}\times\bm{v}$ also includes the acceleration $a$ which is usually not considered in the case of global equilibrium. Since the corresponding quantities enter the Gibbs distribution in combination with the temperature, the latter is taken into account in terms of the so-called thermal vorticity containing the four-dimensional temperature vector $\beta^\mu \equiv v^\mu /T$. The use in computations of essentially quantum objects, such as the Wigner function and the Zubarev density matrix, in essence, allows one to consider the establishment of the local equilibrium as a kind of quantum-mechanical measurement. One can also recall here the proposal found in the classic textbook \cite{LLvol5} (\S 8) that the increase of entropy and the irreversibility of time are related to the processes of the quantum-mechanical measurement.

Another way to describe the effect of rotation on spin is to consider hydrodynamics as an effective theory, with the relativistic invariance leading to the treatment of four-dimensional velocity as a gauge field \cite{Sadofyev:2010is}. Indeed, the presence of a conserved charge with the density $\rho$ and the corresponding chemical potential $\mu$ leads to the appearance of a term in the Lagrangian 
\begin{equation}
\mu \rho  = \mu j^0  \to \mu v_\alpha j^\alpha\,.
\end{equation}
The above analogy between the gauge field and $\mu v^\alpha$ should not be taken literally. By virtue of gauge invariance, in QED and QCD the observable quantities  are the field strengths rather than the field potentials. In contrast to that, $\mu v^\alpha$ is in princile a measureable quantity.

The corresponding vertex leads to the appearance of new diagrams, among which the so often mentioned in the review triangle anomaly \cite{Son:2009tf, Sadofyev:2010is} plays a special role. This is related to its protection against the perturbative (according to the Adler-Bardeen theorem) and the nonperturbative (due to the 't Hooft correspondence principle) corrections. As a result, one can derive the corresponding contribution for hadrons\footnote{We are talking about the anomalous contribution to the current, while there is no corresponding contribution from the effective theory in the expression of its divergence, related to the behavior of the fundamental theory at small distances.} by considering the triangle diagram at the quark level.

In particular, it was proposed \cite{Rogachevsky:2010ys} to apply this effect for the description of the polarization of hyperons in heavy ion collisions, which is a natural analogue of the anomalous gluon contribution \cite{Efremov:1989sn}, manifested in the analysis of the so-called ``spin crisis''. In this case, the four-dimensional velocity of the medium starts to play the role of the gluon field, with the vorticity playing the role of the (color)magnetic field. The emergence of the chemical potential as a coupling constant made it possible to make a qualitative prediction \cite{Rogachevsky:2010ys} about the rapid decrease of the polarization with the energy. Later, the magnitude of this effect was estimated \cite{Baznat:2013zx}    to be of the  order of $1\%$ for the energies at the NICA complex under construction in Dubna. This value is in agreement with the published 4 years later experimental result \cite{STAR:2017ckg}, obtained by the STAR collaboration during the beam energy scanning at the RHIC collider.

We derive the polarization by using \cite{Baznat:2013zx, Sorin:2016smp} the axial charge, calculated at the quark level, 
\begin{equation}
Q_5^s = N_c \int d^3 x \,c_V \gamma^2\,\bm{v}\cdot(\bm{\nabla}\times\bm{v}),
\end{equation}
where the coefficient $c_V$ contains the contribution of the strange chemical potential $\mu_s$: 
\begin{equation}\label{cv}
c_V = {\frac{\mu_s^2}{2 \pi^2}} + {\frac{T^2}{6}}.
\end{equation}

In hydrodynamics, the role of the topological current is played by the hydrodynamic helicity, related \cite{Baznat:2016xiy} to the topological properties of flows and chaos. The quark-hadron duality (or the 't Hooft principle) in this case puts forward an alternative (complementary, in Bohr's sense) possibility to calculate this charge by considering all strange hadrons (hyperons and vector mesons) with spin, among which it is distributed. At the same time, since the axial current and the charge are charge-even, while the number of antihyperons is significantly smaller than the number of hyperons, one finds \cite{Baznat:2017jfj} a natural explanation for an excess \cite{STAR:2017ckg} of the $\bar{\Lambda}$ polarization over the $\Lambda$ polarization. It should be emphasized that in this approach, the local thermodynamic equilibrium is applicable to the charge described by the corresponding chemical potential, whereas the spin dynamics is described by the effective theory and its thermodynamic equilibrium with the medium does not arise.

The anomaly mechanism described above results in a contribution proportional to $\mu_s^2$ in (\ref{cv}), while the term proportional to $T^2$ is associated with the {\it holographic gravitational} anomaly \cite{Landsteiner:2011cp}. The non-renormalization arguments do not apply to it, and in particular, the lattice calculations \cite{Braguta:2014gea} indicate the suppression of the coefficient by an order of magnitude due to collective effects. This makes it possible to explain the smallness of the polarization at high energies and to achieve a better description at low energies \cite{Baznat:2017jfj}.

One of the dynamic mechanisms, realizing the generation of an axial current in a medium, are the quantized vortices in the superfluid pion liquid, near the axis of which baryon degrees of freedom should be excited, which leads to the polarization of baryons \cite{Teryaev:2017wlm}. The dissipation, characteristic for such a process, is analogous to the absorbing phases needed for the polarization generation. Since this effect is (``naively'') $T$-odd, the phases mimic the true $T(CP)$ invariance violation, and a special care must be taken to avoid them in experiments, as in the case of the total cross section for the scattering on tensor-polarized deuterons in Sec.~\ref{CPbeyond}. In QCD, such phases can arise due to contributions from the higher twist \cite{Efremov:1984ip,Qiu:1991pp} or the Wilson lines \cite{Brodsky:2002cx,Collins:2002kn,Boer:2003cm}.

Since the pion field is formally quite similar to the axion field, one can assume that the appearance of vortices in an axion fluid can lead to the polarization of fermions, similar to the effects discussed in section~\ref{axion}.

The existence of different mechanisms for describing the polarization (statistical and anomalous) raises a question of the correspondence between them. In this regard, the problem of calculating the axial current in the statistical approach arises. Using the Wigner function method \cite{Becattini:2020ngo}, one can obtain the expression \cite{Prokhorov:2017atp} corresponding to the anomalous current. Such a relation between the statistical physics and the field theory may seem unexpected. However, it should be noted that the pioneering derivation of the corresponding expression by A. Vilenkin \cite{Vilenkin:1980fu} makes use of Green's functions in a noninertial reference frame and it also does not appear to be directly related to the anomaly. 

The statistical approach to the anomalous current \cite{Prokhorov:2018qhq,Prokhorov:2018bql}, which is based also on Zubarev's density matrix, allows one to consider the angular velocity and acceleration as the real and imaginary chemical potentials, respectively. In this case, the characteristic for the chemical potential threshold effect for the angular velocity indicates a decrease of polarization when the vorticity is of the order of the (effective) mass. At the same time, when the vorticity is much larger than the fermion mass, an anomalous current is reproduced. The anomalous contribution can be therefore compared to the statistical one for the massless quarks.

It should be also emphasized that while the statistical calculation for the polarization yields the result that depends directly on the momentum of the polarized particle, to derive the current one needs an integration over the momentum which yields an expression that depends on the coordinate. This explains the need to integrate also over this coordinate and to use the axial charge to establish the quark-hadron duality. One can therefore say that in addition to the quark-hadron duality (complementarity), there is also a complementarity between the coordinate and momentum pictures. While the statistical method should lead to polarizations of hyperons of the same sign and close magnitude, the sign and magnitude in the anomaly method depend on their quark structure, which opens up fundamental possibilities for the experimental verification, that is part of the research program at the NICA complex.

The statistical approach also allows one to study such a characteristic quantum field-theoretic phenomenon as the Unruh effect \cite{Kharzeev:2005iz,Becattini:2017ljh,Prokhorov:2019cik}. It is interesting that the use of the acceleration (related to the temperature as $T=a/2\pi $) as an independent variable in (\ref{cv}) results in the same degree of the $\pi$ factor in the $T$ and $\mu$ terms. At the same time, and independent treatment of $a$ and $T$ leads to an instability \cite{Prokhorov:2019hif} for $T < a/2\pi$. This is natural, since the equilibrium temperature in the accelerated reference frame must be greater than the Unruh temperature. An unstable state can arise at high accelerations in heavy ions collisions and, in a certain sense, this corresponds to a fall into the black hole, concluded by a decay into thermal states.

The statistical formulas (let us emphasize, in the flat spacetime!) also correspond \cite{Prokhorov:2019yft} to the effects caused by conical singularities, and the instability can be interpreted as a  transformation of a cone into a plane.

Thus, the study of the heavy ion collisions provides an opportunity to indirectly investigate the extremely strong effective gravity and its dual description in statistical physics and effective field theory. One can use the polarization of different hadrons (to compare the field-theoretic and the thermal mechanisms) as an appropriate observable, along with the thermalization dynamics in different regions of the phase space (to study possible instability at high accelerations).

The gravity thereby manifests itself in the accelerator physics as a genuine field (in precision experiments due to the terrestrial gravity and rotation) and as an effective disguise \cite{Kharzeev:2005iz} (due to the rotation and acceleration of the quark-gluon matter), also modeling physics near the black hole horizon \cite{Prokhorov:2019cik} and even a fall into a black hole \cite{Prokhorov:2019hif}. The investigations of these manifestations in heavy-ion collisions experiments present a new challenging problem.

\section{Conclusions}\label{conc}

Precision spin experiments for testing fundamental symmetries provide  unique perspectives for the investigation of the variety of topical physical problems, from solving the riddle of the baryogenesis to establishing the nature of the dark matter in the Universe. The focus is on the search for $CP$- and $T$-noninvariant EDMs of neutral atoms and ultracold neutrons, molecules, as well as charged particles and nuclei. These searches have already achieved a spectacular sensitivity to the EDM that is 12 orders of magnitude in the case of neutrons, and 18 orders of magnitude in the case of electrons, lower than the respective dipole moments allowed by all discrete symmetries. A target of the next generation experiments is a further increase of the sensitivity by another one or two orders of magnitude. The search for the EDM of charged protons and nuclei is possible only in accelerator experiments. After a series of works on the precision spin dynamics by the JEDI collaboration at the COSY storage ring, at the forefront is a construction of the PTR prototype storage ring with both all electric and hybrid bending of protons with the energy of 30-45 MeV. This will be a prologue to the construction of an electrostatic proton storage ring with the spin frozen at the magic energy of 233 MeV, and with a potential sensitivity to the proton EDM of $d_p \sim 10^{-29}\,e\cdot$cm, which will exceed a sensitivity of experiments with ultracold neutrons. An active analysis of possible searches is under way for the EDM of protons and deuterons at the NICA collider at JINR. All these efforts are aimed at searches for new mechanisms of the $CP$-invariance violation beyond the minimal Standard Model, that failed to describe quantitatively the observed baryon asymmetry of the Universe. As a rule, these new mechanisms imply an expansion of the spectrum of particles. At the discussed levels of accuracy, precision searches for EDM can significantly exceed the sensitivity of direct searches for new particles at colliders. The direct influence of the gravity on EDMs is negligible. However, given that our laboratories are located on a gravitating and rotating Earth, the role of gravitational effects in the spin dynamics, at the anticipated levels of accuracy, even exceeds the role of the proton EDM. The similar interweaving of the problems of cosmology and $CP$-{nonconservation} takes place in the physics of an axion as the most probable explanation of the nature of the dark matter, and here the particle spin in accelerators can act as a kind of an axion antenna. It is noteworthy that some of the discussed spin effects have analogues in condensed matter physics. The analysis of these new aspects of the spin dynamics was the main subject in our review. 

{\bf Acknowledgments.}
The authors are grateful to the Russian Federation Foundation for Basic Research for the support of this review with the grant 20-12-50190 ``Expansion''. N.N.N. and A.J.S. thank the participants of the srEDM, JEDI, and CPEDM collaborations for many years of the joint work on the subject of the review. We would like to specially mention the fruitful discussions with V.I. Zakharov, A.A. Starobinsky, F. Rathmann, R. Talman, A. Wirzba, P. Leniza, J. Pretz, Y. Semertzidis, A. Saleev, Y. Senichev and J. Slim. A great stimulus for interest in the topic was the NICA collider program at JINR. The research of Y.N.O. was supported by the Russian Federation Foundation for Basic Research under grant 18-02-40056-mega (NICA); the studies of O.V.T. were supported by the Russian Federation Science Foundation under grant 21-12-00237; the research of A.J.S. was supported by the Chinese Academy of Sciences under the PIFI grant 2019VMA0019. 

\bibliographystyle{apsrev4-1}
\bibliography{Nikolaev_Bibliography_EN_2022nov01}

\end{document}